%% file: main.tex
\documentclass[11pt,letterpaper]{article}

\makeatletter
\def\input@path{{./}{../}{arXiV/}{./sections/}{arXiV/sections/}}
\makeatother

\usepackage[margin=1in]{geometry}
\usepackage[utf8]{inputenc}
\usepackage[T1]{fontenc}
\usepackage{lmodern}
\usepackage{microtype}
\usepackage{mathtools}
\usepackage{amssymb}
\usepackage{amsthm}
\usepackage[dvipsnames]{xcolor}
\usepackage{graphicx}
\usepackage{mdframed}
\usepackage{needspace}
\usepackage{enumitem}
\usepackage{algpseudocode}
\usepackage{dsfont}
\usepackage{stmaryrd}
\usepackage{mleftright}
\usepackage{bm}
\usepackage{xspace}
\usepackage{booktabs}
\usepackage{tabularx}
\usepackage{multirow}
\usepackage{tikz}
\usepackage{wrapfig}
\usepackage{caption}
\usepackage{subfiles}
\usepackage[numbers,sort&compress]{natbib}
\usepackage[hidelinks,pdfusetitle,hypertexnames=false]{hyperref}
\usepackage[capitalize,nameinlink,noabbrev]{cleveref}

\graphicspath{{./}{../}{./figures/}{./sections/}{arXiV/}{arXiV/figures/}{arXiV/sections/}}
\newcommand{\email}[1]{\href{mailto:#1}{\nolinkurl{#1}}}

\numberwithin{equation}{section}
\counterwithin{figure}{section}
\counterwithin{table}{section}

\ifdefined\newcounteralias\else
  \usepackage{thmtools}
\fi
\theoremstyle{plain}
\newtheorem{theorem}{Theorem}[section]
\newtheorem{lemma}[theorem]{Lemma}

\newtheorem{corollary}[theorem]{Corollary}

\newtheorem{observation}[theorem]{Observation}

\newtheorem{claim}[theorem]{Claim}
\theoremstyle{definition}
\newtheorem{definition}[theorem]{Definition}

\theoremstyle{remark}
\newtheorem{remark}[theorem]{Remark}
\usepackage{thmtools}
\AddToHook{cmd/@thm/before}{\Needspace{3\baselineskip}}

\crefname{figure}{Figure}{Figures}
\crefname{table}{Table}{Tables}
\crefname{theorem}{Theorem}{Theorems}
\crefname{lemma}{Lemma}{Lemmas}
\crefname{conjecture}{Conjecture}{Conjectures}
\crefname{proposition}{Proposition}{Propositions}
\crefname{corollary}{Corollary}{Corollaries}
\crefname{coro}{Corollary}{Corollaries}
\crefname{claim}{Claim}{Claims}
\crefname{observation}{Observation}{Observations}
\crefname{obs}{Observation}{Observations}
\crefname{remark}{Remark}{Remarks}
\crefname{definition}{Definition}{Definitions}
\crefname{example}{Example}{Examples}
\crefname{section}{Section}{Sections}
\crefname{appendix}{Appendix}{Appendices}

\definecolor{seagreen}{RGB}{46,139,87}

\newcommand{\eps}{\varepsilon}

\newcommand{\R}{\mathcal{R}}

\newcommand{\N}{\mathcal{N}}

\newcommand{\mayank}[1]{}
\newcommand{\arturo}[1]{}
\newcommand{\vasilis}[1]{}
\newcommand{\pasin}[1]{}
\newcommand{\gibeom}[1]{}
\newcommand{\adarsh}[1]{}

\newcommand{\cA}{{\mathcal A}}
\newcommand{\cB}{{\mathcal B}}
\newcommand{\cbx}{{\mathcal G}}

\newcommand{\cD}{{\mathcal D}}
\newcommand{\cE}{{\mathcal E}}
\newcommand{\cF}{{\mathcal F}}
\newcommand{\cG}{{\mathcal G}}

\newcommand{\cI}{{\mathcal I}}

\newcommand{\cO}{{\mathcal O}}

\newcommand{\cS}{{\mathcal S}}
\newcommand{\cT}{{\mathcal T}}

\newcommand{\EE}{{\mathbb E}}

\newcommand{\NN}{{\mathbb N}}

\newcommand{\PP}{{\mathbb P}}

\newcommand{\RR}{{\mathbb R}}

\DeclareMathOperator{\avg}{avg}
\DeclareMathOperator{\diam}{diam}
\DeclareMathOperator{\conv}{conv}

\DeclareMathOperator{\Ball}{Ball}

\DeclareMathOperator{\poly}{poly}

\DeclareMathOperator{\Var}{Var}
\DeclareMathOperator{\E}{\mathbb{E}}

\let\N\Natural

\let\R\Real
\let\Z\Integer

\newcommand{\dummy}{\mathrm{dummy}}
\newcommand{\dfstprob}{\textsc{Decision Farthest Spanning Trees}\xspace}
\newcommand{\dfsprob}{\textsc{Decision Farthest String}\xspace}
\newcommand{\kclique}{\textsc{$k$-Clique}\xspace}

\def\dif#1{\mathop{d #1}}

\def\floor#1{\left\lfloor #1 \right\rfloor}
\def\ceil#1{\left\lceil #1 \right\rceil}

\def\set#1{\left\{ #1 \right\}}
\def\abs#1{\left| #1 \right|}

\newcommand{\prn}[2][\!]{\mleft({#2}\mright)}%
\def\brk#1{\left[ #1 \right]}
\def\1{\mathds{1}}
\def\indic#1{\1\llbracket #1 \rrbracket}

\def\midd{\:\middle|\:}

\newcommand{\Prob}[1]{\PP\left(   #1   \right)}

\newcommand{\Arb}{\mathsf{Arb}}

\newcounter{question}
\newenvironment{question}
  {%
    \refstepcounter{question}%
    \begin{center}%
    \itshape Question~\thequestion.\ %
    \ignorespaces
  }
  {%
    \end{center}%
  }

\title{The Power of Sampling for Finding Far-Apart Bases in Matroids}

\author{Mayank Goswami\thanks{Department of Computer Science, Queens College, CUNY (\email{Mayank.Goswami@qc.cuny.edu}).}
\qquad Vasilis Livanos\thanks{Department of Computer Science, University of Southern California (\email{vas.livanos@gmail.com}).}
\qquad Pasin Manurangsi\thanks{Google Research (\email{pasin@google.com}).}\\[.4em]
Arturo Merino\thanks{Department of Computer Science, Universidad de Chile (\email{amerino@dcc.uchile.cl}).}
\qquad GiBeom Park\thanks{CUNY Graduate Center (\email{gpark1@gc.cuny.edu}).}
\qquad Adarsh Srinivasan\thanks{Department of Computer Science, Rutgers University (\email{adarsh.srinivasan@rutgers.edu}).}}
\hypersetup{pdfauthor={Mayank Goswami, Vasilis Livanos, Pasin Manurangsi, Arturo Merino, GiBeom Park, Adarsh Srinivasan}}

\date{}

\begin{document}
\maketitle

\begin{abstract}
We study the geometry of matroids via the \emph{diameter} and \emph{dispersion} of the metric space defined by the bases of a matroid. Given a matroid $M$ we consider its basis-exchange graph $\mathcal{G}(M)$ and ask if its diameter can be approximated by sampling two uniformly random bases. As our main result we show that the expected distance of two uniformly sampled bases is always at least $1/16$ of the diameter of $\mathcal{G}(M)$. While this might naturally appear to be a consequence of the edge-expansion properties established in the celebrated work of Anari, Liu, Gharan and Vinzant (STOC'19), we show that this intuition is false in general by constructing a graph $G$ with edge-expansion at least $1$ for which sampling two random vertices does \emph{not} give a constant-factor approximation to the diameter of $G$.

We then turn to the problem of finding $k\geq 2$ maximally-diverse bases of a matroid, namely the $k$ (max--min) diversity problem, where $k$ is part of the input.
When $k = 2$ the problem is solvable via matroid intersection.
However, for $k \geq 3$ no polynomial-time approximation algorithm or hardness result is known, even for the well-studied case of finding diverse spanning trees of a graph (i.e., in a graphic matroid).
We first show that sampling $k\geq 3$ bases does \emph{not} provide an $\mathcal{O}(1)$ approximation to the above problem: in fact, it provides a tight $\Theta(\log k)$ approximation. We then combine structural insights about the local density of $\mathcal{G}(M)$ with classical LP-rounding techniques to obtain a polynomial-time $1/6$-approximation algorithm for the $k$-diversity problem in arbitrary matroids, obtaining the first constant-factor polynomial-time approximation for max--min diversity on any matroid.

A key ingredient for our result is the \emph{Farthest Matroid Basis} problem, in which we are given a collection of bases and seek a new basis that is as far as possible from all of them. We give a polynomial-time $1/3$-approximation algorithm for this problem, which serves as the main algorithmic primitive behind our general result for $k$ max--min diversity. We also show that this problem is computationally hard in a parameterized sense; specifically, that Farthest Matroid Basis is $W[1]$-hard when parameterized by the optimal distance. Finally, we obtain stronger guarantees for important special cases. For \emph{partition matroids} and for \emph{graphic matroids} of constant arboricity, we give polynomial-time $1/2$ and $1/4$ approximation algorithms, respectively, for the $k$-diversity problem.
\end{abstract}

\clearpage
\subfile{sections/intro.tex}
\subfile{sections/preliminaries.tex}
\subfile{sections/diameter.tex}

\subfile{sections/diversity.tex}

\subfile{sections/special_cases.tex}

\subfile{sections/hardness.tex}

\appendix

\subfile{sections/appendix.tex}

\subfile{sections/acknowledgments}
\bibliographystyle{plainnat}
\bibliography{references}

\end{document}

%% file: sections/intro.tex
\section{Introduction}\label{sec:intro}

The  \emph{diameter} of a metric space, defined as the maximum distance between two points, is a fundamental quantity. It has been a source of rich theoretical results, and has been studied both qualitatively and computationally for many metric spaces of interest to computer science:  polytopes in $\mathbb{R}^{d}$ \cite{brieden1998approximation}, weighted and unweighted graphs~\cite{roditty2013fast,backurs2018towards}, flip/exchange graphs of various combinatorial and geometric objects: binary trees~\cite{sleator1986rotation}, spanning trees~\cite{edmonds1965lehman,roskind1985note,arbelaez2018constraint,clausen2009finding}, matchings~\cite{sanita2018diameter,wulf2025computing} and Hamiltonian cycles~\cite{krarup1995peripatetic}, triangulations of point sets~\cite{pournin2014diameter,eppstein2007happy}, and solutions to CNF formulae~\cite{feige2011diameter,crescenzi2002hamming,austrin2025algorithms}. 

A related measure called \emph{dispersion}, or \emph{diversity}, generalizes diameter: given $k \geq 2$, we can consider $k$ points in the metric space that maximize the minimum pairwise distance between them. 
For $k=2$ this distance is the diameter.
Owing to several applications, \emph{computing diverse solutions to optimization problems} has also generated substantial recent interest~\cite{baste2019fpt,baste2022diversity,hanaka2022computing,hanaka2023framework,fomin2023diverse}. Here the metric space of interest is the space of all `solutions' to some optimization problem\footnote{These solutions can range from being just feasible, to optimal, or approximately optimal: e.g., spanning tree, minimum spanning tree, or approximate minimum spanning tree, respectively. The analogous problems are diverse spanning-trees, diverse-MSTs, and diverse approximate MSTs.}, and one wants algorithms that return $k$ most-diverse solutions in this space.
The classical theory of error-correcting codes can also be viewed as studying $k$ most-diverse codewords (albeit with some structure) in the Hamming space $\{0,1\}^n$; we elaborate more on this connection later on.

We will focus on \emph{matroids}.
A matroid $M=(E,\cI)$ is a tuple comprised of a finite ground set $E$ and a down-closed family of subsets $\cI\subseteq 2^E$ called \emph{independent sets} satisfying the augmentation axiom.\footnote{For precise definitions see Section~\ref{sec:preliminaries}.} 
The maximal independent sets in $M$ are called bases.
We will denote $n:=|E|$.

Our metric space of interest is the \emph{basis-exchange graph $\cbx(M)$ of $M$}. 
Its vertices are the bases of a matroid and its edges correspond to single-element exchanges. 
In this graph, the distance between two bases $B$ and $B'$ is $d(B,B') = |B \setminus B'| = |B' \setminus B|$. We remark that if the size of the ground set $E$ of the matroid $M$ is $n$, typically the size of $\cbx(M)$ could be exponential in $n$.
We also study the special case of \emph{graphic} matroids, where the input is a graph $G$, and we consider the diameter and diversity of the single-element exchange graph whose vertices correspond to spanning trees of $G$.
If $G$ is weighted, then the family of minimum-weight bases of the associated graphic matroid is the set of bases of another matroid.
Consequently, all our results apply to the family of minimum spanning trees of a weighted graph.

Given a matroid $M$, computing the diameter of $\cbx(M)$ is exactly a matroid intersection problem between $M$ and its dual $M^*$~\cite{MR2849819}.
This gives an easy polynomial-time algorithm to compute the diameter~\cite{roskind1985note,hanaka2021finding}. 
However, arguably the \emph{simplest thing one could do is to sample two bases of a matroid uniformly at random}. 
The fundamental work of~\cite{anari2019log} shows how to sample a basis from a distribution (arbitrarily) close to uniform in polynomial time, and borrowing the analogy from coding theory, where random codes exhibit good rate-distance tradeoffs, perhaps \emph{just sampling two bases provides an approximation to the diameter}. 
This is the first question we consider.

\begin{question}
    \label{q:how-far}
    How far are two random bases of a matroid?
\end{question}

In other words, what structure must a metric space possess, for the distance between two uniformly-random points in it to approximate its diameter? Clearly this cannot hold for the metric space consisting of $n-1$ points in a small cluster and one point far away, or a clique with a chain attached to it; random sampling will sample in the cluster/clique whp. What about convex bodies in $\mathbb{R}^{n}$? In~\cite{brieden1998approximation} it was shown that while there is a PTAS for approximating the \emph{volume} of a convex body using random sampling~\cite{dyer1991random}, any reasonable approximation of the \emph{Euclidean diameter} must require exponentially (in $n$) many samples. In fact, this lower bound can also be shown to apply to approximating the \emph{combinatorial diameter} (the distance on the edges) of a polytope. Does $\cbx(M)$, whose vertices can be identified with the vertices of the matroid polytope in $\mathbb{R}^{n}$, behave more like the convex bodies in~\cite{brieden1998approximation} and their combinatorial counterparts, or like the Hamming cube?

Next, we consider the \emph{diversity} problem: we want $k \geq 3$ ``most-diverse'' bases in $\cbx(M)$, i.e., $k$ bases that maximize the minimum pairwise distance between them.\footnote{We remark that this is completely different from the \emph{diversity maximization under matroid constraints} problems studied in~\cite{abbassi2013diversity}.}
A classical result by Edmonds~\cite{edmonds1965lehman} characterizes which matroids achieve the best possible diversity of $\mathrm{rank}(M)$ (i.e., disjoint bases, since $\mathrm{rank}(M)$ is the size of a base), and polynomial time algorithms to find $k$-disjoint bases (if they exist) are known~\cite{roskind1985note}. What if the matroid does not admit $k$ disjoint bases, but we want to find the $k$ most-diverse bases in $\cbx(M)$?  

Foundational work on finding the $k$ most-diverse bases of a matroid was conducted by Fomin et al.~\cite{fomin2023diverse} who gave a $2^{\cO(d k^2 \log{(k \cdot d))}} \poly(n)$ exact algorithm: here $d$ is the optimal distance between the $k$ most-diverse bases. 
They also prove that the (decision) version of deciding if there exist $k$ bases with distance at least $d$ is NP-Hard, but only when $k$ is given in binary; therefore, this does not rule out polynomial time algorithms that return the $k$ most-diverse bases in time $\poly(n,k)$. Recently, Golovach, Majumdar and Saurabh~\cite{golovach2026polynomial} consider FPT algorithms for the diverse-bases problem restricted to graphic matroids, and under additional parameters on the number of leaves and internal nodes of the spanning trees. As mentioned, our focus is on polynomial time approximation algorithms.

Another line of work studies $k$ bases maximizing the average distance (as opposed to the minimum distance): this is generally considered an easier problem, and polynomial time approximation algorithms are known~\cite{hanaka2021finding,gao2022obtaining}. However, it has been observed that the solutions for this version may be clustered and therefore not diverse in the true sense: one may return the union of two clusters of size $k/2$ each centered at the two diameter points and obtain an approximation~\cite{baste2019fpt}. 

In this work, we will be concerned with the max-min version of returning the $k$ most-diverse bases of a matroid. Unlike the diameter ($k=2$), \emph{the tractability of this problem when $k\geq 3$ is given as part of the input is famously open}~\cite{fomin2023diverse,hanaka2023framework,gao2022obtaining,borodin2017max} - no hardness result, or even a polynomial-time $\cO(1)$-factor approximation algorithm, is known.
This brings us to our second question.

\begin{question}
    \label{q:diverse}
    Is there a $\poly(n,k)$-time (approximation) algorithm for finding $k \geq 3$ diverse bases of a matroid, or finding $k \geq 3$ Diverse Spanning Trees of a graph?
\end{question}

While NP-completeness of the max-min variant (when $k$ is given in unary) is still open, the following observation hints at the difficulty of computing the \emph{exact} solution: for partition matroids whose parts encode alphabet positions, the problem of finding the $k$ most-diverse bases is closely related to the open problem in coding theory of computing the maximum number of binary codewords with pairwise distance at least~$d$~\cite{ostergard2011size}. 
Therefore, we will focus on polynomial time \emph{approximation} algorithms for returning $k$ most-diverse bases.

\paragraph{Notation and Algorithmic Perspectives.} 
Before we describe our results, we introduce some notation used throughout: $M$ is the given matroid, $\cbx(M)$ denotes its basis exchange graph on $M$, $n$ denotes the number of elements of the matroid, $\cB(M)$ is the collection of all bases of $M$, $U(\cdot)$ denotes the uniform distribution, $\diam_2(M)$ is the diameter of $\cbx(M)$, and $\diam_k(M)$ is the minimum of the $\binom{k}{2}$ pairwise distances between the $k$ most-diverse bases of $M$, also referred to as the $k$-diversity.

We remark that our algorithms assume only \emph{independence-oracle access} to the underlying matroid: given a set $S\subseteq E$, the oracle decides whether $S$ is independent in time $\poly(n)$. 

\subsection{Our Contributions}

Our contributions are summarized in~\Cref{tab:approximation-summary}.

\begin{table}[htbp]
\centering
\small
\renewcommand{\arraystretch}{1.15}
\setlength{\tabcolsep}{4pt}
\begin{tabularx}{\linewidth}{@{}l >{\raggedright\arraybackslash}X c >{\raggedright\arraybackslash}X l@{}}
\toprule
{\bf Setting} 
& {\bf Restriction} 
& {\bf Appx.} 
& {\bf Running time/notes} 
& {\bf Reference+Proof} \\[3pt]
\midrule

Diameter
& all matroids
& $1/16$
& two samples, in expectation
& {\footnotesize Thm.~\ref{thm:constant-diameter-lower}+Sec.~\ref{ss:approx-via-2-sampling}} \\[3pt]

\midrule

\multirow{4}{*}{FMB}

& all matroids
& $1/3$
& $4^{\diam_2(M)}\poly(n,k,\log(1/\delta))$
& {\footnotesize Lem.~\ref{lem:fmb-sampling}+Sec.~\ref{sss:fmb-uniform}} \\[3pt]

& $\diam_2(M)\ge 4\varepsilon^{-2}\log(k/\delta)$
& $1-\varepsilon$
& $\poly(n,k)$
& {\footnotesize Lem.~\ref{lem:fmb-ptas-large-diameter}+Sec.~\ref{sss:ptas-fmb}} \\[3pt]

& all matroids
& $1/3$
& $\poly(n,k,\log(1/\delta))$
& {\footnotesize Thm.~\ref{thm:fmb-approx}+Sec.~\ref{sss:aprox-fmb}} \\[3pt]

& all matroids
& $1-\varepsilon$
& $\poly(n,k,\log(1/\delta))\, n^{\cO(\varepsilon^{-2}\log k)}$
& {\footnotesize Thm.~\ref{thm:fmb-qptas}+Sec.~\ref{sss:aprox-fmb}} \\[3pt]

\midrule

\multirow{5}{*}{DMB}
& all matroids
& $\Theta(\frac{1}{\log k})$
& $k$ samples, no replacement, in expectation
& {\footnotesize Thm.~\ref{thm:k-diversity-upper}+Sec.~\ref{ss:diversity-via-naive-samples}} \\[3pt]

& all matroids
& $1/6$
& $\poly(n,k,\log(1/\delta))$
& {\footnotesize Cor.~\ref{cor:gonzalez-and-FMB}+Sec.~\ref{ss:diverse-consequence}} \\[3pt]

& all matroids
& $(1/2)-\varepsilon$
& $\poly(n,k,\log(1/\delta))\, n^{\cO(\varepsilon^{-2}\log k)}$
& {\footnotesize Cor.~\ref{cor:gonzalez-and-FMB}+Sec.~\ref{ss:diverse-consequence}} \\[3pt]

& partition matroids
& $1/2$
& $\poly(n,k,\log(1/\delta))$
& {\footnotesize Thm.~\ref{thm:special-vertex-tran}+Sec.~\ref{sec:vertex-transitive}} \\[3pt]

& graphic and constant arboricity
& $1/4$
& $\poly(n,k,\log(1/\delta))$
& {\footnotesize Thm.~\ref{thm:bounded-arbocity}+Sec.~\ref{sec:graphic-matroid}} \\[3pt]

\bottomrule
\end{tabularx}
\caption{Summary of our approximation guarantees for diameter, farthest matroid basis (FMB), and Diverse Matroid Bases (DMB).
All results are randomized and the results that depend on $\delta$ succeed with probability at least $1-\delta$.
The column reference+proof indicates the precise statement referenced plus the section where the reader can find its proof.
}
\label{tab:approximation-summary}
\end{table}

\paragraph{Uniform Sampling.} Our first result answers Question~\ref{q:how-far} with a structural result: we prove that, for every matroid, the distance between two random bases is a constant fraction of the matroid's diameter in expectation. 

\begin{restatable}{theorem}{lowerdiameter}
\label{thm:constant-diameter-lower}
For every matroid $M$, we have
\[
    \E_{B_1, B_2 \sim U(\cB(M))}[d(B_1, B_2)] \geq \frac{1}{16} \diam_2(M).
\]
\end{restatable}

We also show a not-so-distant upper bound.

\begin{restatable}{theorem}{higherdiameter}
\label{thm:constant-diameter-upper}
For any $\eps > 0$, there exists a transversal matroid $M = M_\eps$ for which
\[
    \E_{B_1, B_2 \sim U(\cB(M_\eps))}[d(B_1, B_2)] \leq \prn{\frac{4}{9} + \eps} \diam_2(M_\eps).
\]
\end{restatable}
The reader familiar with the groundbreaking results by Anari et al.~\cite{anari2019log} may expect that the first theorem above is a consequence of their resolution of the Mihail-Vazirani conjecture on \emph{edge expansion} of basis exchange graphs of matroids: for any matroid $M$ the edge expansion of $\cbx(M)$ is at least one. 
Could it be that \emph{all} graphs with high edge expansion have the property in Theorem~\ref{thm:constant-diameter-lower}?

Surprisingly, we show that this intuition is misleading: there exists a family of graphs with high edge-expansion for which the distance between two uniformly random vertices is \emph{not} a constant approximation to the graph diameter. 
In fact, it is easy to construct such a graph on $N$ vertices with maximum degree $\Omega(N)$, but such a graph is not a representative basis-exchange graph: 
For an $n$-element matroid the basis-exchange graph has maximum degree at most $\cO(n^2)$ and diameter at most $\cO(n)$.
Thus, for matroid families with $N = \exp(\Omega(n))$ many bases, these quantities are $\cO(\poly\log(N))$.
Can we build a graph with high edge expansion, logarithmic degree and diameter, for which Theorem~\ref{thm:constant-diameter-lower} does not hold?

Let $\Delta(G)$ denote the maximum degree of $G$, and let $h(G)$ denote the edge expansion of $G=(V,E)$, i.e., $ \displaystyle h(G) \coloneqq \min {|\partial S|}/{|S|} $ among all $ S\subseteq V $ such that $ 0<|S| \leq |V|/2 $, where $\partial S$ is the set of edges crossing the cut~$(S,V\setminus S)$.

\begin{restatable}{theorem}{EdgeExCounter}
\label{thm:edge-expansion-counterexample}
    For every constant $c \geq  1$, there exists an increasing sequence of simple connected graphs $\{G_\ell\}_{\ell \in \NN}$ such that for all $\ell\in \NN$: $\Delta(G_\ell) = \cO(\log |V(G_\ell)|)$, $\diam (G_\ell) = \Theta(\log |V(G_\ell)|)$ and $h(G_\ell) \geq 1$ for which if $x_\ell,y_\ell$ are vertices chosen uniformly at random from $G_\ell$, we have
    \[ \PP\left(d(x_{\ell}, y_{\ell}) \geq  \frac1c \diam(G_{\ell})\right) \rightarrow 0 \ \ \text{as} \ \  {\ell} \rightarrow \infty. \]
\end{restatable}

\paragraph{Diversity.}
Theorem~\ref{thm:constant-diameter-lower} suggests a natural extension to address Question~\ref{q:diverse}: just sample $k$ bases uniformly at random. 
More formally, denote by $\avg_k(M)$ the expected minimum distance between $k$ bases sampled uniformly at random without replacement.
We show that this gives a logarithmic approximation to diversity.

\begin{restatable}{theorem}{kdiversityupper}
\label{thm:k-diversity-upper}
If $k \geq 2$ and $M$ is a matroid with at least $k$ bases, we have $\diam_k(M)\leq \cO(\log k) \avg_{k}(M)$.
\end{restatable}

The previous analysis is tight as shown by instances with hypercube structure. 

\begin{restatable}{theorem}{kdiversitylower}
\label{thm:k-diversity-random-tight}
Let $r$ be even, and let $M = \bigoplus_{i=1}^r U_{1,2}$ be the direct sum of $r$ parallel edges.
If $k=2^{r/2}$, then
\[
\diam_k(M) = \Theta(\log k) \avg_k(M).
\]
\end{restatable}
Therefore, a different approach is indeed needed to obtain a $\cO(1)$ approximation for $k \geq 3$. 
Given the limits of uniform sampling above, we consider a more ``geometric'' or distance-aware sampling\footnote{An analogous phenomenon was noticed by~\cite{austrin2025algorithms} where uniform sampling was found inadequate for the purpose of returning diverse satisfying assignments to a CNF formula, and an approximate farthest-first insertion was deployed.}, referred to as the greedy/farthest-first permutation~\cite{gonzalez1985clustering}: Given a set of bases $\cS \subseteq \cB(M)$, the \emph{Farthest Matroid Basis (FMB)} problem asks for a base $B$ maximizing $d(B,\cS) = \min_{A\in\cS} d(B,A)$. Our main algorithmic guarantees are the following.

\begin{restatable}{theorem}{ThmFmbApprox}\label{thm:fmb-approx}
    Let $M$ be a matroid, $\cS \subseteq \cB(M)$ be a set of $k$ bases and $\delta\in(0,1)$.
    There is a randomized algorithm that returns a base $B$ satisfying
    \[
    d(B,\cS) \geq \frac{1}{3} \max_{B' \in \cB(M)} d(B',\cS)
    \]
    with probability at least $1-\delta$ in time $\poly(n,k,\log(1/\delta))$.
\end{restatable}
Combining this with the farthest-insertion paradigm due to Gonzalez~\cite{gonzalez1985clustering}, which loses an extra factor of $2$, gives our first constant-factor, a (1/6)-approximation for $k$-diversity on arbitrary matroids. This is the first polynomial time approximation algorithm for $k$-diversity on matroids (including graphic matroids); in fact, to the best of our knowledge, the only existing result of this type is on the diverse median string selection problem~\cite{chakraborty2026maximizing}.

What about improving the constant $1/6$? Note that if one uses an algorithm for FMB to get $k$-diversity, the best approximation for $k$-diversity one can hope for is a 1/2. 
For the $k$-diversity problem in general metric spaces, the best possible approximation guarantee is $1/2$, under $\text{P}\neq\text{NP}$~\cite{ravi1994heuristic}. 
We also show how to achieve an approximation factor arbitrarily close to $1/2$ in randomized quasi-polynomial-time for arbitrary matroids. These two results are combined into the following corollary.

\begin{restatable}{corollary}{CorGonzalezFmb}\label{cor:gonzalez-and-FMB}
    Let $M$ be a matroid, $\delta,\varepsilon\in(0,1)$, and $k\geq 2$.
\begin{enumerate}
    \item There is a randomized algorithm that returns a set of $k$ bases $B_1,\ldots,B_k$ satisfying
    \[
    \min_{i<j} d(B_i,B_j) \geq \frac{1}{6} \diam_k(M)
    \]
    with probability at least $1-\delta$ in time $\poly(n,k,\log(1/\delta))$.
    \item There is a randomized algorithm that returns a set of $k$ bases $B_1,\ldots,B_k$ satisfying
    \[
    \min_{i<j} d(B_i,B_j) \geq \frac{1-\varepsilon}{2} \diam_k(M)
    \]
    with probability at least $1-\delta$ in time $\poly(n,k,\log(1/\delta))\cdot n^{\cO(\varepsilon^{-2}\log k)}$.
\end{enumerate}
\end{restatable}

\vspace{2mm}\noindent\textbf{Special matroids.} We also design algorithms that achieve improved approximation factors for classes of matroids with more structure. First, we note that if the basis-exchange graph is vertex transitive\footnote{A graph is said to be vertex-transitive if its automorphism group acts transitively on its vertex set~\cite{godsil2013algebraic}. In other words, every vertex is structurally identical.}, a property satisfied, for example, by \emph{partition matroids}, we can achieve an approximation factor of $1/2$ using random sampling again. Another well-studied special case of the Diverse-Matroid Bases problem is the problem of finding Diverse Spanning Trees. We show that we can achieve an improved approximation factor of $1/4$ if the graph in question has bounded arboricity. Bounded arboricity graphs include several well-studied classes of graphs such as \emph{planar graphs, graphs of bounded treewidth and graphs excluding a fixed minor}~\cite{dory2022near}. 
\begin{restatable}{theorem}{ThmVertexTran}\label{thm:special}

    For each $0<\delta<1$, there exist randomized  $\poly(n,k,\log(1/\delta))$ time algorithms that achieve, with success probability at least $1-\delta$:

    \begin{enumerate}
        \item A $1/2$-approximation for the Diverse Matroid Bases problem for matroids with vertex-transitive basis exchange graphs. 

        \item A $1/4$-approximation for the Diverse Spanning Tree for graphs with constant arboricity. 
    \end{enumerate}
\end{restatable}

\noindent\textbf{Hardness of FMB.} On the hardness side, we show that, even for graphic matroids, the problem of deciding whether or not there exists a spanning tree that is at distance at least $d$ from a set of given spanning trees of a graph is $W[1]$-Hard when parametrized by $d$. To do this, we employ a two-step reduction. First, we show that the Farthest String Problem, in which we are given $k$ strings of fixed length from a known alphabet and a target distance $d$ and we want to decide whether or not there exists a string of the same length that is at Hamming distance at least $d$ from all given strings, is $W[1]$-Hard when parametrized by $d$ and the string length. This is shown by a reduction from $k$-Clique. Then, our result follows by reducing Farthest String to Farthest Spanning Tree.

\begin{restatable}{theorem}{FstWHard}\label{thm:fst-w1}
\dfstprob is $W[1]$-Hard parameterized by $d$. Moreover, \dfsprob over an unbounded alphabet is $W[1]$-Hard parameterized by $d$ and the string length.
\end{restatable}

\subsection{Technical Overview}

We first discuss the proof of our main result, \Cref{thm:constant-diameter-lower}. The first idea is to transform the problem from a question about diameter to matroid intersection. 
Specifically, we express the diameter of $\cbx(M)$ as the maximum size of a common independent set of $M$ and the \emph{dual matroid} $M^*$~\cite{MR2849819}, and the average distance between two random bases as $\sum_e p_e(1-p_e)$,where $p_e$ is the marginal probability that an element $e$ belongs to a uniformly random basis. 
We then partition the elements into high-marginal and low-marginal sets called $H$ and $L$, respectively. 
Using matroid intersection, the diameter is at most $r_M(L)+r_{M^*}(H)$ and thus the goal is to show that the total marginal mass of $L$ in $M$ together with the total marginal mass of $H$ in $M^*$ approximates this sum. 
This follows easily from the dual rank formula when the two ranks are very unbalanced. The challenging case is when they are balanced. 
Here, we show that the total mass of $L$ in $M$ can't be too small, as then the deficit $t = r_M(L) - |B \cap L|$ would be large on average for a random basis $B$, and an exchange-counting argument of many missing $L$-elements gives an upper bound on $\E[\binom{t}{j}]$ in terms of $\binom{|A \cap B|}{j}$, where $j$ is a constant fraction of $r_M(L)$ and $A$ is a basis of $L$. 
On the other hand, the line of work by Anari et al. \cite{anari2019log,anari-liu-oveisgharan-vinzant-mason} showing that matroid-base-generating polynomials are ultra-log-concave has the following consequence:
After symmetrizing the distribution of $|A\cap B|$, ultra-log-concavity yields an exchangeable random set whose coordinates are negatively associated.
This translates to strong binomial-moment bounds for $|A \cap B|$. These upper bounds contradict the assumed large expected deficit, proving the $1/16$ factor approximation.

For \Cref{thm:constant-diameter-upper}, we build a sequence of transversals matroids whose elements are partitioned equally into elements which have marginal tending to $1/3$ and those which have marginals tending to $2/3$, which makes the average distance tend to a $4/9$ fraction of the diameter.

The starting intuition for \Cref{thm:edge-expansion-counterexample} is a chain of cliques whose sizes double: almost all vertices lie near the largest layers, so two random vertices are close to each other even though the graph has many layers and hence large diameter. To make this compatible with the logarithmic degree and diameter behavior of basis-exchange graphs, we replace each clique layer by a sparse expander and add a $\lceil \log n\rceil$-ary tree on the same vertex set so that each layer still has small internal diameter. Consecutive layers are connected by a two-to-one map and the resulting graph has edge expansion at least one via an induction over prefixes of layers. The diameter remains $\Theta(\log n)$ because any path from the first to the last layer crosses all layers, but the expected distance of two random vertices is only $
\cO(\log n/\log\log n)$, since their layer indices are concentrated within constant distance of the largest layer.

For \Cref{thm:k-diversity-upper,thm:k-diversity-random-tight}, the upper bound starts from the two-point sampling theorem and adds a lower-tail estimate. If $\mu$ is the expected distance between two random bases, then every fixed basis has expected distance at least $\mu/2$ from a random basis. The same ULC moment bound used in the diameter proof implies a Chernoff-type lower tail for this distance, so a random pair is very unlikely to have distance much below $\mu$. We set a threshold $\tau=\Theta(\mu/\log k)$ and union bound this bad event over the $\binom{k}{2}$ sampled pairs; with constant probability all sampled pairs are at distance at least $\tau$, which gives $\avg_k(M)=\Omega(\mu/\log k)$. Since \Cref{thm:constant-diameter-lower} gives $\diam_k(M)\leq \diam_2(M)=\cO(\mu)$, this is an $\cO(\log k)$ guarantee. The tight example is the direct sum of $r$ copies of $U_{1,2}$, whose bases are the vertices of the $r$-dimensional cube. For $k=2^{r/2}$, a birthday-paradox calculation shows that a random sample has a pair at constant Hamming distance in expectation, while the Gilbert--Varshamov bound~\cite{MR1664228} supplies $k$ cube vertices with pairwise distance $\Omega(r)=\Omega(\log k)$.

The proof of \Cref{thm:fmb-approx} follows by splitting into two regimes: \emph{low}, when $\diam_2(M)=\cO(\log k)$ and \emph{high}, when $\diam_2(M)=\Omega(\log k)$.
In the low diameter regime: 
Let $B^*$ be an optimal farthest basis from the given set $\cS$, let $d^*=d(B^*,\cS)$, and set $r=\lceil d^*/3\rceil$. 
Every base within distance $d^*-r$ of $B^*$ is \emph{good}, because the triangle inequality leaves it at distance at least $r$ from every base in $\cS$. 
Conversely, every bad base lies in a radius-$(r-1)$ ball around some member of $\cS$. This brings us to an interesting question: how different can the volumes of radius $r$ balls centered at $A$ and $B$ be?
The key lemma~\ref{lem:ball-homogeneity} says that for any two centers $A,B$, the ball $\Ball(A,r-1)$ has size at most $4^{\diam_2(M)}|\Ball(B,2(r-1))|\leq 4^{\diam_2(M)}|\Ball(B,d^*-r)|$. 
This is proved by taking a base near $A$ and using strong basis exchange to move the same deletions and insertions onto another basis around $B$. 
Applying this comparison to all centers in $\cS$ shows that a uniformly random base is good with probability at least $1/(k4^{\diam_2(M)}+1)$, so repeated sampling finds a $1/3$-approximate farthest basis.

For the high diameter regime, we solve a linear relaxation over the matroid base polytope and apply Chekuri, Vondrak, and Zenklusen's swap rounding~\cite{chekuri2010dependent}.
Every input basis gives a linear constraint, and the concentration given by the swap rounding ensures that, when the diameter is sufficiently large, all $k$ constraints are simultaneously satisfied within a $(1-\varepsilon)$ factor with high probability. 
Combining the two regimes yields the polynomial-time $\frac{1}{3}$-approximation of Theorem~\ref{thm:fmb-approx}.
If, in the small-diameter regime, we enumerate all bases instead of sampling we obtain a quasi-polynomial-time $(1-\varepsilon)$-approximation stated in Theorem~\ref{thm:fmb-qptas}.

For \Cref{thm:special}, we first show that, given a matroid $M$, when $\cG(M)$ is vertex-transitive, the size of a ball centered at a vertex in $\cG(M)$ does not depend on the center.
Since the balls around the optimal solutions of radius $ \lfloor(\diam_k(M)-1)/2\rfloor $ are disjoint of each other, $ k $ times the volume of any ball of radius $ \lfloor(\diam_k(M)-1)/2\rfloor $ is less than the total number of bases of $M$.
Thus, given $i$ bases, one can always sample a basis which is at distance at least $ \lfloor(\diam_k(M)-1)/2\rfloor $ from all of the given $i$ bases with probability at least $ 1/(i+1) $.
This gives us a polynomial time $1/2$-approximation randomized algorithm for the Diverse Matroid Bases problem for $M$. As a corollary, we also show that the same algorithm applies to partition matroids. Using this corollary, in \Cref{sec:graphic-matroid} we design a randomized polynomial-time $1/4$-approximation algorithm for the Diverse Spanning Tree problem on graphic matroids whose underlying graph has constant arboricity (\Cref{thm:bounded-arbocity}).
Let $\cT(G)= \cbx(M)$ where $M$ is the graphic matroid defined by $G$.
We first note that it is enough to consider $\diam_2(\cT(G)) = \cO(\log k)$ since in the complementary regime when $\diam_2(\cT(G)) = \Omega(\log k)$, \Cref{lem:fmb-ptas-large-diameter} yields a $1/4$-approximation for Diverse Spanning Trees via linear programming followed by Chekuri, Vondrak, and Zenklusen's swap rounding~\cite{chekuri2010dependent}.
The rest of the proof rests on a structural insight.
The edges of $G$ split into a highly structured part $G'$, consisting of all edges incident to a vertex of degree at most $2$, and the remaining ``core,'' spanned by the vertices of degree at least $3$. Since $G'$ is a union of edge-disjoint paths and cycles, once we fix the spanning tree's image on the contracted core, the set of extensions to a spanning tree of $G$ forms a partition matroid.
We then show that when $G$ has bounded arboricity, this bound on the diameter forces the contracted core to have only $\cO(\log k)$ edges, and hence only $\poly(k)$-many spanning trees. We enumerate all of them in $\poly(k)$ time and, for each, exploit the partition-matroid structure of its extensions.

Finally, \Cref{thm:fst-w1} is proved by two reductions. First, we reduce from $k$-\textsc{Clique} to non-binary Farthest String. The string length is $k$, the target distance is $k-1$, and the alphabet consists of the graph vertices together with one dummy symbol. For every pair of coordinates and every ordered pair of vertices $(u,v)$ that is not an edge, we create a string that places $u$ and $v$ in those two coordinates and dummy symbols elsewhere. A string at distance at least $k-1$ from all constructed strings cannot use dummy symbols and cannot choose a non-edge in any coordinate pair; therefore it encodes a $k$-clique, and any clique gives such a string. Second, we reduce Farthest String to Farthest Spanning Tree by using a star-like graph with one bundle of parallel edges per coordinate and one edge in the bundle per alphabet symbol. Every spanning tree chooses exactly one edge from each bundle, so spanning trees are in bijection with strings and the tree distance is exactly the Hamming distance. This transfers the $W[1]$-hardness to Farthest Spanning Tree parameterized by the target distance.

\subsection{Applicability of our results}

We remark that matroids, and in particular their bases, provide a flexible abstraction for many natural combinatorial constraints. We recall a few of the most prominent examples.
(1) Given a graph $G$, its \emph{graphic matroid} has as independent sets the spanning forests of the graph.
(2) A \emph{laminar family} is a collection $\cA\subseteq 2^E$ such that, for any two intersecting sets one is contained in the other.
Given capacities $c\colon \cA \to \NN$, the associated \emph{laminar matroid} has independent sets $\{I \subseteq E : |A\cap I|\leq c(A), \; \text{for all }A \in \cA \}$.
(3) When the sets in the laminar family are pairwise disjoint this specializes to a \emph{partition matroid}.
(4) Given a bipartite graph $G=(L\cup R,E)$, the associated \emph{transversal matroid} is the matroid on ground set $L$ whose independent sets are precisely those subsets of $L$ that can be matched to $R$. 
Note that these respectively model (1) acyclicity constraints, (2) hierarchical quota constraints, (3) group-cardinality constraints, (4) compatibility or assignment constraints. 
There are many other well-studied classes of matroids, such as gammoids, paving matroids, and regular matroids; we do not discuss them further here, and refer the reader to the classic text of Oxley~\cite{MR2849819}.

Another case of interest for us arises from optimal bases. 
Let $M=(E,\cI)$ be a matroid and let $w\colon E\to \RR$ be a weight function. 
It is a folklore fact that the family of $w$-optimal bases of $M$ is itself the family of bases of another matroid on the same ground set; we denote this matroid by $M_w$ 
(see, e.g., Proposition 2 in~\cite{MR2185977}). 
Thus, our results also apply to the problem of finding a farthest $w$-optimal base from a prescribed collection of $w$-optimal bases, as well as to the problem of finding diverse collections of $w$-optimal bases.

We note that our results yield polynomial-time algorithms for all of the aforementioned settings in their standard presentations.
This follows as all matroid families discussed above have independence oracles implementable in polynomial time in their standard presentation. 
Moreover, given independence-oracle access to $M$, one also obtains independence-oracle access to the optimal-base matroid $M_w$ in polynomial time.

\subsection{Future directions}

We believe our results can be explored in several directions, and give rise to interesting open problems. First, we believe it should be possible to close the gap between the lower bound of 1/16 in Theorem~\ref{thm:constant-diameter-lower} and the upper bound of 4/9 in Theorem~\ref{thm:constant-diameter-upper}. As mentioned earlier, a major open problem is to investigate hardness for $k$-diversity for finding $k$ optimally diverse bases of a matroid when $k$ is given in unary. 

Furthermore, our QPTAS in~\Cref{thm:fmb-qptas} strongly suggests the existence of a PTAS for Farthest Matroid Basis. Finally, while our results apply to the \emph{diverse optimal bases} (minimum or maximum weight) of a matroid problem, another interesting variant, studied in~\cite{gao2022obtaining}, is where we are given an approximation factor $c \geq 1$, and the task is to return $k$ most diverse $c$-optimal bases. 

\noindent\textbf{Roadmap.} Section~\ref{sec:preliminaries} introduces necessary preliminaries. 
Section~\ref{sec:diameter} relates the diameter of a matroid to the expected distance between random bases, constructs the transversal-matroid upper bound, and shows that edge expansion alone does not explain the sampling guarantee.
Section~\ref{sec:diversity} studies direct sampling for $k$-diversity, develops the FMB approximation algorithms in the small- and large-diameter regimes, and derives the general diverse-basis guarantees.
Section~\ref{sec:special-cases} provides improved guarantees for partition and graphic matroids with bounded-arboricity.
Section~\ref{sec:hardness} proves parameterized hardness results for Farthest Matroid Basis.

%% file: sections/preliminaries.tex
\section{Preliminaries}\label{sec:preliminaries}

\paragraph{Matroids.}

A \emph{set system} $M = (E, \cF)$ consists of a collection of subsets $\cF \subseteq 2^E$ of a finite ground set of elements $E = \set{e_1, \dots, e_n}$. A set system $M = (E, \cF)$ is called \emph{downwards-closed} if $A \in \cF \implies B \in \cF$ for any $B \subseteq A \subseteq E$. A downwards-closed set system is also called a \emph{simplicial complex}. A \emph{matroid} is a downwards-closed set system $M = (E, \cI)$, where $\cI \subseteq 2^E$ denotes the set of \emph{independent sets} of $M$ and satisfies: (i) $\emptyset \in \cI$, (ii) $A \in \cI \implies B \in \cI$ for all $B \subseteq A$, and (iii) for any $A, B \in \cI$ with $|A| < |B|$ there exists $e \in B \setminus A$ with $A \cup \set{e} \in \cI$. The maximally feasible sets of $M$ are called \emph{bases} and we usually denote them by $\cB(M)$. In matroids, all bases of $M$ have the same cardinality, called the \emph{rank} of $M$ and denoted by $r(M)$.  For $S\subseteq E$, we write $r_M(S)$ for the maximum size of an independent subset of $S$.

For $X \subseteq E$, the \emph{deletion} $M \setminus X$ is the matroid on ground set $E \setminus X$ whose independent sets are $\set{I \in \cI \midd I \cap X = \emptyset}$. The \emph{contraction} $M/X$ of $M$ by an independent set $X$ is the matroid on $E \setminus X$ whose independent sets are $\set{I \subseteq E \setminus X \midd I \cup X \in \cI}$. A matroid obtained from $M$ by a sequence of deletions and contractions is called a \emph{minor} of $M$.

The \emph{dual matroid} $M^*$ of $M$ has the same ground set $E$. Its set of bases is $\cB(M^*) = \set{E \setminus B \midd B \in \cB(M)}$ and its rank function is $r_{M^*}(S) = |S| - r(M) + r_M(E\setminus S)$. The matroid intersection theorem of Edmonds \cite{edmonds} implies that if $M_1,M_2$ are matroids on the same ground set $E$, then
\[
  \max\set{|I|:\ I\text{ is independent in both }M_1\text{ and }M_2}
  =
  \min_{U\subseteq E} \set{r_{M_1}(U) + r_{M_2}(E\setminus U)}.
\]

A special subclass of matroids that is of interest to us are \emph{graphic matroids}. For a given graph $G = (V, E)$, the graphic matroid $M(G) = (E, \cI)$ associated with $G$ is the matroid on the ground set of edges $E$, where $\cI$ denotes the set of all forests of $G$. In other words, a set $I$ is independent in $M(G)$ if and only if it is acyclic. The bases of $M(G)$ are the spanning trees of $G$ and its rank $r(M(G))$ is $|V| - 1$.

We use $\1_A$ to denote the indicator vector of a set $A$, i.e. the vector where the $i$-th position is $1$ if $e_i \in A$ and $0$ otherwise, and $A \triangle B$ to denote the \emph{symmetric difference} of $A$ and $B$, i.e. $A \triangle B = \set{e \in E \midd e \in A \setminus B \text{ or } e \in B \setminus A}$. 
For brevity, we use $A + e$ and $A - e$ to denote $A \cup \{e\}$ and $A \setminus \{e\}$, respectively.

\paragraph{Diameter and Diversity.}

Given a graph $G = (V, E)$, the \emph{diameter} of $G$ is the maximal shortest-path distance between any two vertices in $G$. 
In this paper, we focus on a special case of graphs $G$, namely \emph{basis-exchange graphs}. 
An undirected graph $G = (V(G), E(G))$ is a basis-exchange graph of a matroid $M = (E, \cI)$ with bases $\cB$ if there exists a bijection $\pi : \cB \to V(G)$ such that, for any two distinct bases $A, B \in \cB$, there exists an edge between $\pi(A)$ and $\pi(B)$ if and only if there exist elements $e \in A, f \in B$ such that $B = A - e + f$, i.e. if $B$ can be obtained from $A$ via an exchange swap.

The \emph{(combinatorial) distance} between two bases $A, B \in \cB(M)$ of a matroid $M$ is defined as
\[
    d(A,B) \coloneqq |A \setminus B| = |B \setminus A| = \frac{1}{2}|A \triangle B|.
\]
Notice that $d(A, B)$ is the shortest-path distance between the vertices corresponding to $A$ and $B$ int he basis-exchange graph of $M$. 

The main object of study of this paper is the \emph{$k$-diversity}, or simply the \emph{diversity}.
\begin{definition}[Diversity]
For a matroid $M$ with a set of bases $\cB(M)$, let $\diam_k$ be the smallest pairwise distance between the $k$ farthest bases; i.e.,
\[
    \diam_k (M) \coloneqq \max_{B_1,\dots, B_k \in \cB(M)} \min_{i \neq j \colon i,j \in [k]} d(B_i, B_j). 
\]
\end{definition}

We use $\diam_2(M)$ to refer to the \emph{(combinatorial) diameter of a matroid} $M$, which is the diameter of the basis-exchange graph of $M$. 
Note that the diversity generalizes the diameter for matroids, and that $\diam_k(M) \leq \diam_\ell(M)$ for $k \geq \ell$.

A lot of our results rely on sampling bases of $M$ uniformly at random and comparing their expected combinatorial distance to $\diam_k(M)$. This motivates the following definition.

\begin{definition}[Average Distance]
For a matroid $M$ with a set of bases $\cB(M)$, let $U(\cB(M))$ denote the uniform distribution on $\cB(M)$. The average minimum pairwise distance between $k$ uniformly random bases $\avg_k$ is defined as
\[
    \avg_k(M) \coloneqq \E_{B_1,\dots, B_k \sim U(\cB(M))} \brk{\min_{i \neq j : i,j \in [k]}\frac{1}{2}|B_i \Delta B_j|}.
\]
\end{definition}

Here, we assume the sampling is done without replacement. Let $B \sim \mu$, and $X_e \coloneqq \indic{e \in B}$ and $p_e \coloneqq \Pr_{B \sim \mu}[e \in B]$ denote the indicator of the event $e \in B$ and its probability, respectively, for $e \in E$. For a subset $S\subseteq E$, let $p(S) \coloneqq \sum_{e \in S} p_e$ and $(1-p)(S) \coloneqq \sum_{e\in S} (1 - p_e)$.

\subsection{Negative Correlation}

In some of our results, we make use of the following definition of \emph{Negative Association (NA)}.

\begin{definition}[Negative Association (NA) \cite{neg-assoc-1,neg-assoc-2}]
A set of random variables $X_1, X_2, \ldots, X_n$ is said to be \emph{negatively associated (NA)} if for any two disjoint index sets $I, J \subseteq [n]$ and two functions $f, g$ that are both monotone increasing or both monotone decreasing, we have
\[
    \E\brk{f\prn{X_i : i \in I} \cdot g\prn{X_j : j \in J}} \leq \E\brk{f\prn{X_i : i \in I}} \cdot \E\brk{g\prn{X_j : j \in J}}.
\]
\end{definition}

%% file: sections/diameter.tex
\section{Diameter in Matroids}\label{sec:diameter}

In this section, we study how well the distance between two uniformly random bases approximates the diameter of the basis-exchange graph of a matroid. Our main positive result shows that, for every matroid $M$, $\avg_2(M) \geq (1/16) \diam_2(M)$. We complement this with an example of a matroid $M$ with $\avg_2(M) \leq (4/9) \diam_2(M)$. Our final result indicates that the matroid structure is essential for such a guarantee. Indeed, high edge-expansion by itself is not enough, which we show by constructing an expander graph with edge expansion at least $1$ for which the distance between two uniformly random vertices is not a constant-factor approximation to the diameter.

\input{sections/diameter-lower}

\input{sections/diameter-higher}

\input{sections/diam-edge-expansion-hard}

%% file: sections/diameter-lower.tex
\subsection{A Constant-Factor Approximation via Random Sampling}
\label{ss:approx-via-2-sampling}
Here, we show the following theorem.

\lowerdiameter*

The proof of Theorem~\ref{thm:constant-diameter-lower} follows from the following three ideas. First, we replace the distance between two bases in $M$ by the intersection of a basis of $M$ with an independently drawn basis of $M^*$. Second, we partition the ground set into elements $e$ with marginal $p_e \leq 1/2$ and those with marginal $p_e > 1/2$. The idea then is that, if the elements with low marginals also have too little total marginal mass, then this forces many exchanges from a fixed low-marginal basis into a uniformly random basis $B$. Third, the size sequence of the intersection $S$ of a fixed set and a uniformly random basis of a matroid is \emph{ultra-log-concave (ULC)}, as shown in the seminal work of Anari, Liu, Oveis Gharan, and Vinzant \cite{anari-liu-oveisgharan-vinzant-mason}. We then tweak the distribution to make it exchangeable and, combined with the fact that it is ULC, Pemantle's work on negative dependence \cite{pemantle-negative-dependence} implies that $S$ satisfies \emph{negative association (NA)}; this prevents those exchanges from overlapping too much. The precise constant we prove is a little better ($\approx 1/15.02$) but we present the weaker constant for clarity of exposition.

We start by rewriting the expected distance in a useful form.

\begin{lemma}\label{lem:variance}
For every matroid $M$,
\[
  \EE(d(A,B)) = \sum_{e\in E} p_e(1-p_e) = \sum_{e\in E}\Var(X_e).
\]
\end{lemma}
\begin{proof}
Let $B,B'$ be bases of $M$ drawn independently from $U(\cB(M))$ and let $X'_e=\mathbf{1}\{e\in B'\}$. For each element $e$,
\[
    \Pr[X_e \ne X'_e] = p_e(1-p_e) + (1-p_e)p_e = 2 p_e(1-p_e).
\]
Also
\[
  |B\triangle B'|=\sum_{e\in E}\mathbf{1}\{X_e\ne X'_e\}.
\]
The first equality follows by linearity of expectation along with the fact that all bases of $M$ have the same size, implying $|B \setminus B'| = |B' \setminus B| = |B \triangle B'| / 2$. The second equality is the usual identity $\Var(X_e)=p_e(1-p_e)$ for an indicator.
\end{proof}

Our first idea is to reformulate the problem into a question about matroid intersection, using the dual matroid.

\begin{lemma}\label{lem:dual-formulation}
For every matroid $M$,
\[
	\avg_2(M) = \E_{B,C} \brk{|B\cap C|},
\]
where $B$ and $C$ are sampled independently and uniformly from $\cB(M)$ and $\cB(M^*)$.  Moreover,
\[
	\diam_2(M) = \max\set{|I| \midd I \text{ is independent in both } M \text{ and } M^*}.
\]
\end{lemma}
\begin{proof}
If $B$ and $B'$ are bases of $M$ drawn independently and uniformly at random, then $C = E\setminus B'$ is a basis of $M^*$ drawn independently from $B$ and uniformly at random from $\cB(M^*)$.  Hence
\[
	\E_{B,C}[|B\cap C|] = \E_{B,B'}[|B\setminus B'|] = \avg_2(M).
\]
	
If $P,Q$ are bases of $M$, then $P\setminus Q$ is contained in the basis $P$ of $M$ and in the dual basis $E\setminus Q$ of $M^*$.  Thus $P\setminus Q$ is a common independent set of $M$ and $M^*$, and the maximum common independent set has size at least $\diam_2(M)$. Conversely, let $I$ be independent in both $M$ and $M^*$. Extend $I$ to a basis $P$ of $M$ and to a basis $C$ of $M^*$. Since $C = E \setminus Q$ for some basis $Q$ of $M$, we have $I \subseteq P \setminus Q$. Hence $|I| \leq \diam_2(M)$.
\end{proof}

Fix a set $A \subseteq E$, and let $B \sim U(\cB(M))$, $Z = |A \cap B|$ and $r = r(M)$ denote the rank of $M$. The polynomial $g_U(x) = \sum_{B \in \cB(M)} \prod_{i \in B} x_i$ is called the \emph{generating polynomial} of the uniform distribution $U(\cB(M))$. By the work of Anari, Liu, Oveis Gharan and Vinzant, $g_U$ is an ULC polynomial (\cite{anari-liu-oveisgharan-vinzant-mason}, Theorem 4.1). For a fixed set $A$, we can symmetrize $g_U$ by substituting $x_i$ with $y$ for $i \in A$ and $x_i$ with $z$ for $i \notin A$, to obtain
\[
q_A(y, z) = \sum_{B \in \cB(M)} y^{|A \cap B|} z^{r - |A \cap B|} = |\cB(M)| \sum_{i = 0}^{|A|} \zeta_i y^i z^{r - i},
\]
where $\zeta_i = \Pr[Z = i]$ and the last equality follows since $B$ is drawn uniformly from the bases of $M$. Let
\[
p_A(y,z) \coloneqq \frac{q_A(y,z)}{|\cB(M)|} = \sum_{i = 0}^{|A|} \zeta_i y^i z^{r - i}.
\]
By (\cite{anari-liu-oveisgharan-vinzant-mason}, Lemma 2.2), ULC polynomials are closed under affine transformations, and since $g_U$ is ULC it follows that $p_A$ is also ULC. Furthermore, using Gurvits' machinery (\cite{gurvits}, Proposition 2.7), it follows that the coefficient sequence of $p_A(y,z)$ is a ULC sequence, i.e. for all $1 \leq i \leq r - 1$,
\[
\prn{\frac{\zeta_i}{\binom{r}{i}}}^2 \geq \frac{\zeta_{i-1}}{\binom{r}{i-1}} \cdot \frac{\zeta_{i+1}}{\binom{r}{i+1}}.
\]
Next, we will turn the distribution of $A \cap B$ into an exchangeable distribution. Consider a random set $S$ distributed as follows: first draw a random variable $K$ distributed exactly as $Z = |A \cap B|$ and then, conditioned on $K = i$, draw $S$ uniformly at random from the subsets of $\set{1, \ldots, r}$ of size $i$. Since $B \sim U(\cB(M))$, it follows that $|S|$ has the same distribution as $Z$. Furthermore, for all $T \subseteq \set{1, \ldots, r}$,
\[
\Pr[S = T] = \frac{\zeta_{|T|}}{\binom{r}{|T|}}.
\]
For $i \in [r]$, let $X_i = \indic{i \in S}$. Notice that the $X_i$'s are exchangeable random variables, i.e. their joint distribution is invariant to permutations of $\set{1, \ldots, r}$. Indeed, for any $x \in \set{0,1}^r$ where the number of $1$'s in $x$ is $i$, we have
\[
\Pr[(X_1, \ldots, X_r) = x] = \frac{\zeta_i}{\binom{r}{i}}.
\]
Also, $\sum_{i = 1}^r X_i = |S| = K$, and so $\Pr\brk{\sum_{j = 1}^r X_j = i} = \zeta_i$, for all $i \in [r]$. Furthermore,
\begin{align*}
\E[X_i] &=\Pr[i \in S] = \sum_{j = 0}^r \Pr[K = j] \Pr\brk{i \in S \midd K = j} \\
&= \sum_{j = 0}^r \zeta_j \frac{\binom{r-1}{j-1}}{\binom{r}{j}} = \sum_{j = 0}^r \zeta_j \frac{j}{r} = \frac{\E[K]}{r} \\
&= \frac{\E[Z]}{r}.
\end{align*}
The exchangeability of the $X_i$'s together with the ultra-log-concavity of the $\zeta_i$'s imply that the $X_i$'s satisfy negative association (NA) (\cite{pemantle-negative-dependence}, Theorem 2.7). The following lemma is a straightforward application of a well-known property of negatively associated random variables.
\begin{lemma}\label{lem:pemantle-ncd}
For any $R \subseteq \set{1,\ldots, r}$, we have
\[
\Pr[S \supseteq R] \leq \prod_{i \in R} \Pr[i \in S] = \prn{\frac{\E[Z]}{r}}^{|R|}.
\]
\end{lemma}
\begin{proof}
We have that $\Pr[S \supseteq R] = \E\brk{\prod_{i \in R} X_i}$. Let $R = \set{e_1, \ldots, e_{|R|}}$, $f(\bm{X}) = \prod_{j = 1}^{|R|-1} X_{e_j}$ and $g(\bm{X}) = X_{e_{|R|}}$. Applying the NA condition on $f$ and $g$ yields
\[
\E[f(\bm{X}) g(\bm{X})] \leq \E[f(\bm{X})] \E[g(\bm{X})] = \E\brk{\prod_{j = 1}^{|R|-1} X_{e_j}} \E[X_{e_{|R|}}].
\]
The claim then follows by induction and the fact that $\E[X_i] = \E[Z]/r$ for all $i \in [r]$.
\end{proof}

Using Lemma~\ref{lem:pemantle-ncd}, we can show the following strong moment bound for $Z$.

\begin{lemma}\label{lem:ulc-moment}
Let $B \sim U(\cB(M))$ and consider a fixed set $A\subseteq E$. Also, let $Z = |A \cap B|$ with $z = \E[Z]$. Then, for every $j \geq 1$,
\[
    \E\brk{\binom{Z}{j}} \leq \binom{r}{j} \prn{\frac{z}{r}}^j \leq \frac{z^j}{j!}.
\]
\end{lemma}
\begin{proof}
If $j > r$, then the left-hand side is zero, so let $1\le j\le r$. Notice that $\binom{Z}{j}$ counts how many subsets of $A$ of size $j$ are contained in $A \cap B$ and is identically distributed to $\binom{|S|}{j}$. Thus, for $\Omega = \set{1, \ldots, r}$,
\[
	\binom{Z}{j} = \binom{|S|}{j} = \sum_{\substack{R \subseteq \Omega \\ |R| = j}} \indic{S \supseteq R}.
\]
Taking expectations and using Lemma~\ref{lem:pemantle-ncd}, we get that
\[
	\E\brk{\binom{Z}{j}} = \sum_{\substack{R\subseteq \Omega \\ |R| = j}} \Pr[S \supseteq R] \leq \binom{r}{j} \prn{\frac{z}{r}}^j = \frac{r!}{j! (r-j)!} \cdot \frac{z^j}{r^j} = \frac{z^j}{j!} \prod_{\ell = 0}^{j-1} \frac{r - \ell}{r} \leq \frac{z^j}{j!}. \qedhere
\]
\end{proof}
Intuitively, $\E\brk{\binom{Z}{j}}$ is the expected number of fixed subsets of $A$ of size $j$ captured by the random basis. If the elements of the symmetrized model were independent with density $z/r$, each fixed subset of size $j$ would appear with probability exactly $(z/r)^j$. The inequality of Lemma~\ref{lem:pemantle-ncd} implied by NA essentially says that, under an exchangeable ultra-log-concave distribution, each such fixed subset is no more likely to appear than in that independent model.

Next, recall that $\avg_2(M) = \sum_e p_e (1-p_e)$. We lower bound each term by
\begin{equation}\label{eq:avg2-min-lower-bound}
    p_e(1-p_e) \geq \frac{1}{2} \min(p_e, 1-p_e).
\end{equation}
Let $E = L \sqcup H$ be any partition of the elements such that $p_e \leq 1/2$ for all $e \in L$ and $p_e \geq 1/2$ for all $e \in H$. Combining Lemma~\ref{lem:variance} and \eqref{eq:avg2-min-lower-bound} we get
\begin{equation}\label{eq:avg2-lower-bound}
    \avg_2(M) \geq \frac{1}{2} \sum_e \min(p_e, 1-p_e) = \frac{1}{2} \prn{\sum_{e \in L} p_e + \sum_{e \in H} (1-p_e)} = \frac{p(L) + \bar{p}(H)}{2},
\end{equation}
where we use $p(S) = \sum_{e \in S} p_e$ and $\bar{p}(S) = \sum_{e \in S} (1-p_e)$ for brevity. Furthermore, notice that, by Lemma~\ref{lem:dual-formulation}, $\diam_2(M)$ is the maximum size of a common independent set of $M$ and $M^*$. By Edmonds' matroid intersection theorem, we have that
\[
    \diam_2(M) = \min_{U \subseteq E} \set{r_M(U) + r_{M^*}(E \setminus U)}.
\]
Thus, if we choose $U = L$ in the above equation, we get
\begin{equation}\label{eq:diam2-upper-bound}
	\diam_2(M) \le r_M(L) + r_{M^*}(H)
\end{equation}
Given \eqref{eq:avg2-lower-bound} and \eqref{eq:diam2-upper-bound}, to complete the proof of Theorem~\ref{thm:constant-diameter-lower} it suffices to show that
\begin{equation}\label{eq:symmetrized-mass-rank}
    p(L) + \bar{p}(H) \geq \frac{r_M(L) + r_{M^*}(H)}{8}.
\end{equation}

First, notice that, if $r_{M^*}(H) \geq (9/7) r_M(L)$ or $r_M(L) \geq (9/7) r_{M^*}(H)$, then a simple argument shows \eqref{eq:symmetrized-mass-rank}.

\begin{claim}\label{clm:rank-mass-easy-inequalities}
If $r_{M^*}(H) \geq (9/7) r_M(L)$ or $r_M(L) \geq (9/7) r_{M^*}(H)$, then
\[
    p(L) + \bar{p}(H) \geq \frac{r_M(L) + r_{M^*}(H)}{8}.
\]
\end{claim}
\begin{proof}
First, assume that $r_{M^*}(H) \geq (9/7) r_M(L)$. Then,
\[
    p(L) + \bar{p}(H) = p(L) + |H| - p(H) = p(L) + |H| - (r - p(L)) = 2 p(L) + |H| - r,
\]
and the dual rank formula gives
\[
    r_{M^*}(H) = |H| - r + r_M(L) \iff |H| - r = r_{M^*}(H) - r_M(L).
\]
Combining the two equalities, we get that
\[
    p(L) + \bar{p}(H) = 2 p(L) + r_{M^*}(H) - r_M(L) \geq r_{M^*}(H) - r_M(L),
\]
and since
\begin{align*}
    r_{M^*}(H) &\geq (9/7) r_M(L) & \iff \\
    7 r_{M^*}(H) &\geq 9 r_M(L) & \iff \\
    8r_{M^*}(H) - 8 r_M(L) &\geq  r_M(L) + r_{M^*}(H) & \iff \\
    r_{M^*}(H) - r_M(L) &\geq \frac{r_M(L) + r_{M^*}(H)}{8},
\end{align*}
the claim follows.

Next, assume that $r_M(L) \geq (9/7) r_{M^*}(H)$. Then, 
\[
    p(L) + \bar{p}(H) = r - p(H) + \bar{p}(H) = r - |H| + 2 \bar{p}(H),
\]
and the dual rank formula gives
\[
    r_{M^*}(H) = |H| - r + r_M(L) \iff r - |H| = r_M(L) - r_{M^*}(H).
\]
Combining the two equalities, we get that
\[
    p(L) + \bar{p}(H) = r_M(L) - r_{M^*}(H) + 2 \bar{p}(H) \geq r_M(L) - r_{M^*}(H),
\]
and since
\begin{align*}
    r_M(L) &\geq (9/7) r_{M^*}(H) & \iff \\
    7 r_M(L) &\geq 9 r_{M^*}(H) & \iff \\
    8r_M(L) - 8 r_{M^*}(H) &\geq r_M(L) + r_{M^*}(H) & \iff \\
    r_M(L) - r_{M^*}(H) &\geq \frac{r_M(L) + r_{M^*}(H)}{8},
\end{align*}
the claim follows.
\end{proof}

Thus, we can assume that $7 r_{M^*}(H) < 9 r_M(L)$ and $7 r_M(L) < 9 r_{M^*}(H)$. Next, suppose that we could prove $p(L) \geq r_M(L)/8$. In this case, $\bar{p}(H) \geq r_{M^*}(H)/8$ would follow by duality.
\begin{lemma}\label{lem:dual-mass-rank-inequality}
Let $E = L\sqcup H$ be any partition of the elements such that $p_e \leq 1/2$ for all $e \in L$ and $p_e \geq 1/2$ for all $e \in H$. If, for every matroid $M$ on $E$ with rank function $r_M$ and dual matroid $M^*$ such that $7 r_{M^*}(H) < 9 r_M(L)$ and $7 r_M(L) < 9 r_{M^*}(H)$, we have
\[
    p(L) \geq \frac{r_M(L)}{8},
\]
where $p(L) = \sum_{e \in L} p_e$, then
\[
    p(L) + \bar{p}(H) \geq \frac{r_M(L) + r_{M^*}(H)}{8}.
\]
\end{lemma}
\begin{proof}
If $B$ is a uniformly random basis of $M$, then $E \setminus B$ is a uniformly random basis of $M^*$. This implies that, if the marginal of an element $e$ in $U(\cB(M))$ is $p_e = \Pr[e \in B]$, then if we draw $B^* = E \setminus B$ uniformly at random from $\cB(M^*)$, we have $p^*_e = \Pr[e \in B^*] = \Pr[e \notin B] = 1 - p_e$.

Let $L^* = H$ and $H^* = L$, since $p^*_e \leq 1/2 \implies p_e \geq 1/2$. Therefore, the same argument applied to the dual matroid $M^*$ would yield
\[
    p^*(L^*) \geq \frac{r_{M^*}(L^*)}{8} \iff \bar{p}(H) \geq \frac{r_{M^*}(H)}{8}.
\]
\end{proof}

Therefore, it remains to show
\begin{equation}\label{eq:one-sided-mass-rank}
    p(L) \geq \frac{r_M(L)}{8},
\end{equation}
under the assumption
\begin{equation}\label{eq:one-sided-assumption}
    7 r_{M^*}(H) < 9 r_M(L) \quad \text{and} \quad 7 r_M(L) < 9 r_{M^*}(H) \iff \frac{7}{9} r_M(L) < r_{M^*}(H) < \frac{9}{7} r_M(L),
\end{equation}
since then \eqref{eq:symmetrized-mass-rank} follows by Lemma~\ref{lem:dual-mass-rank-inequality}.

We prove \eqref{eq:one-sided-mass-rank} under the assumption of \eqref{eq:one-sided-assumption} via contradiction. First, we present the following lemma, which essentially lower-bounds the expected rank of $B$ inside $L$ and is the main tool we use for our proof.
\begin{lemma}\label{lem:main-coordinate-shadow}
Let $B \sim U(\cB(M))$ and $E = L\sqcup H$ be any partition of the elements such that $p_e \leq 1/2$ for all $e \in L$ and $p_e \geq 1/2$ for all $e \in H$. Also, let $A$ be a basis of $M_{|L} = M \setminus (E \setminus L)$ and $t = r_M(L) - |B \cap L|$. Then, for every $j \geq 1$,
\[
	\E\brk{\binom{t}{j}} \leq \E\brk{\binom{|A \cap B|}{j} \binom{|H \setminus B|}{j}}.
\]
\end{lemma}
\begin{proof}
If $H = \emptyset$, then $L = E$ and $r_M(L) = r$. Hence, every basis $B \in \cB(M)$ satisfies $|B \cap L| = r_M(L)$, and thus $t = 0$. Also, $|H \setminus B| = 0$. Since $j \geq 1$, both sides
of the inequality are zero.

Assume that $H \neq \emptyset$ and fix a realization of $B$. Let $M'$ denote the minor of $M_{|L}$ where we have contracted $B \cap L$, i.e. $M' = (M \setminus H) / (B \cap L)$. The idea is to count the ways that we can exchange $j$ new elements from a fixed basis $A'$ of $M'$ into an arbitrary basis $B$.

Since $A$ spans $M_{|L}$, the set $A \setminus (B \cap L)$ spans $M'$ and thus contains a basis $A'$ of $M'$. Furthermore, the rank of $A'$ is exactly $r_M(L) - |B \cap L| = t$. In other words, $A'$ is a basis of $M'$ of size $t$ and therefore contains at least $\binom{t}{j}$ independent subsets $R$ of size $j$.

For each such independent set $R \subseteq A'$ of size $j$, the set $B_R = (B \cap L) \cup R$ is independent in $M$ and we can further extend it to a basis of $M$ using elements of $B \setminus B_R$, by the exchange axiom for matroids. Recall that $B \cap L$ and $B \cap H$ partition $B$. Since $B \cap L \subseteq B$ and $R \cap (B \cap L) = \emptyset$, this extension uses exactly $|B \cap H| - j$ elements of $B\cap H$; the remaining $j$ elements are the elements in $R$. In other words, there exists a set $J \subseteq B \cap H$ of size $j$ such that $B' = (B \setminus J) \cup R$ is a basis of $M$. For each pair $(B,R)$, we choose one such $J$, and define the map
\[
\Phi(B,R) = (B',R,J),
\]
where $B' = (B \setminus J) \cup R$. The domain of $\Phi$ is
\[
\cD = \set{(B,R) \midd B \text{ is a basis of } M \text{ and }R \subseteq A'}.
\]
See Figure~\ref{fig:coordinate-shadow-sets} for a visual representation of all the sets.

\begin{figure}[ht]
\centering
\begin{tikzpicture}[scale=0.88, every node/.style={font=\small}]
  % Ambient box
  \draw[rounded corners=2pt, thick] (0,0) rectangle (12,4.2);

  % B first, so middle line supersedes it
  \fill[green!12] (6.35,1.95) ellipse (4.25 and 1.15);
  \draw[green!50!black, thick] (6.35,1.95) ellipse (4.25 and 1.15);
  \node[green!40!black] at (6.25,3.35) {$B$};
  \node[green!40!black] at (5.0,1.85) {$B\cap L$};
  \node[green!40!black] at (7.35,1.85) {$B\cap H$};

  % A
  \fill[blue!10, opacity=.75] (2.35,1.85) ellipse (1.45 and 1.35);
  \draw[blue!65!black, thick] (2.35,1.85) ellipse (1.45 and 1.35);
  \node[blue!65!black] at (1.85,3.35) {$A$};
  \node[blue!65!black] at (1.65,2.35) {$A'$};

  % Z = A cap B
  \begin{scope}
    \clip (2.35,1.85) ellipse (1.45 and 1.35);
    \fill[teal!25] (6.35,1.95) ellipse (4.25 and 1.15);
  \end{scope}
  \draw[green!50!black, thick] (6.35,1.95) ellipse (4.25 and 1.15);

  % R = the part of A below/outside B, clipped away from Z
  \begin{scope}
    \clip (2.35,1.85) ellipse (1.45 and 1.35);
    \clip (0,0) rectangle (12,1.95);

    \begin{scope}[even odd rule]
      \path[clip]
        (0,0) rectangle (12,4.2)
        (6.35,1.95) ellipse (4.25 and 1.15);

      \fill[yellow!25] (2.10,1.95)
        .. controls (1.90,1.55) and (1.78,1.00) .. (1.92,0.56)
        .. controls (2.75,0.35) and (3.55,0.75) .. (3.72,1.65)
        .. controls (3.15,1.75) and (2.55,1.88) .. (2.10,1.95);

      \draw[yellow!70!black, thick] (2.10,1.95)
        .. controls (1.90,1.55) and (1.78,1.00) .. (1.92,0.56)
        .. controls (2.75,0.35) and (3.55,0.75) .. (3.72,1.65)
        .. controls (3.15,1.75) and (2.55,1.88) .. (2.10,1.95);
    \end{scope}
  \end{scope}
  \node[yellow!50!black] at (2.75,1.05) {$R$};

  % J, inside B cap H near boundary
  \begin{scope}
    \clip (10.35,1.95) ellipse (1.45 and 1.35);
    \fill[red!25] (6.35,1.95) ellipse (4.25 and 1.15);
  \end{scope}
  \draw[green!50!black, thick] (6.35,1.95) ellipse (4.25 and 1.15);
  \node[red!50!black] at (9.65,1.9) {$J$};

  % B' enclosing (B \setminus J) together with R
  \draw[orange!85!black, ultra thick, dashed, line join=round]
    (2.10,1.95)
    .. controls (1.90,1.55) and (1.78,1.00) .. (1.92,0.56)
    .. controls (2.5536,0.3997) and (3.1697,0.5949) .. (3.5014,1.0965)
    arc[start angle=227.9137, end angle=312.5059, x radius=4.25, y radius=1.15]
    arc[start angle=-141.0979, end angle=-218.9021, x radius=1.45, y radius=1.35]
    arc[start angle=47.4941, end angle=180, x radius=4.25, y radius=1.15];
    \node[orange!85!black, align=center] at (7.75,0.4) {$B' = (B\setminus J)\cup R$};

  % Partition line drawn last
  \draw[thick] (6,0) -- (6,4.2);
  \node[font=\small\bfseries] at (3.5,3.9) {$L$};
  \node[font=\small\bfseries] at (8.5,3.9) {$H$};
  \node[align=center] at (0.8,3.7) {$L\setminus B$};  
  \node[align=center] at (11.2,3.7) {$H\setminus B$};  
\end{tikzpicture}
\caption{$E$ is partitioned into $L$ and $H$. $B$ and $B' = (B \setminus J) \cup R$ are bases of $M$. $A \subseteq L$ is a basis of $M_{|L}$. $R$ is a subset of $A' = A \setminus (B \cap L)$. Both $J \subseteq B \cap H$ and $R \subseteq A'$ have size $j$.}
\label{fig:coordinate-shadow-sets}
\end{figure}
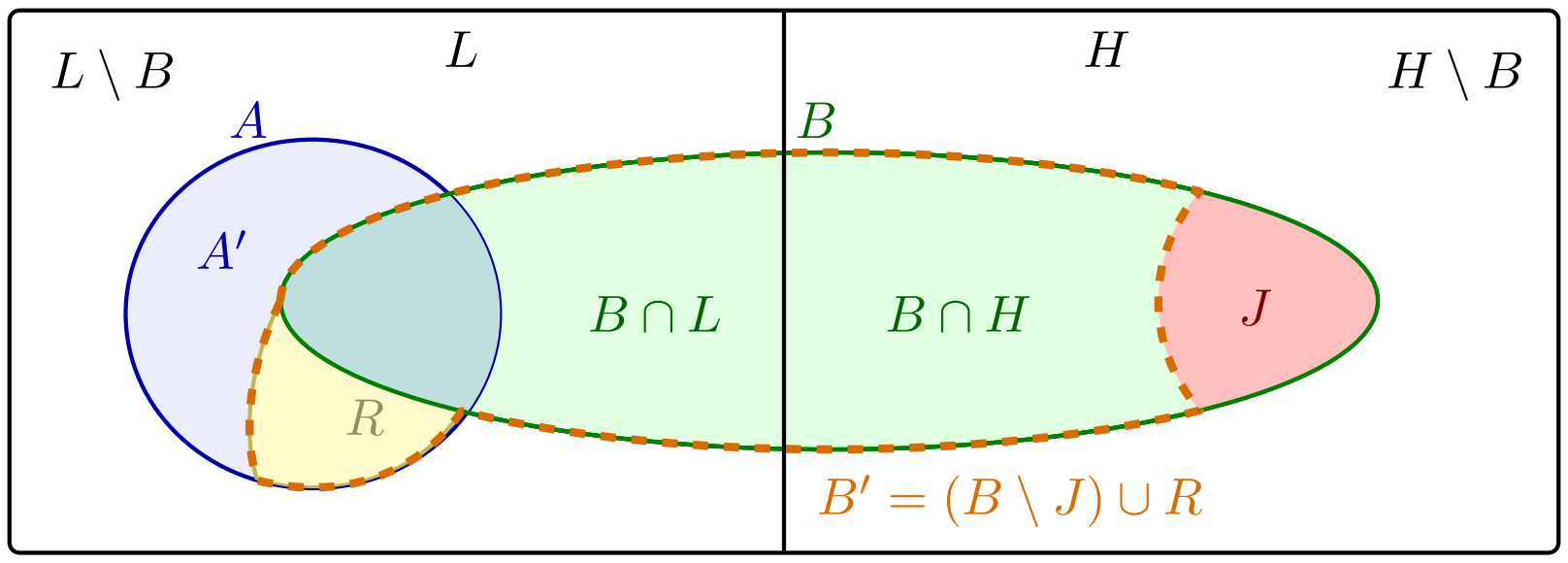

Note that $\Phi$ is injective since, from a triple $(B',R,J)$ we recover $B$ as $B = (B' \setminus R) \cup J$. Next, we uncondition our fixed choice of $B$. We can count all subsets of $A \setminus B$ of size $j$, which correspond to possible sets $R$, by counting all possible pairs $(B', J)$ that, together with a choice of $R$, yield a basis $B$ of $M$ via the map $\Phi$. In other words, if we denote the image of $\Phi$ by $\textsc{Im}(\Phi)$, we have
\[
\sum_{B \in \cB(M)} \binom{t}{j} \leq \abs{\textsc{Im}(\Phi)}.
\]
It remains to bound the size of the image of $\Phi$, by summing over the target bases $B'$. Fix a target basis $B'$. For an triple $(B', R, J)$ in the image of $\Phi$, we have
\begin{enumerate}
    \item $R \subseteq A \cap B'$, since $R$ is chosen from the subsets of $A$ and also $R \subseteq B'$,
    \item $J \subseteq H \setminus B'$, since $J$ was removed from $B'$, and
    \item $|R| = |J| = j$.
\end{enumerate}
Hence, for this fixed target basis $B'$, the number of possible pairs $(R,J)$ such that $(B', R, J) \in \textsc{Im}(\Phi)$ is at most
\[
\binom{|A \cap B'|}{j} \binom{|H \setminus B'|}{j}.
\]
Finally, summing over all target bases $B'$ gives
\[
\sum_{B \in \cB(M)} \binom{t}{j} \leq \sum_{B' \in \cB(M)} \binom{|A \cap B'|}{j} \binom{|H \setminus B'|}{j},
\]
and if we divide both sides by $|\cB(M)|$, we obtain that
\[
\E\brk{\binom{t}{j}} \leq \E\brk{\binom{|A \cap B|}{j} \binom{|H \setminus B|}{j}}. \qedhere
\]
\end{proof}

We will also need the following lemma which follows by applying Jensen's inequality to an appropriate convex extension of the binomial function.

\begin{lemma}\label{lem:convex-binomial-lower}
Let $W$ be an integer-valued random variable in $\set{0, \ldots, s}$. For any $j \geq 1$ and $w \ge j-1$, if $\E[W] \geq w$, then
\[
	\E\brk{\binom{W}{j}} \geq \binom{w}{j}.
\]
\end{lemma}
\begin{proof}
The idea is to extend the binomial function from the integers to the reals by linear interpolation. Specifically, let $f(k) = \binom{k}{j}$ and $\phi$ be the piecewise-linear interpolation of $f$ between the integer values. In other words,
\[
    \phi(x) = (\floor{x} + 1 - x) \binom{\floor{x}}{j} + (x - \floor{x}) \binom{\ceil{x}}{j}.
\]
Since $\phi$ is piece-wise linear, $\floor{x}$ and $\ceil{x}$ do not change between consecutive integers, and also it is differentiable in $\R \setminus \Z$. Therefore, its derivative is
\[
    \phi'(x) = \binom{\ceil{x}}{j} - \binom{\floor{x}}{j} = \binom{\floor{x}}{j-1},
\]
which is non-decreasing in $x$. Thus, $\phi$ is convex. By Jensen's inequality, we have
\[
    \E\brk{\binom{W}{j}} = \E[\phi(W)] \geq \phi(\E[W]).
\]
Also, since $\phi'(x) \geq 0$, $\phi$ is non-decreasing, and thus $\phi(\E[W]) \geq \phi(w)$. Finally, the polynomial $\binom{w}{j}$ is convex in $w$ and since $\phi$ is the linear interpolation between the integer points, we get that $\phi(w) \geq \binom{w}{j}$.
\end{proof}

We are finally ready to prove Theorem~\ref{thm:constant-diameter-lower}.

\begin{proof}[Proof of Theorem~\ref{thm:constant-diameter-lower}]
Combining Claim~\ref{clm:rank-mass-easy-inequalities} with \eqref{eq:avg2-lower-bound} and \eqref{eq:diam2-upper-bound} yields the theorem when $r_{M^*}(H) \geq (9/7) r_M(L)$ or $r_M(L) \geq (9/7) r_{M^*}(H)$. Assume that $(7/9) r_M(L) < r_{M^*}(H) < (9/7) r_M(L)$. We show that $p(L) \geq (1/8) r_M(L)$. The theorem then follows from Lemma~\ref{lem:dual-mass-rank-inequality}, again combined with \eqref{eq:avg2-lower-bound} and \eqref{eq:diam2-upper-bound}.

First, note that, if $r_M(L) = 0$ the inequality holds trivially. Also, if $1 \leq r_M(L) \leq 3$ then, since $r_M(L)$ is an integer, by our assumption, $(7/9) r_M(L) < r_{M^*}(H) < (9/7) r_M(L)$, we obtain that $\max\set{1, r_M(L)} \leq r_{M^*}(H) \leq \min\set{3, r_M(L)}$. By Lemma~\ref{lem:main-coordinate-shadow} for $j = 1$, we have
\[
    \E[t] = r_M(L) - p(L) \leq \E[Z \cdot |H \setminus B|].
\]
Furthermore, by the definition of the dual rank function,
\begin{equation}\label{eq:dual-rank-definition}
    r_{M^*}(H) = |H| - r + r_M(L),
\end{equation}
where $r = r(M)$ is the rank of $M$. Also,
\begin{equation}\label{eq:bookkeeping}
    |H \setminus B| = |H| - |B \cap H| = |H| - (r - |B \cap L|) \iff |H \setminus B| - |B \cap L| = |H| - r.
\end{equation}
By the above two equations, we get that $|H \setminus B| = |B \cap L| + r_{M^*}(H) - r_M(L) \leq |B \cap L|$, where the inequality follows from $r_{M^*}(H) \leq r_M(L)$. Since this inequality holds point-wise for all $B$, we obtain
\[
    \E[t] = r_M(L) - p(L) \leq \E[Z \cdot |B \cap L|] \leq r_M(L) \E[Z] \leq r_M(L) p(L).
\]
Re-arranging the above inequality, we obtain
\[
    p(L) \geq \frac{r_M(L)}{1 + r_M(L)} > \frac{r_M(L)}{8},
\]
for all $r_M(L) \in \set{1,2,3}$.

Next, we assume that $r_M(L) \geq 4$. Fix a constant $\lambda \in [0,1]$ to be decided later. Assume, towards contradiction, that $p(L) < \lambda \; r_M(L)$, and let $A$ be a basis of $M_{|L} = M \setminus H$ that minimizes $p(A)$. Also let $Z = |A \cap B|$ and $z = \E[Z]$. Then,
\[
    z = \E[|A \cap B|] = \sum_{e \in A} p_e \leq \sum_{e \in L} p_e = p(L) < \lambda \; r_M(L),
\]
where the last inequality follows from our assumption. Let $j = \ceil{(3/4) r_M(L)}$ and note that $j \geq 1$. Also, let $t \coloneqq r_M(L) - |B \cap L|$, with $\E[t] = r_M(L) - p(L) > (1-\lambda) r_M(L)$. Thus, by Lemma~\ref{lem:convex-binomial-lower}, we have
\begin{equation}\label{eq:contradiction-1}
    \E\brk{\binom{t}{j}} > \binom{\floor{(1 - \lambda) r_M(L)}}{j},
\end{equation}
for $\floor{(1 - \lambda) r_M(L)} \geq j - 1 = \ceil{(3/4) r_M(L)} - 1$, which holds for $r_M(L) \geq 4$ if $\lambda \leq 1/2$. The inequality is strict, because $\E[t] > \floor{(1-\lambda) r_M(L)}$ and the interpolation is strictly increasing beyond $j-1$. We show that \eqref{eq:contradiction-1} contradicts Lemma~\ref{lem:main-coordinate-shadow}, which says that
\[
    \E\brk{\binom{t}{j}} \leq \E\brk{\binom{Z}{j} \binom{|H \setminus B|}{j}}.
\]
First, we deal with $|H \setminus B|$. Since $B \in \cB(M)$, we have that $E \setminus B \in \cB(M^*)$, which implies that $H \setminus B$ is an independent set of $M^*$. Thus, $|H \setminus B| = r_{M^*}(H \setminus B) \leq r_{M^*}(H)$. By \eqref{eq:bookkeeping} and \eqref{eq:dual-rank-definition} we obtain that
\[
    r_{M^*}(H) - r_M(L) = |H \setminus B| - |B \cap L| = \bar{p}(H) - p(L).
\]
By our assumption, we have that $r_M(L) > (7/9) \; r_{M^*}(H)$. Since $|H \setminus B| \leq r_{M^*}(H) \leq (9/7) r_M(L)$ and the function $f(x) = \binom{x}{j}$ is increasing in $x$ for $x \in [j-1, +\infty)$, we obtain that
\[
    \binom{|H \setminus B|}{j} \leq \binom{\ceil{(9/7) \: r_M(L)}}{j},
\]
where the inequality holds point-wise for any $|H \setminus B|$. Thus,
\[
    \E\brk{\binom{Z}{j} \binom{|H \setminus B|}{j}} \leq \binom{\ceil{(9/7) \: r_M(L)}}{j} \E\brk{\binom{Z}{j}}.
\]
Next, we deal with $\E\brk{\binom{Z}{j}}$. Recall that $z = \E[Z] < \lambda r_M(L)$. Using Lemma~\ref{lem:ulc-moment}, we get that
\[
    \E\brk{\binom{Z}{j} \binom{|H \setminus B|}{j}} \leq \binom{\ceil{(9/7) \: r_M(L)}}{j} \frac{z^j}{j!} \leq \binom{\ceil{(9/7) \: r_M(L)}}{j} \frac{\prn{\lambda r_M(L)}^j}{j!}.
\]
Note that, by Lemma~\ref{lem:main-coordinate-shadow}, this implies that
\begin{equation}\label{eq:contradiction-2}
\E\brk{\binom{t}{j}} \leq \binom{\ceil{(9/7) \: r_M(L)}}{j} \frac{\prn{\lambda r_M(L)}^j}{j!}.
\end{equation}
We claim that \eqref{eq:contradiction-1} contradicts \eqref{eq:contradiction-2}. In other words, we claim that, for an appropriate choice of $\lambda \in (0,1)$,
\begin{equation}\label{eq:contradicting-inequality}
\binom{\floor{(1 - \lambda) r_M(L)}}{j} \geq \binom{\ceil{(9/7) \: r_M(L)}}{j} \frac{\prn{\lambda r_M(L)}^j}{j!}.
\end{equation}
First, we simplify \eqref{eq:contradicting-inequality}.
\begin{align}
\binom{\floor{(1 - \lambda) r_M(L)}}{j} &\geq \binom{\ceil{(9/7) \: r_M(L)}}{j} \frac{\prn{\lambda r_M(L)}^j}{j!} &\iff \nonumber \\
\prod_{k = 0}^{j-1} \floor{(1-\lambda) r_M(L) - k} &\geq \frac{\prn{\lambda r_M(L)}^j}{j!} \prod_{k = 0}^{j-1} \ceil{(9/7) \: r_M(L) - k} &\iff \nonumber \\
\label{eq:lambdas} \prod_{k = 0}^{j-1} \frac{\floor{(1-\lambda) r_M(L) - k}}{\ceil{(9/7) \: r_M(L) - k}} &\geq \frac{\prn{\lambda r_M(L)}^j}{j!}. &
\end{align}

For the factors above to be all positive, we need that $\ceil{(9/7) \: r_M(L) - j + 1} > 0$ which is true for $r_M(L) \geq 4$, and $\floor{(1-\lambda) r_M(L) - (j - 1)} > 0 \iff \floor{(1-\lambda) r_M(L) - \ceil{(3/4) r_M(L)} + 1} > 0$. The latter implies
\[
    (1-\lambda) r_M(L) - \ceil{(3/4) r_M(L)} + 1 \geq 1 \iff (1-\lambda) r_M(L) \geq \ceil{(3/4) r_M(L)} \iff \lambda \leq 1 - \frac{\ceil{\frac{3 r_M(L)}{4}}}{r_M(L)},
\]
and since $r_M(L) - \ceil{(3/4) r_M(L)} = \floor{(1/4) r_M(L)}$, we get $\lambda \leq \floor{(1/4) r_M(L)}/r_M(L)$. Since $\lambda > 0$, this implies $r_M(L) \geq 4$, which is true; thus all terms of \eqref{eq:lambdas} are positive. Also, the left-hand side is decreasing in $\lambda$, while the right-hand side in increasing in $\lambda$. Furthermore, note that, for $\lambda$ close to $0$, the left-hand side is positive, while the right-hand side tends to $0$. Thus, there exists a $\lambda^* \in (0,1)$ such that \eqref{eq:lambdas} holds for $\lambda \in (0, \lambda^*)$. The proof of the following claim can be found in Appendix~\ref{app:proofs}.

\begin{restatable}{claim}{lambdaAsymptotic}\label{clm:lambda-asymptotic}
Let
\begin{align*}
    F(\lambda) &= (1-\lambda)\log(1-\lambda) -\prn{\frac14-\lambda}\log\prn{\frac14-\lambda} \\
    &\qquad -\frac97\log\frac97+\frac{15}{28}\log\frac{15}{28} +\frac34\log\frac{3}{4\lambda}-\frac34,
\end{align*}
$\lambda_F \approx 0.13525$ be the unique root of $F$ in $(0,1/4)$ and $\lambda^* = \frac{2}{15} < \lambda_F$. Then, for every $\lambda\in(0,\lambda^*)$ and every integer $\ell \geq 4$, if
$j=\ceil{3\ell/4}$, then
\[
    \prod_{k = 0}^{j-1} \frac{\floor{(1-\lambda)\ell-k}}{\ceil{(9/7)\ell-k}} \geq \frac{(\lambda\ell)^j}{j!}.
\]
\end{restatable}

Choose $\lambda = 1/8 \in (0, \lambda^*)$ in the above claim. We conclude that \eqref{eq:lambdas} holds, contradicting our assumption that $p(L) < (1/8) r_M(L)$. Therefore,
\[
p(L) \geq \frac{r_M(L)}{8},
\]
and the theorem follows.
\end{proof}

\begin{remark}
It is possible to optimize the constants further to obtain that
\[
\avg_2(M) \geq \frac{\diam_2(M)}{c},
\]
for $c \approx 15.02$, but we choose to present a cleaner number for the sake of clarity and to avoid further complicated calculations.
\end{remark}

%% file: sections/diameter-higher.tex
\subsection{A Transversal Upper Bound}

Here, we show the following $4/9$ upper bound on the approximation obtained by sampling bases at random.

\higherdiameter*

To show the above theorem, we construct a sequence of transversal matroids $\set{F_n}_{n \geq 1}$. The construction is defined below.

\begin{definition}[The transversal matroid $F_n$]
Let $n \geq 1$. We define a matroid $F_n = (E, \cF)$ as follows. The ground set $E$ consists of four $n$-element sets $A_F, A_R, D_F$ and $D_R$, where
\[
    A_F = \set{a_i^F}, \qquad
    A_R = \set{a_i^R}, \qquad
    D_F = \set{d_i^F}, \qquad
    D_R = \set{d_i^R}.
\]
A bipartite representation of $F_n$ consists of a bipartite graph $G = (E \cup O, A)$ where $O = \set{f_1, \ldots,f_n} \cup \set{r_1,\ldots,r_n}$, and the edge set $A$ consists of the edges
\[
    A =
    \prn{\bigcup_{i = 1}^n \set{a_i^F, f_i}} \cup
    \prn{\bigcup_{i = 1}^n \set{a_i^R, r_i}} \cup
    \prn{\bigcup_{i = 1}^n \set{d_i^F, f_i}} \cup
    \prn{\bigcup_{i = 1}^n \set{d_i^R, r_i}} \cup
    \prn{\bigcup_{i,j = 1}^n \set{d_i^R, f_j}}.
\]
The independent sets $\cF$ of $F_n$ are the subsets of $E$ that admit a perfect matching in $G$.
\end{definition}

Observe that the sets $A_F \cup A_R$ and $D_F \cup D_R$ are disjoint bases, since the former can be matched to $O$ using the $\set{a^F_i, f_i}$ and $\set{a^R_i, r_i}$ edges, while the latter via the $\set{d^F_i, f_i}$ and $\set{d^R_i, r_i}$ edges. Hence $F_n$ has rank $2n$ and $\diam_2(F_n) = 2n$.

\begin{lemma}\label{lem:ferrers-count}
The marginal $p_L(n)$ of every element of $A_F \cup D_F$ with respect to $U(\cB(M))$ and the marginal $p_H(n)$ of every element of $A_R \cup D_R$ satisfy
\[
    \lim_{n \to \infty} p_L(n) = \frac{1}{3} \qquad \text{and} \qquad \lim_{n \to \infty} p_H(n) = \frac{2}{3}.
\]
\end{lemma}
\begin{proof}
We first count the number of bases of $F_n$ via a generating function method. For a polynomial $P(z)$ we write $[z_0] P(z)$ for the coefficient of $z_0$ in $P(z)$. Fix a basis $B$. For a vertex $f_i \in O$, there are three possibilities: it is matched to either $a_i^F$, $d_i^F$, or to some element of $D_R$. The first two choices fill $f_i$ locally and the third leaves $f_i$ open, to be filled by some element of $D_R$ later on, so we denote the third case by $z$. Thus, the local generating function for the $F$ side at index $i$ is $2+z$. Similarly, for a vertex $r_i \in O$, we either match it to $d_i^R$, or to $a_i^R$ and let $d_i^R$ be unmatched, or we match $r_i$ to $a_i^R$ while matching $d_i^R$ to some $f_j$. The first two choices fill $r_i$ locally and the third fills $r_i$ by $a_i^R$ and leaves $d_i^R$ matched to some $f_j$, so we mark this extra element of $D_R$ by $z^{-1}$. Thus, the local generating function for the $R$ side at index $i$ is $2+z^{-1}$. Note that we have $n$ independent terms for each case, so we multiply them and expand both factors. We have
\[
    (2+z)^n = \sum_{k = 0}^n \binom{n}{k} 2^{n-k} z^k,
\]
where $k$ denotes the number of locally-unfilled vertices $f_i$, and
\[
    \prn{2+z^{-1}}^n = \sum_{\ell = 0}^n \binom{n}{\ell} 2^{n-\ell} z^{-\ell},
\]
where $\ell$ denotes the number of extra vertices of $D_R$ to be matched to some $f_j$. Multiplying both yields
\[
    (2+z)^n prn{2+z^{-1}}^n = \sum_{k,\ell = 0}^n \binom{n}{k} \binom{n}{\ell} 2^{2n-k-\ell} z^{k-\ell}.
\]
Notice that we get a basis only when the number of locally-unfilled vertices $f_i$ is equal to the number of extra vertices of $D_R$ to be matched to some $f_j$ and thus, when $k = \ell$ in the sum above. Therefore,
\[
    |\cB(F_n)| = \sum_{k = 0}^n \binom{n}{k}^2 4^{n-k}.
\]
In other words, for $Q(z) = (2+z)\prn{2+z^{-1}}$, the number of bases of $F_n$ is equal to the constant term of $(Q(z))^n$, i.e. $|\cB(F_n)| = [z^0] (Q(z))^n$.

For a fixed element of $A_F$ or $D_F$, the count with that element present is
\[
  [z^0](2+z)^{n-1}(2+z^{-1})^n = [z^0](2+z^{-1})Q(z)^{n-1}.
\]
For a fixed element of $A_R$ or $D_R$, the count is
\[
  (1+z^{-1})(2+z)^n(2+z^{-1})^{n-1} = \prn{1+z^{-1}}(2+z)Q(z)^{n-1}.
\]
For every fixed polynomial $R$,
\[
  \frac{R(z)(Q(z))^{n-1}}{(Q(z))^n}
  \longrightarrow \frac{R(1)}{Q(1)}.
\]
Applying this with $R(z)=2+z^{-1}$ and $R(z)=(1+z^{-1})(2+z)$ gives the claimed marginals.
\end{proof}

\Cref{thm:constant-diameter-upper} now follows from \Cref{lem:ferrers-count}. Indeed, $F_n$ has $2n$ elements with marginal $p_L(n)$ and $2n$ elements with marginal $p_H(n)$. Since $\diam_2(F_n) = 2n$,
\begin{align*}
  \frac{\avg_2(F_n)}{\diam_2(F_n)} &= \frac{1}{2n} \sum_{e \in E} p_e(1 - p_e) = \frac{1}{2n} \sum_{e \in A_F \cup D_F} p_L(n)(1 - p_L(n)) + \frac{1}{2n} \sum_{A_R \cup D_R} p_H(n)(1 - p_H(n)) \\
  &= p_L(n)(1 - p_L(n)) + p_H(n)(1 - p_H(n)) \\
  &\to \frac{1}{3} \cdot \frac{2}{3} + \frac{2}{3} \cdot \frac{1}{3} = \frac{4}{9}.
\end{align*}
Therefore, for any $\eps > 0$, there exists an $n = n(\eps)$ such that $p_L(n)(1 - p_L(n)) + p_H(n)(1 - p_H(n)) \leq 4/9 + \eps$.

%% file: sections/diam-edge-expansion-hard.tex
\subsection{Diameter in General Graphs with High Edge-Expansion}

For a graph $G=(V,E)$, its edge expansion is defined as
\[
    h(G)
    \coloneqq
    \min_{\substack{S\subseteq V \\ 0<|S| \leq  |V|/2}}
    \frac{|\partial S|}{|S|},
\]
where $\partial S$ is the set of edges crossing the cut $(S,V\setminus S)$.

In this section, we show that distances between randomly sampled pairs of vertices can significantly underestimate the diameter of a graph, even when the graph has a high edge expansion.
A trivial example where random sampling fails is a chain of cliques constructed as follows.
The first clique consists of a single vertex, and the size of the $i$-th clique is twice that of the previous clique.
Moreover, every pair of consecutive cliques is connected by a complete bipartite graph.

Since the resulting graph consists of $ \Theta(\log n)$ cliques, it has diameter $\Theta(\log n)$.
However, because the largest clique contains a half of all vertices, two uniformly sampled vertices are likely to lie among the largest few cliques, and hence their distance is typically much smaller than the diameter.
In fact, for any constant $ c \geq 1 $ and two uniformly sampled vertices $x$ and $y$, $ d(x,y) < \diam_2(G)/c $ with high probability.

However, the example above does not reflect the behavior of basis-exchange graphs of matroids since the maximum degree $\Delta(G)$ is $\Theta(n)$.
To show that random sampling still fails even when $\Delta(G) = \cO(\log n)$, we replace the cliques by sparse expanders.
More precisely, on each sufficiently large
layer, we first place a 10-regular graph with edge expansion at least $2$ (See \Cref{obs:bollobas}).
We then add a $\lceil \log n\rceil$-ary complete tree on the same vertex set to keep the diameter of each layer is still small. Consequently, the $i$-th layer has diameter $ \cO(\log_{\lceil\log n\rceil}|V_i| ) = \cO(\frac{\log n}{\log\log n} )  $, where $V_i$ denote the vertex set of the $i$-th layer, whereas the diameter of the entire graph remains $\Theta(\log n)$. However, two uniformly sampled vertices are still likely to lie among the largest few layers. To prove this claim more formally, we begin with a simple observation, which follows immediately from~\cite{bollobas1988isoperimetric}.

\begin{observation}[\cite{bollobas1988isoperimetric}]\label{obs:bollobas}
    For all sufficiently large $n$, there exists a 10-regular graph $G$ on $n$ vertices such that $h(G) \geq  2$.
\end{observation}
\begin{proof}
    Let $G(n, r)$ be a uniformly random $r$-regular graph on $n$ vertices.
    By~\cite{bollobas1988isoperimetric}, if $r \geq  3$ and $\eta \in (0,1)$ satisfy
    \begin{equation}\label{eq:bollobas-condition}
        (1-\eta)\log_2(1-\eta) + (1+\eta)\log_2(1+\eta) > \frac{4}{r},
    \end{equation}
    then
    \begin{equation}\label{eq:bollobas-prob}
        \Prob{h(G(n,r)) \geq  (1-\eta)r/2} \to 1
         \text{ as } n \to \infty.
    \end{equation}
    Applying this with $r=10$ and $\eta=0.6$ immediately proves our claim.
\end{proof}

Now, we formally state our theorem.

\EdgeExCounter*
\noindent\textit{Proof.} Fix a constant $ c \geq  1$. For each integer $\ell \geq  1$, let $ n_{\ell} = 2^{\ell+2}-4$.

\setlength{\columnsep}{5pt}
\begin{wrapfigure}[15]{r}{0.4\textwidth}
  \vspace{-\baselineskip}
  \centering
  \includegraphics[
  width=0.7\linewidth,
  trim=0 10cm 0 0,
  clip
]{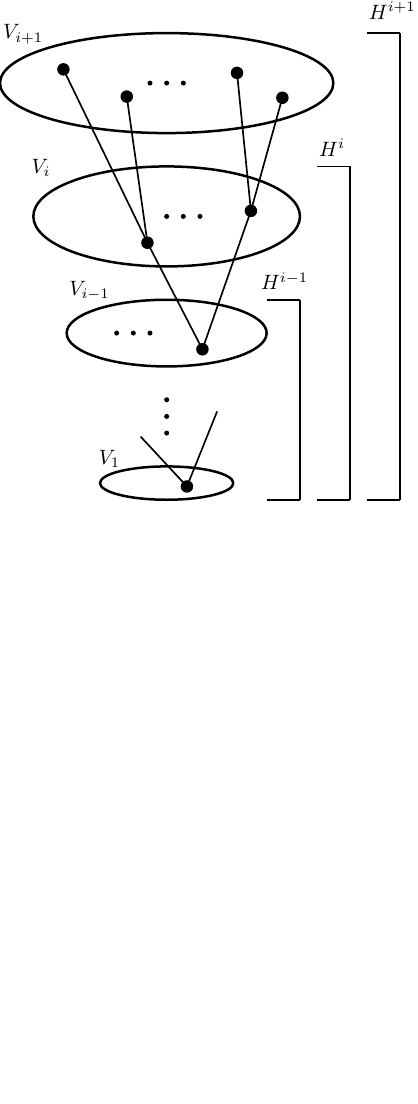}
  \caption{ }
  \label{fig:G-construction}
  \vspace{-\baselineskip}
\end{wrapfigure}
We build a graph $G_{\ell}$ on $n_{\ell}$ vertices such that $ \Delta(G_{\ell}) = \cO(\log {n_\ell})$, $\diam[G_{\ell}]=\Theta(\log {n_\ell}) $ and $ h(G_{\ell}) \geq  1 $.
For notational simplicity, given $\ell$, we write $G$ in place of $ G_{\ell} $ and $n$ in place of $n_\ell$ as long as the intended meaning is clear from the context.

\vspace{2mm}\textbf{(Construction of $G$.)}
Let $V_1, V_2, \ldots, V_{\ell}$ be collections of vertices such that $ |V_1| = 4 $ and $ |V_i| = 2|V_{i-1}|  $ for every $ i \geq  2 $.
Note that $ |V_i| = 2^{i+1} $ and $ \sum_{i=1}^{\ell} |V_i| = 2^{\ell+2}-4 = n $.
We build an expander graph on each $V_i$, which we will call the $i$-th layer, then connect every consecutive layers to build our graph $G$.

Define a graph $R_i$ on each $i\in[\ell]$ as follows. If $ |V_i| $ is sufficiently large, $R_i$ is a 10-regular graph with edge-expansion at least 2 due to \Cref{obs:bollobas}. Otherwise, $R_i$ is a complete graph.
To bound the diameter of this layer, we add additional edges to it.
Let $ F_i $ be a $ \lceil \log n \rceil $-ary complete tree on $V_i$. Define the $i$-th layer $L_i$ as the graph $ L_i \coloneqq \left( V_i, E(R_i) \cup E(F_i) \right) $.
Note that $L_i$ has edge expansion at least 2 due to $R_i$ and diameter $ \cO(\log_{\lceil \log n \rceil} |V_i|) =  \cO\left(\frac{\log n}{\log\log n}\right)$ due to $F_i$.

Now, connect $ L_i $ and $L_{i+1}$ for every $i\in[\ell-1]$ by a surjection $ \pi_i: V_{i+1} \rightarrow V_i $ such that $ |\pi^{-1}_i(u)| = 2 $ for every $u \in V_i $. In other words, every vertex in $V_i$ corresponds to two distinct vertices in $V_{i+1}$ and every vertex in $V_{i+1}$ corresponds to exactly one vertex in $V_i$. This mapping exists since $V_{i+1}$ contains exactly twice as many vertices as $V_i$ for every $i\in[\ell-1]$.

Now, define $ H^i = (V^i, E^i) $ as
\begin{equation}
    V^i \coloneqq \bigcup_{j=1}^i V_j \qquad \text{and} \qquad E^i \coloneqq \left(\bigcup_{j=1}^i E(L_j)\right) \cup \left(\bigcup_{j=1}^{i-1}E(L_j, L_{j+1})\right),
\end{equation}
where $ E(L_j, L_{j+1}) $ denotes the edges defined by $\pi_j $ between the layers $L_j$ and $L_{j+1}$. In other words, $ H^i $ comprises all vertices and edges from the first layer up to the $i$-th layer; see \Cref{fig:G-construction}.
We take $ H^{\ell} $ as our graph $ G = (V,E) $.
By construction, it is trivial to show that the maximum degree of a vertex of $G$ is $\cO(\log n)$.
For the diameter of $G$, note that $ \operatorname{diam}_2(G) \geq  \ell - 1 $ because any path from a vertex in $V_1$ to a vertex in $V_{\ell}$ must cross at least one inter-layer edge for each consecutive pair of layers. Thus, $ \operatorname{diam}_2(G) \geq  \ell - 1 $. On the other hand, for any $ x \in V_i $ and $ y \in V_ j$, where $ i \leq  j $,
\[ d(x, y) \leq  (j-i) + \operatorname{diam}_2(L_i) \leq  \ell + \cO\left(\frac{\log n}{\log\log n}\right) = \cO(\log n). \]
Therefore, $ \operatorname{diam}_2(G) = \Theta(\ell) = \Theta(\log n) $.

Now, we show that $ h(G) \geq  1$.

\vspace{2mm}\textbf{(Edge-expansion of $G$ is at least 1.)}
We prove by induction on $i$ that $h(H^i)\geq 1$. The base case is clear since $H^1=L_1$ and $h(L_1)\geq 2$.

\setlength{\columnsep}{5pt}
\begin{wrapfigure}[18]{r}{0.5\textwidth}
  \vspace{-\baselineskip}
  \vspace{-\baselineskip}
  \centering
  \includegraphics[width=0.7\linewidth]{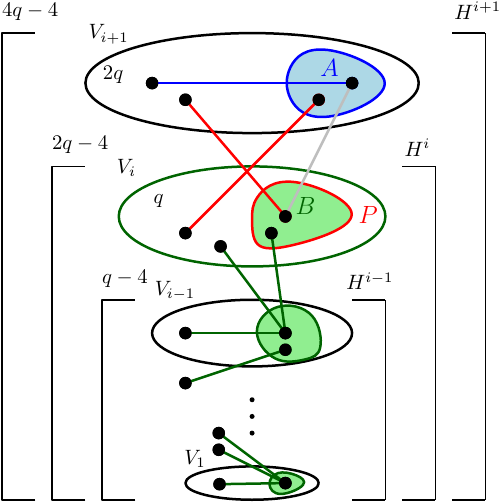}
  \caption{}
  \label{fig:S-illustration}
\end{wrapfigure}

Assume $h(H^i)\geq 1$. Let $S\subseteq V^{i+1}$ satisfy $0<|S|\leq |V^{i+1}|/2$. Put
\[
A=S\cap V_{i+1},\qquad B=S\cap V^i,\qquad P=S\cap V_i,
\]
and let $a=|A|$, $b=|B|$ and $p=|P|$.
Also let $q=|V_i|$. Then
\[
|V_{i+1}|=2q,\quad |V^i|=2q-4,\quad |V^{i+1}|=4q-4,
\]
and therefore
\begin{equation}\label{eq: ub: a + b}
    a+b=|S|\leq 2q-2.
\end{equation}
The boundary of $S$ has three parts: edges inside $V_{i+1}$, edges
between $V_{i+1}$ and $V_i$, and edges inside $H^i$; see \Cref{fig:S-illustration} for illustration. Therefore
\begin{equation}\label{eq: lb: partialS}
    |\partial S| \geq 2\min\{a,2q-a\} + |a-2p| + \min\{b,2q-4-b\},
\end{equation}
where the first term follows from $ h(L_{i+1}) \geq 2 $, the second term from the fact that the bipartite edges between $V_{i+1}$ and $V_i$ are defined by a 2-to-1 map, and the last from the induction hypothesis $ h(H^i) \geq 1 $.
We also have that
\begin{equation}\label{eq: bound: p}
    b-(q-4) \leq p \leq b
\end{equation}
since $ P \subseteq B $ and  $ V^{i-1} $ has $q-4$ vertices in it.

We now show that $|\partial S|\geq a+b$.
First suppose $a>q$. Then by \eqref{eq: ub: a + b}, $ b \leq 2q - a - 2 < q - 2 $.
Therefore, $ \min\{  b, 2q - 4 - b \} = b $.
Hence,
\begin{equation}\label{eq: a > q}
\begin{split}
    |\partial S| &\geq 2(2q - a) + (a-2p) + b
    = 4q - a - 2p + b \\
    &\overset{\eqref{eq: bound: p}}{\geq} 4q - a - b
    \geq 2(2q-2) - a - b \overset{\eqref{eq: ub: a + b}}{\geq} a + b.
\end{split}
\end{equation}

Now suppose $a\leq q$. Then \eqref{eq: lb: partialS} together with \eqref{eq: bound: p} give
\begin{equation}\label{eq: a <= q}
\begin{split}
    |\partial S| &\geq 2a + (2p - a) + \min\{b,2q-4-b\}
    = a + 2p + \min \{ b, 2q-4-b \} \\
    &\overset{\eqref{eq: bound: p}}{\geq} a + 2\max\{0, b-(q-2)\} + \min \{ b, 2(q-2)-b \} \geq a + b,
\end{split}
\end{equation}
where the last inequality is immediate by checking $ b\leq q-2 $ and $ b>q-2 $.
Combining \eqref{eq: a > q} and \eqref{eq: a <= q}, we obtain $ |\partial S|\geq a+b=|S| $, hence
$ h(H^{i+1}) \geq 1 $. By induction, \[ h(G) = h(H^\ell) \geq 1, \]
as we desired.

Finally, we show that a uniform random pair of vertices does not approximate the diameter with constant probability.

\vspace{2mm}\textbf{(Inapproximability of $\operatorname{diam}_2(G)$.)} We first show that $\mathbb{E}[d(x, y)] = \cO\left(\frac{\log n}{\log\log n}\right)$ for any two random vertices $ x $ and $y$ drawn independently and uniformly at random from $V$. We then apply Markov inequality to prove our claim.

Let $ I(v) $ denote the unique layer index of $v$, so that $ v \in V_{I(v)} $. Let $ Z(x) = \ell - I(x) $, the number of layers between $v$ and the largest layer $L_{\ell}$.
Since the sizes of the layers double, for $x$ uniform on $V(G)$,
\[ \Prob{Z(x)\geq  t} \leq  2^{1-t}. \]
Therefore,
\[ \mathbb{E}[Z(x)] = \sum_{t \geq  1} \Prob{ Z(x) \geq  t } \leq  \sum_{t \geq  1} 2^{1-t} = 2, \]
and thus,
\[ \mathbb{E}[|I(x) - I(y)|] = \mathbb{E}[|Z(x) - Z(y)|] \leq  \mathbb{E}[Z(x)] + \mathbb{E}[Z(y)] \leq  4. \]
Therefore,
\[ \mathbb{E}[d(x,y)] \leq  \mathbb{E}[| I(x) - I(y) |] + \max_i \operatorname{diam}_2(L_i) \leq  4 + \cO\left(\frac{\log n}{\log\log n}\right) = \cO\left(\frac{\log n}{\log\log n}\right). \]
Moreover, since $\operatorname{diam}_2(G) \geq \ell-1 = \Theta(\log n)$, we finally have
\[ \Prob{d(x,y) \geq  \frac{1}{c}\operatorname{diam}_2(G)} \leq  \frac{c \mathbb{E}[d(x,y)]}{\operatorname{diam}_2(G)} = \cO\left(\frac{1}{\log\log n}\right) , \] 
which tends to zero as $n \rightarrow \infty $. This completes the proof.
\qed

%% file: sections/diversity.tex
\section{Diversity}
\label{sec:diversity}

In this section we prove algorithmic results for finding far-apart bases of an arbitrary matroid.
We begin with the most basic algorithm: choose bases at random.
Using this algorithm, we obtain an $\cO(\log k)$-approximation for the Diverse Matroid Basis problem, and we show that this is tight.
We then give a polynomial-time $3$-approximation for the Farthest Matroid Basis problem and a QPTAS for the same problem.
These results imply a $\frac{1}{6}$-approximation in polynomial time and a $\frac{1}{2}-\varepsilon$ approximation in quasi-polynomial time for the Diverse Matroid Basis problem via the standard farthest-insertion argument.

\subsection{Diversity via Random Sampling}
\label{ss:diversity-via-naive-samples}

In this subsection, we show that sampling $k$ bases uniformly without replacement gives an $\cO(\log k)$-approximation for the Diverse Matroid Basis problem, and that this logarithmic loss is tight for random sampling.
More formally, let $\avg_k(M)$ denote the expected minimum pairwise distance between $k$ distinct bases of $M$ sampled uniformly at random; i.e.,
\[
    \avg_k(M) = \EE_{B_1,\ldots,B_k}\left[ \min_{i<j} d(B_i,B_j) \right].
\]
Here, the expectation is over bases $B_1,\ldots,B_k$ sampled uniformly without replacement.
Thus, in this language, we show that $\avg_k(M)$ is an $\cO(\log k)$-approximation to $\diam_k(M)$, and that this upper bound is tight for some matroids.

\subsubsection{A logarithmic sampling guarantee}

We begin by relating the average distance between two random bases to the average distance from any fixed base.

\begin{lemma}
    \label{lem:fixed-base-average}
    Let $A \in \cB(M)$ be a fixed base of $M$, then $2\EE_B[d(A,B)] \geq \EE_{B_1,B_2} [d(B_1,B_2)]$.
\end{lemma}
\begin{proof}
    For $e \in E$, let $p_e$ be the probability that $e$ belongs to a uniformly random base of $M$.
    By linearity of expectation, we have
    \begin{align*}
      \EE_{B_1,B_2} [d(B_1,B_2)]  & = \sum_{e \in E} p_e (1-p_e) \\
        & = \sum_{e \in A} p_e (1-p_e) + \sum_{e \notin A} p_e (1-p_e) \\ 
        & \leq \sum_{e \in A} (1-p_e) +  \sum_{e \notin A} p_e \\ 
        & = \EE_{B} [|A \setminus B|] + \EE_{B}[|B \setminus A|] \\
        & = 2\EE_{B} [d(A,B)] \qedhere
    \end{align*}
\end{proof}

The previous lemma says that, from every fixed base, the expected distance to a random base is large whenever the average pairwise distance is large.
We next use the moment bounds from Lemma~\ref{lem:ulc-moment} to show that this distance is unlikely to fall far below its mean.
\begin{lemma}
    \label{lem:concentration}
    Let $A \in \cB(M)$ be a fixed base of $M$.
    Define the random variable $d_A = d(A,Y)$, where $Y$ is drawn uniformly at random from the set of bases of $M$, and let $\mu_A = \EE[d_A]$.
    Then, for every $\alpha \in (0,1)$,
    \[
    \PP_{Y} \left[d_A \leq \alpha \cdot \mu_A\right] \leq \exp\left(-\frac{(1-\alpha)^2}{2}\mu_A\right). 
    \]
\end{lemma}
\begin{proof}
    Since $A$ and $Y$ are both bases of $M$, we have that $d_A = d(A,Y) = |A\setminus Y| = r - |A \cap Y|$, where $r$ is the rank of $M$.
    Set $Z = |A \cap Y| = r - d_A$ and note that $\EE[Z] = r - \mu_A$.
    We apply Lemma~\ref{lem:ulc-moment} to $A$ and $Z$, obtaining that for every $j \in \NN$,
    \begin{equation}
        \EE\left[ \binom{Z}{j} \right] \leq \binom{r}{j} \left(\frac{\EE[Z]}{r}\right)^j = \binom{r}{j} \left(1 - \frac{\mu_A}{r}\right)^j.
        \label{eq:moment-bound-1}    
    \end{equation}
    
    We now use the moment bounds to obtain a bound on the moment generating function of $d_A$.
    Let $\lambda > 0$, and note that
    \begin{align*}
        \EE[e^{-\lambda d_A}] & = e^{- \lambda r} \EE[e^{\lambda Z}] \\
        & = e^{- \lambda r} \EE[(1+(e^{\lambda}-1))^Z] \\
        & = e^{- \lambda r} \sum_{j\geq 0} (e^{\lambda}-1)^j \EE\left[ \binom{Z}{j} \right] \\
        & \overset{\eqref{eq:moment-bound-1}}{\leq} e^{- \lambda r} \sum_{j\geq 0} (e^{\lambda}-1)^j \binom{r}{j} \left(1 - \frac{\mu_A}{r}\right)^j \\
        & = \bigg[e^{- \lambda} \left(1 + (e^{\lambda}-1)\left(1 - \frac{\mu_A}{r}\right)\right) \bigg]^r \\
        & = \left(1 - \frac{\mu_A}{r}(1-e^{-\lambda})\right)^r \\ 
        & \leq \exp\left(-\mu_A(1-e^{-\lambda})\right).
    \end{align*}

    We now apply Markov's inequality to $e^{-\lambda d_A}$ to obtain a tail bound for $d_A$.
    By Markov's inequality, we have that for every $\alpha \in (0,1)$,
    \begin{align*}
        \PP[d_A \leq \alpha \cdot \mu_A] & = \PP[e^{-\lambda d_A} \geq e^{-\lambda \alpha \mu_A}] \\
        & \leq \frac{\EE[e^{-\lambda d_A}]}{e^{-\lambda \alpha \mu_A}} \\
        & \leq \exp\left(-\mu_A(1-e^{-\lambda}) + \lambda \alpha \mu_A\right).
    \end{align*}
    Setting $\lambda = -\log(\alpha)$, we obtain that $e^{-\lambda} = \alpha$, and thus,
    \[
    \PP[d_A \leq \alpha \cdot \mu_A] \leq \exp\left(-\mu_A(1-\alpha+\alpha\log\alpha)\right) \leq \exp\left(-\frac{(1-\alpha)^2}{2}\mu_A\right),
    \]
    where the last inequality uses $1-\alpha+\alpha\log\alpha \geq (1-\alpha)^2/2$ for $\alpha\in(0,1)$.
\end{proof}

The concentration lemma has the following useful consequence over distances between two random bases without replacement.

\begin{lemma}\label{lem:close-pair-from-average}
Let $M$ be a matroid with at least two bases.
Let $X,Y$ be independent uniformly random bases of $M$, and set $\mu=\EE_{X,Y}[d(X,Y)]$.
Let $B_1,B_2$ be two bases sampled uniformly without replacement.
Then, for every $\tau>0$ with $\tau\leq \mu/4$,
\[
    \PP[d(B_1,B_2)<\tau] \leq 2\exp\left(-\frac{\mu}{16}\right).
\]
\end{lemma}
\begin{proof}
Fix a base $A$ and write $\mu_A=\EE_Y[d(A,Y)]$, where $Y$ is a uniformly random base.
By Lemma~\ref{lem:fixed-base-average}, $\mu_A\geq \mu/2$.
Since $\tau\leq \mu/4$, we have $\tau/\mu_A\leq 1/2$.
Applying Lemma~\ref{lem:concentration} with $\alpha_A=\tau/\mu_A$, we get
\[
    \PP_Y[d(A,Y)<\tau]
    \leq
    \exp\left(-\frac{\mu_A}{8}\right)
    \leq
    \exp\left(-\frac{\mu}{16}\right).
\]
Averaging over $A$ gives
\[
    \PP[d(X,Y)<\tau]\leq \exp\left(-\frac{\mu}{16}\right).
\]
The pair $(B_1,B_2)$ has the same distribution as $(X,Y)$ conditioned on $X\neq Y$.
Since $M$ has at least two bases, $\PP[X\neq Y]\geq 1/2$, and hence
\[
    \PP[d(B_1,B_2)<\tau]
    \leq
    \PP[d(X,Y)<\tau \mid X\neq Y]
    \leq
    2\exp\left(-\frac{\mu}{16}\right).\qedhere
\]
\end{proof}

We can now prove the sampling guarantee.
The idea is to choose a threshold $\tau$ that is a logarithmic factor below the average pairwise distance, apply Lemma~\ref{lem:close-pair-from-average} to each sampled pair, and then union bound over the $\binom{k}{2}$ pairs.

\kdiversityupper*
\begin{proof}
Let $X,Y$ be independent uniformly random bases of $M$.
Set
\[
\mu = \EE_{X,Y}[d(X,Y)] = \avg_2(M)
\qquad\text{and}\qquad
\tau = \frac{\mu}{64\log k}.
\]
By Theorem~\ref{thm:constant-diameter-lower}, $\diam_2(M) \leq 16\mu$, and clearly $\diam_k(M)\leq \diam_2(M)$.
Thus, it is enough to show that $\avg_k(M) \geq \Omega(\tau)$.
If $\tau < 1$, then the claim follows immediately: the bases in the definition of $\avg_k(M)$ are sampled distinctly, so every pair of sampled bases has distance at least $1$, and hence $\avg_k(M)\geq 1 > \tau$.
Assume from now on that $\tau \geq 1$.
Then $\mu \geq 64\log k$.

Let $B_1,\ldots,B_k$ be the bases sampled uniformly without replacement in the definition of $\avg_k(M)$.
Since $\tau\leq \mu/4$, Lemma~\ref{lem:close-pair-from-average} implies that, for every fixed pair $i<j$, $\PP[d(B_i,B_j)<\tau] \leq 2\exp\left(-\frac{\mu}{16}\right)$.
A union bound over all pairs $i<j$ gives
\[
\PP\left[\min_{i<j} d(B_i,B_j)<\tau\right]
\leq \sum_{i<j}\PP[d(B_i,B_j)<\tau]
\leq 2\binom{k}{2}\exp\left(-\frac{\mu}{16}\right)
\leq k^2\exp\left(-\frac{\mu}{16}\right).
\]
Since $\mu\geq64\log k$, the last expression is at most $k^2\exp(-4\log k)\leq k^2 \cdot k^{-4}\leq \frac{1}{4}$.
It follows that
\[
\avg_k(M)
\geq \tau \cdot \PP\left[\min_{i<j}d(B_i,B_j)\geq\tau\right]
\geq \frac{3}{4}\tau,
\]
concluding the proof.
\end{proof}

\subsubsection{Tightness of the analysis}\label{subsec:random-sampling-lower-bound}

We now show that the logarithmic loss is inherent for random sampling.
The example is the direct sum of $r$ rank-one matroids on two parallel elements, whose bases are naturally the vertices of the $r$-dimensional cube.

First, a birthday-paradox-like computation shows that a random sample of $2^{r/2}$ cube vertices is likely to contain a close pair, so its minimum pairwise distance remains constant in expectation.
\begin{lemma}
    \label{lem:cube-avg-distance}
    Let $r$ be even, and let $M = \bigoplus\limits_{i=1}^r U_{1,2}$ be the direct sum of $r$ parallel edges.
    If $k=2^{r/2}$, then $\avg_k(M) = \cO(1)$.
\end{lemma}
\begin{proof}
    We identify the bases of $M$ with vertices of the $r$-cube $Q_r$ in the natural way.
    Let $S\subseteq \{0,1\}^r$ be $k$ randomly sampled vertices of the $r$-cube without replacement, and let $D(S)$ be the smallest pairwise Hamming distance in $S$.
    Let $\mathcal{E}$ be the event that every pair of vertices in $S$ has Hamming distance at least $3$.
    Thus,
    \[
    \avg_k(M)=\EE[D(S)] = \sum_{d\geq 1} d \cdot \PP[D(S)=d] \leq 2 + r\cdot \PP[\mathcal{E}].
    \]
    We bound $\PP[\mathcal{E}]$ by exposing the sampled vertices as an ordered sample $v_1,\ldots,v_k$ without replacement.
    Write $N=2^r$.
    Suppose that $v_1,\ldots,v_t$ have pairwise Hamming distance at least $3$.
    Then their neighborhoods in $Q_r$ are disjoint.
    Hence, among the $N-t$ remaining vertices, exactly $tr$ vertices are at distance $1$ from one of $v_1,\ldots,v_t$.
    Choosing any of these vertices as $v_{t+1}$ creates a pair at distance $1$, and so prevents $\mathcal{E}$ from occurring.
    Therefore,
    \[
    \PP[\mathcal{E}]
    \leq
    \prod_{t=1}^{k-1}\left(1-\frac{tr}{N-t}\right)
    \leq
    \exp\left(-\sum_{t=1}^{k-1}\frac{tr}{N-t}\right).
    \]
    Since $k=2^{r/2}=\sqrt{N}$, for all sufficiently large even $r$,
    \[
    \sum_{t=1}^{k-1}\frac{tr}{N-t}
    \geq \frac{r}{N}\binom{k}{2}
    = \frac{r}{2}\left(1-\frac{1}{k}\right)
    \geq \frac{r}{4}.
    \]
    Thus $\PP[\mathcal{E}] \leq e^{-r/4}$, and consequently $\avg_k(M) \leq 2 + r e^{-r/4} = \cO(1)$.
\end{proof}

The previous lemma controls the value achieved by random sampling.
To compare it with the optimum, we need to show that the cube also contains a large set whose pairwise distances are linear in $r$.
We use the classical Gilbert--Varshamov bound~\cite{MR95090,MR1664228}, stated here in its standard coding-theoretic form.

\begin{lemma}
    \label{lem:GV-bound}
    Let $A(r,d)$ be the maximum size of a binary code $C\subseteq \{0,1\}^r$ with minimum Hamming distance at least $d$.
    Then,
    \[
    A(r,d) \geq \frac{2^r}{\sum_{j=0}^{d-1}\binom{r}{j}}.
    \]
\end{lemma}

We will apply this bound with distance about $r/10$.
The following estimate says that the corresponding Hamming ball is small enough to leave at least $2^{r/2}$ codewords.

\begin{lemma}
    \label{lem:sum-binom-estimate}
    For every $r\in \NN$, we have that $\sum\limits_{j=0}^{\lfloor r/10 \rfloor} \binom{r}{j} \leq 2^{r/2}$.
\end{lemma}
\begin{proof}
    We have that
    \[
    9^{-r/10}\sum_{j=0}^{\lfloor r/10 \rfloor} \binom{r}{j} \leq \sum_{j=0}^{\lfloor r/10 \rfloor}\binom{r}{j} 9^{-j} \leq (1+9^{-1})^r= \frac{10^r}{9^r}.
    \]
    Thus,
    \[
    \sum_{j=0}^{\lfloor r/10 \rfloor} \binom{r}{j} \leq 9^{r/10} \frac{10^r}{9^r} = \left(9^{1/10} \cdot \frac{10}{9} \right)^r \leq (\sqrt{2})^{r} = 2^{r/2}.\qedhere  
    \]
\end{proof}

We can now put the two sides together: random sampling finds only constant diversity in this matroid, while the optimal choice of $k$ bases has diversity $\Theta(r)=\Theta(\log k)$.
\kdiversitylower*
\begin{proof}
Choose $d = \lfloor r/10 \rfloor+1$.
By Lemmas~\ref{lem:GV-bound} and~\ref{lem:sum-binom-estimate} we obtain that
\[
A(r,d) \geq \frac{2^r}{2^{r/2}}=2^{r/2} =k.
\]
Thus, there are $k$ vertices on the $r$-cube with pairwise Hamming distances at least $d$.
This implies that $\diam_k(M) \geq d \geq r/10$.
Since $\diam_k(M)\leq r$, we obtain that $\diam_k(M) = \Theta(r) = \Theta(\log k)$.
On the other hand, $\avg_k(M)\geq 1$ because the sampled bases are distinct, and $\avg_k(M)\leq \cO(1)$ by Lemma~\ref{lem:cube-avg-distance}.
Thus $\avg_k(M)=\Theta(1)$, concluding the proof.
\end{proof}

\subsubsection{Random Sampling in the Large-Diameter Regime}

We end the section by noting that despite the logarithmic loss in the previous theorem, random sampling is still a good approximation when the diameter is sufficiently large.
Below, we show that if $\diam_2(M)$ is a sufficiently large constant multiple of $\log k$, then random sampling gives a constant-factor approximation for the Diverse Matroid Basis problem.

Interestingly, we cannot replace the previous condition with $\diam_k(M)\geq \Omega(\log k)$, as the cube example shows: $\diam_k(M)=\Theta(\log k)$, but random sampling loses a logarithmic factor.

\begin{theorem}\label{thm:k-diversity-large-diameter}
Let $k \geq 2$ and let $M$ be a matroid with at least $k$ bases.
If $\diam_2(M) \geq 768 \log k$, then $\diam_k(M) \leq 128\,\avg_k(M)$.
\end{theorem}
\begin{proof}
Let $X,Y$ be independent uniformly random bases of $M$, and set $\mu = \EE_{X,Y}[d(X,Y)]$.
By Theorem~\ref{thm:constant-diameter-lower}, $\diam_2(M) \leq 16\mu$.
Hence, we have that $\mu \geq 48 \log k$.
Set $\tau = \mu/4$.

Let $B_1,\ldots,B_k$ be sampled uniformly without replacement.
By Lemma~\ref{lem:close-pair-from-average}, for every fixed pair $i<j$,
\[
    \PP[d(B_i,B_j)<\tau]
    \leq
    2\exp\left(-\frac{\mu}{16}\right).
\]
Taking a union bound over the $\binom{k}{2}$ pairs and recalling that $\mu \geq 48\log k$, we obtain
\[
    \PP\left[\min_{i<j}d(B_i,B_j)<\tau\right]
    \leq
    k^2\exp\left(-\frac{\mu}{16}\right)
    \leq
    \frac{1}{k}
    \leq
    \frac{1}{2}.
\]
Consequently,
\[
    \avg_k(M)
    \geq
    \tau \cdot \PP\left[\min_{i<j}d(B_i,B_j)\geq \tau\right]
    \geq
    \frac{\mu}{8}.
\]
Finally, since $\diam_k(M)\leq \diam_2(M)\leq 16\mu$, we get $\diam_k(M) \leq 128\,\avg_k(M)$, as claimed.
\end{proof}

\subsubsection{Efficient Random Sampling}

One may wonder whether the random sampling algorithm can be implemented efficiently for arbitrary matroids with independence oracle access.
The answer is another celebrated consequence of the work of Anari et al.~\cite{anari2019log}, which shows that one can sample a base of a matroid in polynomial time using an independence oracle with small \emph{total variation distance} to the uniform distribution in polynomial time.
More formally, we say that two distributions $\nu$ and $\mu$ have \emph{total variation distance} at most $\delta$ if on every event $\cE$ we have that $|\nu(\cE)-\mu(\cE)|\leq \delta$.
We can now state the sampling result from Anari et al.
\begin{lemma}[Corollary 1.3 in~\cite{anari2019log}]
    \label{lem:single-base-sampling}
Let $M$ be a matroid.
For every $\delta\in(0,1)$, there is a randomized algorithm that outputs a base whose distribution is within total variation distance $\delta$ from the uniform distribution on $\cB(M)$.
The algorithm uses $\poly(n,\log(1/\delta))$ time.
\end{lemma}

We use Lemma~\ref{lem:single-base-sampling} to sample various bases with and without replacement in the following form.

\begin{restatable}{lemma}{efficientsampling}
\label{lem:efficient-base-sampling}
Let $L\geq 1$, $\delta\in(0,1)$, and let $M$ be a matroid.
\begin{enumerate}
    \item There is a randomized algorithm that outputs an ordered $L$-tuple of bases whose joint distribution is within total variation distance $\delta$ of $L$ independent uniformly random bases of $M$.
    The algorithm runs in time $L\cdot\poly(n,\log(L/\delta))$.
    \item If $M$ has at least $L$ bases, then there is a randomized algorithm that outputs an ordered $L$-tuple of distinct bases whose joint distribution is within total variation distance $\delta$ of the uniform distribution over ordered $L$-tuples of distinct bases of $M$.
    The algorithm runs in time $\poly(n,L,\log(1/\delta))$.
\end{enumerate}
\end{restatable}
We defer the details to Appendix~\ref{app:efficient-sampling}.

As a consequence, the random-sampling approximation from Theorem~\ref{thm:k-diversity-upper} can be implemented efficiently.

\begin{corollary}\label{cor:efficient-random-diversity}
Let $M=(E,\cI)$ be a matroid with $n=|E|$ given by an independence oracle.
Assume that $M$ has at least $k$ bases, and let $\delta\in(0,1)$.
There is a randomized algorithm in $\poly(n,k,\log(1/\delta))$-time that outputs $k$ distinct bases $B_1,\ldots,B_k$ such that
\[
    \min_{i<j} d(B_i,B_j)
    \geq
    \Omega\left(\frac{1}{\log k}\right)\diam_k(M)
\]
with probability at least $1-\delta$.
\end{corollary}
\begin{proof}[Proof sketch]
The proof is essentially the same as the proof of Theorem~\ref{thm:k-diversity-upper}.
We sketch the modifications: 
Instead of sampling exactly uniformly without replacement, invoke the second assertion of Lemma~\ref{lem:efficient-base-sampling} with $L=k$ and (small) constant accuracy.
The exact distribution in Theorem~\ref{thm:k-diversity-upper} succeeds with constant probability and this is only decreased by a small constant given by the total variation.
Thus, we obtain a constant-success algorithm for finding $k$ distinct bases with minimum pairwise distance $\Omega(1/\log k)\diam_k(M)$.

To boost the success probability, repeat the constant-success algorithm $\cO(\log(1/\delta))$ times and return the sampled tuple maximizing $\min_{i<j}d(B_i,B_j)$. 
Standard probability amplification shows that this succeeds with probability at least $1-\delta$.
\end{proof}

\subsection{Approximating Farthest Matroid Basis}
\label{sec:farthest-matroid-basis}

We now turn to the Farthest Matroid Basis problem, which is a natural generalization of the Farthest Pair problem in metric spaces.
For a set of bases $\cS\subseteq \cB(M)$, we define the distance of a base $B$ to $\cS$ as
\[
    d(B,\cS) = \min_{B' \in \cS} d(B,B').
\]
Thus, the goal of the Farthest Matroid Basis problem is to find a base $B$ that maximizes $d(B,\cS)$ for a given set of bases $\cS$.
\cite{eisenbrand2024sensitivity} 

The main result of this section is a randomized polynomial-time $\frac{1}{3}$-approximation algorithm for the Farthest Matroid Basis problem.
The algorithm works by separating the analysis into two regimes: low-diameter and high-diameter.
In the low-diameter regime, random sampling gives a $\frac{1}{3}$-approximation in polynomial time.
In the high-diameter regime, an LP-rounding argument gives a PTAS.
Finally, replacing the low-diameter sampling step by exact enumeration gives a QPTAS for the Farthest Matroid Basis problem in general.

\subsubsection{Farthest Matroid Basis via Uniform Sampling}
\label{sss:fmb-uniform}

The algorithm for the low-diameter regime is based on comparing the sizes of balls in the matroid base graph.
For a matroid $M$ and a base $B \in \cB(M)$, we define the ball of radius $r$ around $B$ as
\[
\Ball(B,r) = \{A \in \cB(M) \mid d(B,A) \leq r\} .
\]
The following lemma shows that balls around different bases are comparable in size, up to a factor that depends on the diameter of the matroid.

\begin{lemma}
    \label{lem:ball-homogeneity}
    For every matroid $M$, every two bases $A,B \in \cB(M)$, and every $r \geq 0$, we have
    \[
    |\Ball_M(A,r)| \leq 4^{\diam(M)} |\Ball_M(B,2r)|. 
    \]
\end{lemma}
\begin{proof}
Fix bases $A,B\in\cB(M)$ and a radius $r\geq 0$.
We construct a map from $\Ball_M(A,r)$ to $\Ball_M(B,2r)$ and then bound the size of every fiber of this map.

Let $X\in\Ball_M(A,r)$.
Write $O=A\setminus X$ and $I=X\setminus A$.
Then $|O|=|I|=\ell\leq r$, and the pair $(O,I)$ uniquely determines $X$.
We will start from $B$, remove all elements of $O$, and then add all elements of $I$.

First order the elements of $O$ as $o_1,\ldots,o_\ell$.
Suppose that after $j-1$ steps we have a base $B_{j-1}$ that avoids $o_1,\ldots,o_{j-1}$.
If $o_j\notin B_{j-1}$, set $B_j=B_{j-1}$.
Otherwise $o_j\in B_{j-1}\setminus X$.
By the strong basis-exchange property applied to the bases $B_{j-1}$ and $X$, there is an element $e\in X\setminus B_{j-1}$ such that $B_{j-1}-o_j+e$ is a base.
Set $B_j=B_{j-1}-o_j+e$.
Since $X$ is disjoint from $O$, this operation does not introduce any element of $O$.
After at most $\ell$ exchanges, we obtain a base $B_\ell$ satisfying $B_\ell\cap O=\emptyset$ and $d(B,B_\ell)\leq \ell$.

Next order the elements of $I$ as $i_1,\ldots,i_\ell$.
Starting from $B'_0=B_\ell$, we construct bases $B'_0,\ldots,B'_\ell$ such that $B'_j$ contains $i_1,\ldots,i_j$ and avoids $O$.
Suppose $B'_{j-1}$ has already been constructed.
If $i_j\in B'_{j-1}$, set $B'_j=B'_{j-1}$.
Otherwise $i_j\in X\setminus B'_{j-1}$.
Apply the strong basis-exchange property to the bases $X$ and $B'_{j-1}$.
Since $i_j\in X\setminus B'_{j-1}$, there is an element $f\in B'_{j-1}\setminus X$ such that $B'_{j-1}+i_j-f$ is a base.
Set $B'_j=B'_{j-1}+i_j-f$.
This operation does not remove any of the already inserted elements $i_1,\ldots,i_{j-1}$, because those elements lie in $X\cap B'_{j-1}$.
It also does not introduce an element of $O$, because the only inserted element is $i_j\in I$ and $I\cap O=\emptyset$.
After at most another $\ell$ exchanges, we obtain a base $Y=B'_\ell$ with $I\subseteq Y$ and $Y\cap O=\emptyset$.
Moreover, $d(B,Y)\leq 2\ell\leq 2r$, so $Y\in\Ball_M(B,2r)$.
This defines the desired map $X\mapsto Y$.

It remains to bound the number of bases $X$ that can map to a fixed $Y\in\Ball_M(B,2r)$.
If $X=(A\setminus O)\cup I$ maps to $Y$, then the construction guarantees that $I\subseteq Y\setminus A$ and $O\subseteq A\setminus Y$.
Thus there are at most $2^{|Y\setminus A|}$ choices for $I$ and at most $2^{|A\setminus Y|}$ choices for $O$.
Since $A$ and $Y$ are bases, $|Y\setminus A|=|A\setminus Y|=d(A,Y)$.
Therefore every fiber has size at most $2^{|Y\setminus A|+|A\setminus Y|}=4^{d(A,Y)}\leq 4^{\diam(M)}$.
Since every $X\in\Ball_M(A,r)$ is mapped to some $Y\in\Ball_M(B,2r)$, we conclude that
\[
    |\Ball_M(A,r)|\leq 4^{\diam(M)}|\Ball_M(B,2r)|.\qedhere
\]
\end{proof}

The previous lemma turns uniform sampling into an approximation algorithm for the Farthest Matroid Basis problem.

\begin{restatable}{lemma}{FmbSampling}
\label{lem:fmb-sampling}
    Let $M$ be a matroid, $\cS \subseteq \cB(M)$ be a set of $k$ bases and $\delta\in(0,1)$.
    There is a randomized algorithm that returns a base $B$ satisfying
    \[
    d(B,\cS) \geq \frac{1}{3} \max_{B' \in \cB(M)} d(B',\cS)
    \]
    with probability at least $1-\delta$ in time $4^{\diam(M)}\poly(n,k,\log(1/\delta))$.
\end{restatable}
\begin{proof}
Let $d^*=\max_{B'\in\cB(M)}d(B',\cS)$, and let $B^*$ be a base achieving this maximum.
Further, we suppose that $d^*\geq 1$, as otherwise the Lemma holds trivially.
Set $r=\lceil d^*/3 \rceil$ and $\rho =d^*-r = \lfloor 2d^*/3\rfloor$. 
We call a base good if its distance from $\cS$ is at least $r$. 

Let $N=|\cB(M)|$ and let
\[
    q=\frac{|\Ball_M(B^*,\rho)|}{N}.
\]
Every base in $\Ball_M(B^*,\rho)$ is good.
Indeed, for every $B\in\Ball_M(B^*,\rho)$ and every $A\in\cS$, the triangle inequality gives
\[
    d(B,A)
    \geq d(B^*,A)-d(B,B^*)
    \geq d^*-\rho 
    =r.
\]
Hence a uniformly random base is good with probability at least $q$.

We also upper bound the probability of being bad.
If a base is not good, then it lies in $\Ball_M(A,r-1)$ for some $A\in\cS$.
By Lemma~\ref{lem:ball-homogeneity} and the fact that $2(r-1)\leq \rho$, for every $A\in\cS$ we have
\[
    |\Ball_M(A,r-1)| \leq 4^{\diam(M)}|\Ball_M(B^*,2(r-1))|\leq 4^{\diam(M)}|\Ball_M(B^*,\rho)|. 
\]
Therefore, by an union bound, the probability that a uniformly random base is bad is at most
\[
    \frac{\sum_{A\in\cS}|\Ball_M(A,r-1)|}{N} \leq k 4^{\diam(M)}q.
\]
Equivalently, if $p$ is the probability that a uniformly random base is good, then $p\geq q$ and $p\geq 1-k4^{\diam(M)}q$.
These two inequalities imply that
\[
    p\geq \frac{1}{k4^{\diam(M)}+1}.
\]
Indeed, if $q\geq 1/(k4^{\diam(M)}+1)$, then $p\geq q$ gives the claim.
Otherwise $p\geq 1-k4^{\diam(M)}q>1/(k4^{\diam(M)}+1)$.

Now take $L=\left\lceil (k4^{\diam(M)}+1)\log(2/\delta)\right\rceil$ independent uniformly random bases, and return the sampled base maximizing $d(B,\cS)$.
The probability that none of these samples is good is at most
\[
    \left(1-\frac{1}{k4^{\diam(M)}+1}\right)^L
    \leq
    \exp\left(-\frac{L}{k4^{\diam(M)}+1}\right)
    \leq
    \frac{\delta}{2}.
\]
Thus exact uniform sampling succeeds with probability at least $1-\delta/2$.

The preceding calculation analyzed the ideal distribution of $L$ independent uniform bases.
To implement the algorithm with independence-oracle access, invoke the first part of Lemma~\ref{lem:efficient-base-sampling} with accuracy $\delta/2$ and $L$ samples.
This produces an ordered $L$-tuple whose joint distribution is within total variation distance $\delta/2$ of the uniform distribution.
Since total variation distance bounds the change in probability of every event, the probability of the success event changes by at most $\delta/2$.
Consequently, the implemented algorithm succeeds with probability at least $1-\delta$.
By the first assertion of Lemma~\ref{lem:efficient-base-sampling}, this uses $L\cdot\poly(n,\log(L/\delta))$ independence-oracle calls.
As $L=4^{\diam(M)}\poly(k,\log(1/\delta))$, this is $4^{\diam(M)}\poly(n,k,\log(1/\delta))$, as claimed.
\end{proof}
We remark that combining the previous lemma with Theorem~\ref{thm:k-diversity-large-diameter} already gives a polynomial-time constant approximation for Diverse Matroid Basis (albeit with a rather large constant).

\subsubsection{A PTAS for Farthest Matroid Basis in the Large-Diameter Regime}
\label{sss:ptas-fmb}

In this subsection, we show that the Farthest Matroid Basis problem admits a PTAS in the large-diameter regime.
Let $B_1,\ldots,B_k \in \cB(M)$ be the given bases.
We relax the problem by replacing the unknown basis with a point in the matroid base polytope.
For a vector $x\in \R^E$ and a set $T\subseteq E$, write $x(T)=\sum_{e\in T}x_e$.
Recall that the base polytope of $M$ is
\[
P(M)=\conv\{\1_B : B\in\cB(M)\}
=\{x \in \R^E_{\geq 0} : x(S) \leq r_M(S) \text{ for all } S \subseteq E,\ x(E)=r(M)\}.
\]
For an integral point $x=\1_B$, the quantity $\sum_{e\in E\setminus B_i} x_e$ is exactly $|B\setminus B_i|=d(B,B_i)$.
Thus, the natural relaxation asks for a point $x\in P(M)$ that maximizes the minimum fractional distance to the bases $B_1,\ldots,B_k$.
Introducing a variable $\lambda$ for this minimum gives the following linear program.

\begin{equation}
\tag{FMB-LP}\label{eq:fmb-lp}
\begin{array}{r@{\quad}r@{\;}c@{\;}l@{\qquad}l}
\max       & \multicolumn{4}{l}{\lambda} \\
\text{s.t.}& x                                      & \in  & P(M),    & \\
           & \sum\limits_{e\in E\setminus B_i} x_e
                                                    & \geq & \lambda  & \forall i\in[k], \\
           & \lambda                               & \geq & 0.       &
\end{array}
\end{equation}

We first record that the relaxation can be solved in polynomial time using standard separation over the matroid base polytope.

\begin{lemma}
    \label{lem:lp-solution}
    Let $M$ be a matroid, $\cS \subseteq \cB(M)$ be a set of $k$ bases.
    There is an algorithm that returns a tuple $(x^*,\lambda^*)$ that is optimal for \eqref{eq:fmb-lp} in time $\poly(n,k)$.
\end{lemma}
\begin{proof}
    We solve the linear program using the ellipsoid method, so it suffices to give a polynomial-time separation oracle for its feasible region.
    Given a candidate point $(x,\lambda)$, we first check the explicit constraints $\lambda\geq 0$ and
    \[
        \sum_{e\in E\setminus B_i}x_e \geq \lambda
        \qquad\text{for all }i\in[k].
    \]
    If one of these inequalities is violated, it gives a separating hyperplane.

    It remains to separate over the base polytope $P(M)$.
    We check $x_e\geq 0$ for all $e\in E$ and $x(E)=r(M)$.
    If either condition fails, we again obtain a violated linear constraint.
    Finally, we need to decide whether $x(T)\leq r_M(T)$ forall $T\subseteq E$.

    This is equivalent to checking whether $\min_{T\subseteq E} \{r_M(T)-x(T)\}\geq 0$.
    The function $T\mapsto r_M(T)-x(T)$ is submodular, since $r_M$ is submodular and $x(T)$ is modular.
    Moreover, $r_M(T)$ can be evaluated in polynomial time by the greedy algorithm using the independence oracle.
    Hence this minimum, and a minimizer if the value is negative, can be found in polynomial time by submodular-function minimization (see e.g. Chapter 45 of Schrijver's book~\cite{MR1956926}).
    A negative minimizer gives the violated inequality $x(T)\leq r_M(T)$; otherwise $x\in P(M)$.

    Thus \eqref{eq:fmb-lp} has a polynomial-time separation oracle.
    Since $0\leq \lambda\leq r(M)\leq n$ for every feasible solution, the feasible region is polynomially bounded.
    The standard equivalence between separation and optimization via the ellipsoid method therefore returns an optimal solution $(x^*,\lambda^*)$ in time $\poly(n,k)$.
\end{proof}

We need the following consequences of Chekuri, Vondrak, and Zenklusen's swap rounding~\cite{chekuri2010dependent} to round the fractional solution of \eqref{eq:fmb-lp} to an integral solution while preserving marginals and obtaining concentration bounds.

\begin{lemma}
    \label{lem:rounding}
    Let $M$ be a matroid, $x \in P(M)$ be a point in the base polytope and $S\subseteq E$.
    There is a randomized polynomial-time algorithm that returns a base $B_R$ such that
    \begin{enumerate}
        \item $\PP[e \in B_R] = x_e$ for all $e \in E$,
        \item For every $\epsilon \in (0,1)$,
        \[
        \PP\left[|B_R \cap S| \leq (1-\epsilon) \sum_{e \in S} x_e\right] \leq \exp\left(-\frac{\epsilon^2}{2} \sum_{e \in S} x_e \right)
        \]
    \end{enumerate}
\end{lemma}
\begin{proof}
    For the first property, Theorem~1.1 of Chekuri, Vondrak, and Zenklusen~\cite{chekuri2010dependent} implies that there is a randomized polynomial-time algorithm that returns a base $B_R$ such that $\PP[e\in B_R]=x_e$ for all $e\in E$.

    The second property is implied by Corollary~1.2 of~\cite{chekuri2010dependent}.
    For any vector $a\in[0,1]^E$, let
    \[
        Z=\sum_{e\in B_R}a_e
        \qquad\text{and}\qquad
        \mu=\E[Z]=\sum_{e\in E}a_e x_e.
    \]
    Corollary~1.2 of~\cite{chekuri2010dependent} gives, for every $\epsilon\in [0,1]$,
    \[
        \PP[Z\leq (1-\epsilon)\mu]
        \leq
        \exp\left(-\frac{\epsilon^2}{2}\mu\right).
    \]
    Taking $a_e=\indic{e\in S}$ gives $Z=|B_R\cap S|$ and $\mu=\sum_{e\in S}x_e$, which is exactly the desired bound.
    This completes the proof.
\end{proof}

We now show that swap-rounding the optimal solution to \eqref{eq:fmb-lp} gives a PTAS for the Farthest Matroid Basis problem in the large-diameter regime.

\begin{restatable}{lemma}{BmbPtasLargeDiameter}\label{lem:fmb-ptas-large-diameter}
    Let $M$ be a matroid, $\cS \subseteq \cB(M)$ be a set of $k$ bases, $\delta\in(0,1)$ and $\varepsilon>0$.
    If $\diam_2(M) \geq 4\varepsilon^{-2}\ln(k/\delta)$, then there is a randomized algorithm that returns a base $B$ satisfying
    \[
        d(B,\cS) \geq (1-\varepsilon) \max_{B' \in \cB(M)} d(B',\cS)
    \]
    with probability at least $1-\delta$ in time $\poly(n,k)$.
\end{restatable}

\begin{proof}
    Let $B_1,\ldots,B_k \in \cB(M)$ be the given bases, $B^* \in \cB(M)$ be a base that maximizes $d(B^*,\cS)$, and let $d^* = d(B^*,\cS)$.

    We begin by solving the linear program \eqref{eq:fmb-lp} via Lemma~\ref{lem:lp-solution} to obtain an optimal solution $(x^*,\lambda^*)$.
    Note that $\lambda^* \geq d^*$, since the point $(\1_{B^*},d^*)$ is a feasible solution to \eqref{eq:fmb-lp} with objective value $d^*$.

    We next show that $\lambda^* \geq \frac{1}{2} \diam_2(M)$.
    Let $B_1^*,B_2^* \in \cB(M)$ be two bases such that $d(B_1^*,B_2^*) = \diam_2(M)$.
    Define
    \[
    y = \frac{1}{2}(\1_{B_1^*} + \1_{B_2^*}) \in P(M).
    \]
    Note that for every $B_i \in \cS$, we have
    \[
    \sum_{e \in E \setminus B_i} y_e = \frac{1}{2} \left( |B_1^* \setminus B_i| + |B_2^* \setminus B_i| \right) \geq \frac{1}{2} d(B_1^*,B_2^*) = \frac{1}{2} \diam_2(M).
    \]
    Thus, $(y,\frac{1}{2}\diam_2(M))$ is a feasible solution to \eqref{eq:fmb-lp}, and hence $\lambda^* \geq \frac{1}{2} \diam_2(M)$.

    Apply the rounding algorithm of Lemma~\ref{lem:rounding} to $x^*$, and let $B_R$ be the resulting base.
    For each $i\in[k]$, define $S_i = E\setminus B_i$ and $\mu_i = \sum_{e\in S_i} x^*_e$.
    Since $(x^*,\lambda^*)$ is feasible for \eqref{eq:fmb-lp}, we have $\mu_i\geq \lambda^*$ for every $i\in[k]$.
    By Lemma~\ref{lem:rounding}, for every fixed $i\in[k]$,
    \[
    \PP\left[|B_R\cap S_i| < (1-\varepsilon)\mu_i\right]
    \leq
    \exp\left(-\frac{\varepsilon^2}{2}\mu_i\right)
    \leq
    \exp\left(-\frac{\varepsilon^2}{2}\lambda^*\right).
    \]
    Using $\lambda^* \geq \diam_2(M)/2$ and the assumption $\diam_2(M) \geq 4\varepsilon^{-2}\ln(k/\delta)$, we get
    \[
      \PP\left[|B_R\cap S_i| < (1-\varepsilon)\mu_i\right]\leq \exp\left(-\frac{\varepsilon^2}{2}\lambda^*\right)
    \leq
    \exp\left(-\frac{\varepsilon^2}{4}\diam_2(M)\right)
    \leq \frac{\delta}{k}.
    \]
    Thus, a union bound over all $i\in[k]$ gives
    \begin{align*}
        \PP\left[d(B_R,\cS) < (1-\varepsilon)\lambda^*\right] \leq \sum_{i=1}^k \PP\left[|B_R\cap S_i| < (1-\varepsilon)\mu_i\right] \leq k \cdot \frac{\delta}{k} = \delta.
    \end{align*}
    Using that $\lambda^* \geq d^*$, we conclude that
    \[
    \PP[d(B_R,\cS) < (1-\varepsilon)d^*] \leq \PP[d(B_R,\cS) < (1-\varepsilon)\lambda^*] \leq \delta,
    \]
    as desired.

    The running-time bound follows from the fact that the algorithms from Lemma~\ref{lem:lp-solution} and Lemma~\ref{lem:rounding} run in polynomial time.
\end{proof}

\subsubsection{Approximating Farthest Matroid Basis}
\label{sss:aprox-fmb}

Naturally, we can combine the two regimes to obtain a $\frac{1}{3}$-approximation for the Farthest Matroid Basis problem in polynomial time.
\ThmFmbApprox*
\begin{proof}
    We first describe an algorithm that succeeds with probability at least $2/3$.
    Compute $\diam_2(M)$ using matroid intersection in time $\poly(n)$.
    If $\diam_2(M) \geq 16\ln(3k)$, then we apply the algorithm from Lemma~\ref{lem:fmb-ptas-large-diameter} with $\varepsilon=1/2$ and $\delta=1/3$ to obtain a $1/2$ approximation with probability at least $2/3$ in time $\poly(n,k)$.
    Otherwise, we apply the algorithm from Lemma~\ref{lem:fmb-sampling} with $\delta=1/3$ to obtain a $3$-approximation with probability at least $2/3$ in time $4^{\diam(M)}\poly(n,k)\leq4^{16\ln(3k)}\poly(n,k)\leq\poly(n,k)$.

    To boost the success probability, repeat the constant-success algorithm $\lceil \log_3(1/\delta)\rceil$ times and return the candidate base maximizing $d(B,\cS)$.
    Standard probability amplification shows that the success probability is at least $1-\delta$.
\end{proof}

We can also obtain a QPTAS for the Farthest Matroid Basis problem by replacing the $3$-approximation in the low-diameter regime with a quasi-polynomial-time exact algorithm.
This algorithm follows by noting that for a given base $A \in \cB(M)$, we have that all other bases are within distance $\diam(M)$, and hence we can enumerate all bases in time $n^{\cO(\diam(M))}$.

\begin{theorem}
    \label{thm:fmb-qptas}
    Let $M$ be a matroid, $\cS \subseteq \cB(M)$ be a set of $k$ bases and $\varepsilon,\delta\in(0,1)$.
    There is a randomized algorithm that returns a base $B$ satisfying
    \[
    d(B,\cS) \geq (1-\varepsilon) \max_{B' \in \cB(M)} d(B',\cS)
    \]  
    with probability at least $1-\delta$ in time $\poly(n,k,\log(1/\delta))\cdot n^{\cO(\varepsilon^{-2}\log k)}$.
\end{theorem}
\begin{proof}
    We first describe an algorithm that succeeds with probability at least $2/3$.
    Compute $\diam_2(M)$ using matroid intersection in time $\poly(n)$.
    If $\diam_2(M) \geq 4\varepsilon^{-2}\ln(3k)$, then we apply the algorithm from Lemma~\ref{lem:fmb-ptas-large-diameter} with $\delta=1/3$ to obtain a $(1-\varepsilon)$ approximation with probability at least $2/3$ in time $\poly(n,k)$.
    Otherwise, we have that $\diam_2(M) < 4\varepsilon^{-2}\ln(3k)$.
    Compute some base $A \in \cB(M)$ of the matroid, and note that every base $B \in \cB(M)$ has the form $B=(A\setminus O) \cup I$, where $O \subseteq A$, $I \subseteq E\setminus A$, and $|O|=|I| \leq \diam(M) < 4\varepsilon^{-2}\ln(3k)$.

    Thus, we can enumerate all possible pairs $(I,O)$ in time $n^{\cO(\varepsilon^{-2}\log k)}$, check whether the corresponding set is a base, and return the base that maximizes $d(B,\cS)$.
    
    To boost the success probability, repeat the constant-success algorithm $\lceil \log_3(1/\delta)\rceil$ times and return the candidate base maximizing $d(B,\cS)$.
    Standard probability amplification shows that the success probability is at least $1-\delta$.
\end{proof}

We remark that Eisenbrand et al.~\cite{eisenbrand2024sensitivity} study the problem of finding a matroid basis whose weight vector is equal to a prescribed target, where the element weights are specified by an $ m \times n $ integer matrix.
They give an FPT algorithm parameterized by $m$ and $\Delta$, where $\Delta$ denotes the maximum absolute value of an entry in the weight matrix. 
By using their algorithm as a black box with parameters $ m := k $ and $\Delta := 1 $, the FMB problem can be solved in time $ C^{C^{O(k)}}\operatorname{poly}(n) $ for some constant $C>0$.
This establishes that FMB is fixed-parameter tractable with respect to $k$, but the dependence on $k$ is too large to fit our needs.

\subsection{Consequences for Diverse Matroid Basis}
\label{ss:diverse-consequence}

We end the section by noting that the previous results for Farthest Matroid Basis can be used to obtain approximation algorithms for Diverse Matroid Basis.
This is the farthest-insertion argument of Gonzalez~\cite{gonzalez1985clustering}; closely related farthest-insertion reductions for diversity over exponentially large solution spaces appear, for example, in the work of Gao et al.~\cite{gao2022obtaining}.

\begin{lemma}
    \label{lem:gonzalez}
    Suppose that, for every set $\cS\subseteq\cB(M)$ and every failure parameter $\eta\in(0,1)$, there is a randomized $\alpha$-approximation algorithm for Farthest Matroid Basis that, with probability at least $1-\eta$, returns a base $B$ satisfying
    \[
        d(B,\cS)\geq \alpha\max_{B'\in\cB(M)} d(B',\cS),
    \]
    and runs in time polynomial in $n$, $|\cS|$, and $\log(1/\eta)$.
    Then, for every $k\geq 2$ and $\delta\in(0,1)$, there is a randomized $\frac{\alpha}{2}$-approximation algorithm for Diverse Matroid Basis that succeeds with probability at least $1-\delta$ and runs in time polynomial in $n$, $k$, and $\log(1/\delta)$.
\end{lemma}
\begin{proof}
Let $D=\diam_k(M)$, and let $B^*_1,\ldots,B^*_k$ be bases with pairwise distances at least $D$.
Start from an arbitrary base $B_1$.
For $t=2,\ldots,k$, run the Farthest Matroid Basis algorithm on the current set $\cS_{t-1}=\{B_1,\ldots,B_{t-1}\}$ with failure parameter $\delta/(k-1)$, and let $B_t$ be the returned base.
Finally output $\{B_1,\ldots,B_k\}$.

By a union bound, with probability at least $1-\delta$ all $k-1$ calls to the Farthest Matroid Basis algorithm succeed.
Condition on this event.
We claim that before each insertion step $t$,
\[
    \max_{B\in\cB(M)} d(B,\cS_{t-1})\geq \frac{D}{2}.
\]
To see this, first suppose that every $B^*_j$ has distance strictly less than $D/2$ from $\cS_{t-1}$.
Then, we can choose for each $j$, a base $C_j\in\cS_{t-1}$ with $d(B^*_j,C_j)<D/2$.
Since $|\cS_{t-1}|<k$, two of these choices coincide, say $C_a=C_b=C$.
The triangle inequality would give
\[
    d(O_a,O_b)\leq d(O_a,C)+d(C,O_b)<D,
\]
contradicting the definition of $D$.
Thus, there exist a base at distance at least $D/2$ from $\cS_{t-1}$.
Since the call at step $t$ succeeds,
\[
    d(B_t,\cS_{t-1})\geq \alpha \max_{B'\in \cB(M)}d(B',\cS_{t-1})  \geq \frac{\alpha D}{2}.
\]
Therefore every inserted base is at distance at least $\alpha D/2$ from all previously inserted bases, and hence the final set has minimum pairwise distance at least $\alpha D/2$.
This is exactly an $\alpha/2$-approximation for Diverse Matroid Basis.
The running-time bound follows from the $k-1$ oracle calls with failure parameter $\delta/(k-1)$.
\end{proof}

We have the following immediate corollary of the previous lemma and the results in Section~\ref{sec:farthest-matroid-basis}.

\CorGonzalezFmb*
\begin{proof}
    Immediate from combining Lemma~\ref{lem:gonzalez} with Theorems~\ref{thm:fmb-approx} and~\ref{thm:fmb-qptas}.
\end{proof}

%% file: sections/special_cases.tex
\section{Special Cases}\label{sec:special-cases}

In this section, we present algorithms with improved approximation factors for special classes of matroids.
We begin by recalling the main results of this section.

\ThmVertexTran*

We prove \Cref{thm:special} over two subsections.
In \Cref{sec:vertex-transitive}, we prove part 1, and we prove part 2 in \Cref{sec:graphic-matroid}.

\subsection{Matroids with Vertex-Transitive Basis-Exchange Graphs}\label{sec:vertex-transitive}

The result we prove in this section is closely related to \emph{vertex-transitivity} of graphs.
For a simple graph $ G = (V, E) $, an automorphism of $G$ is a permutation $\sigma$ of its vertex set which maps edges to edges and nonedges to nonedges~\cite{bondy2008graph}.
An immediate consequence of this is that automorphisms preserve distances between vertices.
A graph $G=(V,E)$ is said to be \emph{vertex-transitive} if for any $u,v \in V$ there exists a graph automorphism $ \sigma \in \operatorname{Aut}(G)$ such that $ \sigma(u) = v $~\cite{bondy2008graph}.
\begin{theorem}\label{thm:special-vertex-tran}
    For any matroids with vertex-transitive basis-exchange graph and for every $\delta\in(0,1)$, there exists a $\operatorname{poly}(n,k,\log(1/\delta))$-time $1/2$-approximation randomized algorithm (that succeeds with probability $1-\delta$) for the Diverse Matroid Base problem.
\end{theorem}

To prove \Cref{thm:special-vertex-tran}, we begin by defining \emph{growth-regular} graph.
\begin{definition}[Growth-Regular Graph]
    A graph $G=(V,E)$ is said to be \emph{growth-regular} if, for any two vertices $ u,v \in V $ and $r \in \NN$, $|\Ball(u, r)|=|\Ball(v, r)|$. That is, the volume of a ball around a vertex depends only on the radius, not the vertex. 
\end{definition}

We now show that every vertex-transitive graph is growth-regular. Although this fact is folklore, we include the proof for completeness.
\begin{lemma}(folklore)\label{lem:vertex-tran-growth-regular}
    Any vertex-transitive graph is growth-regular.
\end{lemma}
\begin{proof}
    Let $G = (V,E) $ be a vertex-transitive graph, and let $ u$ and $v$ be two fixed arbitrary vertices of $G$.
    Then, by the definition for vertex-transitivity, there exists a graph automorphism $ \sigma \in \operatorname{Aut}(G) $ such that $ \sigma(u) = v $.
    Given, $ r \in \NN $, let $ x \in \Ball(u,r) $.
    Since graph automorphisms preserve the distance,
    \[ d(u, x) = d(\sigma(u),\sigma(x)) = d(v, \sigma(x)) \leq r. \]
    Therefore, $ \sigma(x) \in \Ball(v,r) $.
    Conversely, let $ y \in \Ball(v, r) $. Since $ \sigma^{-1} $ is also a graph automorphism, the same logic as above applies, and thus we have $ \sigma^{-1}(y) \in \Ball(u, r) $.
    Therefore, $ \sigma: \Ball(u,r) \rightarrow \Ball(v,r) $ is a bijection, and this completes the proof.
\end{proof}

Now, we present the algorithm for \Cref{thm:special-vertex-tran} below. 

\medskip
\begin{mdframed}[frametitle={Algorithm: Diverse Bases via Random Sampling},nobreak=true]
\textbf{Input:} Matroid $M$ with vertex-transitive $\cG(M)$; integer $k$; failure probability $\delta$.

\noindent\textbf{Output:} A list $L_{k+1} = \{B_1,\dots,B_k\} \subseteq \cB(M)$ with $d(B_i,B_j) \geq \diam_k(M)/2 $ for all $i \neq j$. \\
\\
We run the following algorithm for each $\lambda=n$ down to 0. 
\begin{enumerate}
    \item Initialize $L_1 \gets \emptyset$.
    \item \textbf{For} $i = 1, \dots, k$:
    \begin{enumerate}
            \item Repeat the following up to $k\ln(k/\delta)$ times:
            \begin{enumerate}
                \item Sample $B \sim \cB(M)$ uniformly at random.
                \item If $d(B, B_j) > \lfloor \frac{\lambda - 1}{2} \rfloor $ for all $1 \leq j \leq i-1$, set $B_i \gets B$, update $L_{i+1} \gets L_i \cup \{B_i\}$, and proceed to iteration $i+1$.
            \end{enumerate}
            \item If no such $B$ is found within $k\ln(k/\delta)$ trials, continue with a smaller $\lambda$.
    \end{enumerate}
    \item Return $L_{k+1}$.
\end{enumerate}
\end{mdframed}

    \vspace{2mm}\paragraph{Correctness.}
    Let $M$ be a matroid with vertex-transitive basis-exchange graph. 
    \Cref{lem:vertex-tran-growth-regular} implies that we can unambiguously define $B_{M,r}$ to be the cardinality of $\Ball(B,r)$ for every base $B \in \cB(M)$.
    Since $\Ball(B_i,\floor{(\diam_k(M)-1)/2})$ and $\Ball(B_j,\floor{(\diam_k(M)-1)/2})$ are pairwise disjoint for every $i\neq j$, we have 

\begin{equation} \label{eq:part-base-disj}
    \sum_{i=1}^k|\Ball(B_i,\floor{(\diam_k(M)-1)/2})| =k \cdot B_{M,\floor{(\diam_k(M)-1)/2}} \leq |\cB(M)|  \; .
\end{equation}

Now, for any $i < k$, let $B_1, \dots, B_i \in \cB(M)$. \Cref{eq:part-base-disj} implies that
\begin{align*}
    &\Pr_{B \sim \cB(M)}[ d(B, B_j) > \floor{(\diam_k(M)-1)/2}, \forall 1 \leq j \leq i ] \\
    &\qquad = 1-\frac{|\cup_{j=1}^i \Ball(B_j, \floor{(\diam_k(M)-1)/2})|}{|\cB(M)|} \\
    &\qquad \geq 1-\frac{i \cdot B_{M,\floor{(\diam_k(M)-1)/2}}}{|\cB(M)|} \geq 1-\frac{i}{k} \geq 1-\frac{1}{k} \; .
\end{align*}
Now, fix $\lambda$ and let $L= k \cdot \ln(\frac{k}{\delta})$ and sample $L$ bases $B^{(1)}, \dots, B^{(L)}$ uniformly at random. Then, the probability that none of them are at distance greater than $\floor{(\diam_k(M)-1)/2}$ from each of $B_1, \dots, B_i$ is at most 

$$ \left(1-\frac{1}{k} \right)^L \leq \exp(-L/k) \leq \frac{\delta}{k} \; .$$

Hence, with probability at least $1-\frac{\delta}{k}$, for each $1 \leq i \leq k$ round succeeds in adding a base $B_i$ to the list which has distance at least $\floor{(\diam_k(M)-1)/2}+1 \geq \diam_k(M)/2$ from each base already present in the list. Union bounding over each $1 \leq i \leq k$ shows that the algorithm outputs, with probability at least $ 1- \delta$ a set of bases with minimum pairwise distance at least $\diam_k(M)/2$.

The vertices of the Johnson graph $ J(n,r) $ are all $r$-subsets of $[n]$, where two vertices $u$ and $v$ are adjacent if and only if $ |u \cap v| = r-1 $.
Therefore, $ J(n,r) $ is the basis-exchange graph of the uniform matroid with rank $r$ on $n$ elements. Since the Johnson graph is vertex-transitive~\cite{godsil2013algebraic}, so is the basis-exchange graph of every uniform matroid. Moreover, the Cartesian product of vertex-transitive graphs is again vertex-transitive~\cite{pisanski2002growth}. Therefore, \Cref{thm:special-vertex-tran} applies to all partition matroids, and hence we have the following corollary.
\begin{corollary}\label{cor:partition-diversity}
    For every $\delta\in(0,1) $, there exists a $\operatorname{poly}(n,k,\log(1/\delta))$-time $1/2$-approximation algorithm for the $k$-diversity problem for partition matroids.
\end{corollary}

\input{sections/graphic.tex}

%% file: sections/graphic.tex
\subsection{Graphic Matroids}\label{sec:graphic-matroid}

Throughout this section, unless otherwise stated, $G=(V,E)$ is a simple connected graph on $n=|V|$ vertices, and $\cT(G)\subseteq 2^E$ is its set of spanning trees, i.e.\ the bases of the graphic matroid of $G$. For $T,T'\in\cT(G)$ we define the basis exchange distance to be $d(T,T')=\tfrac12|T\triangle T'|$, and
\[
    \diam_k(\cT(G))=\max_{T_1,\dots,T_k\in\cT(G)}\ \min_{i\ne j}\, d(T_i,T_j)
\]
is the optimal $k$-diversity. The \emph{Diverse Spanning Tree} (DST) problem asks, given $G$ and $k\ge2$, for $k$ spanning trees attaining $\diam_k(\cT(G))$. This is a specialization of Diverse Matroid Basis to graphic matroids. In this section, we design an algorithm that achieves an improved approximation factor for the DST problem in graphs with bounded arboricity. Constant arboricity captures many well-studied sparse graph classes such as planar graphs ($\Arb\le3$), graphs of bounded genus, graphs of bounded treewidth. 
\begin{definition}[Arboricity] \label{def:arbocity}
    Let $G=(V,E)$ be a graph. The \emph{arboricity} $\Arb(G)$ is the minimum number of spanning forests of $G$ needed to cover the entire edge set $E$ of $G$. Equivalently, it can be shown~\cite{nash1964decomposition} that
    \[ \Arb(G) = \left\lceil \max_{U\subseteq V,\,|U|\geq2 } \frac{ |E(U)| }{ |V(U)| - 1 } \right\rceil. \]
\end{definition}
\begin{theorem} \label{thm:bounded-arbocity}
    There exists a randomized $\poly(n,k,\log(1/\delta))$ time $1/4$-approximation algorithm that succeeds with probability $1-\delta$ for the Diverse Spanning Tree in graphs of constant arboricity.
\end{theorem}

The proof of \Cref{thm:bounded-arbocity} uses a very different technique than the random-sampling-based approaches used earlier. Essentially, we reduce the problem of finding diverse spanning trees in $G$ to the problem of finding diverse spanning trees in a much smaller ``core graph,'' and then brute-force search over it. We start by defining the core graph.

\paragraph{Construction of the core graph $H$.} We define $ V_c \coloneqq \{v\in V : \deg_G(v)\geq 3\} $ to be the set of \emph{core vertices} of $G$. We construct a multigraph $H$ on the vertex set $V_c$ as follows.
\setlength{\columnsep}{4pt}
\begin{wrapfigure}{R}{0.5\textwidth}
  \centering
  \includegraphics[width=.9\linewidth]{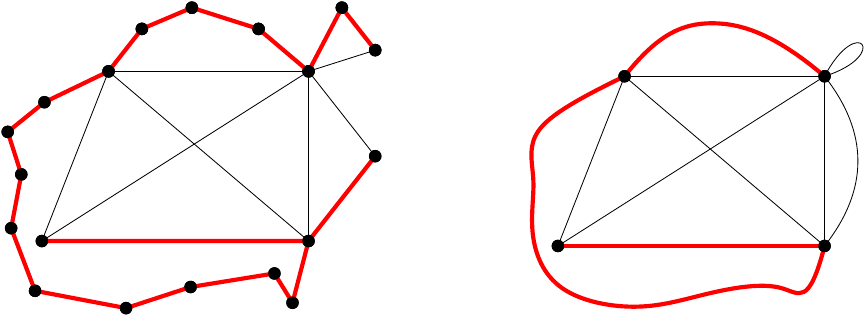}
  \captionsetup{width=0.8\linewidth}
  \caption{\small{An example of a graph $G$ and its associated core graph $H$. Trees $T_G$ and $f(T_G)$ are illustrated in thick red edges.}}
  \label{fig:h-construction}
\end{wrapfigure}
First, for every edge $uv\in E$ with $u,v\in V_c$, we add an edge $uv$ to $H$. Such an edge is called a \emph{core edge}. Next, for every maximal simple path $ P = u ,x_1, x_2, \cdots, x_i, v $, where $i \geq 1 $, whose endpoints $u,v$ are distinct core vertices in $V_c$ and whose internal vertices $x_1,\ldots,x_i$ all lie in $V\setminus V_c$, we add an edge $uv$ to $H$. We call such an edge a \emph{thread}. Finally, for each simple cycle in $G$ that contains exactly one core vertex $u$ and whose remaining vertices are all non-core vertices, we add a loop at $u$ in $H$.
Note that the edges of $H$ are precisely the core edges, threads, and loops obtained in this way; see \Cref{fig:h-construction}.

\paragraph{Mapping spanning trees from $G$ to $H$.} We can now define a map $f: \cT(G) \to \cT(H)$. Let $T_G$ be any spanning tree of $G$. The spanning tree $T_H=f(T_G)$ is defined as follows. Any core edge $e$ is present in $T_H$ if and only if $e$ is present in $T_G$. A thread edge $(u,v)$ created by a maximal path $P=u, x_1, x_2, \cdots ,x_i, v $  is included in $T_H$ if and only if every edge in $P$ is present in $T_G$. It is easy to verify that 1) every $f(T_G)$ contains no loop, 2) it is connected on the core vertices, 3) it is acyclic, and for every $T_H \in \cT(H)$ there exists some $T_G \in \cT(G)$ such that $f(T_G)=T_H$. Define $f^{-1}(T_H):=\{T_G \in \cT(G): f(T_G)=T_H\}$.

\begin{lemma}\label{lem: structure-H}
    $H$ has $ \cO( \diam_2(\mathcal{T}(G)) ) $ vertices, loops and threads.
\end{lemma}

\begin{proof}
    We first upper bound the number of loops.
    \paragraph{Loops.} Fix a spanning tree $T_H$ of $H$, and let $\ell$ be the number of loops of $H$. Each loop corresponds to a simple cycle in $G$ on at least $3$ vertices (as $G$ is simple), and distinct loops of $H$ correspond to distinct  cycles, which are pairwise edge-disjoint (call them $C_1,\dots,C_\ell$).

    For any $T_G \in f^{-1}(T_H)$ and each $i$, $T_G$ must contain all edges of $C_i$ but one: it cannot contain all of $C_i$ (that would create a cycle), and it cannot exclude two or more edges of $C_i$ (that would disconnect $G$, since $f(T_G)=T_H$ already fixes a spanning tree on the rest of $H$). Since this excluded edge may be chosen independently for each $i$, we may pick $T_G^{(1)}, T_G^{(2)} \in f^{-1}(T_H)$ that exclude \emph{different} edges $e_i \ne e_i'$ of $C_i$ for every $i$, and agree elsewhere. Hence, $C_i$ contributes exactly the pair $\{e_i,e_i'\}$ to $T_G^{(1)} \triangle T_G^{(2)}$. As the $C_i$ are pairwise edge-disjoint, these contributions add up:
    \[
        d\big(T_G^{(1)}, T_G^{(2)}\big) = \frac{\big|T_G^{(1)} \triangle T_G^{(2)}\big|}{2} \geq \ell.
    \]
    Since $d\big(T_G^{(1)}, T_G^{(2)}\big) \le \diam_2(\mathcal{T}(G))$, $\ell \leq \diam_2(\mathcal{T}(G))$.

    \paragraph{Vertices.} Fix the same spanning tree $T_H$ of $H$, and let $H^\circ$ denote $H$ with all loops deleted. Define $R$ to be the subgraph of $H^\circ$ with edge set $E(R) := E(H^\circ) \setminus E(T_H)$, vertex set $V(R)$ the endpoints of $E(R)$. Let $\rho$ be the \emph{rank} of $R$: $\rho := |V(R)| - c(R)$, where $c(R)$ is the number of connected components of $R$. 

    \emph{Step 1: $\rho \leq \diam_2(\cT(G))$.} Since every vertex of $R$ is incident to an edge of $R$, each component of $R$ has at least $2$ vertices. This implies that $|c(R)| \leq |V(R)|/2$ and $|V(R)| \leq 2 \rho$, and hence, $|c(R)| \leq \rho$. Let $F_R$ be a spanning forest of $R$. It has exactly $\rho$ edges and can be extended to a spanning tree $T_H'$ of $H$ with $E(F_R) \subseteq E(T_H')$. Since $E(F_R) \cap E(T_H) = \emptyset$ by construction, and hence, $|T_H' \Delta T_H| \geq 2|E(F(R))|=2 \rho$. Lifting $T_H, T_H'$ to $T_G \in f^{-1}(T_H)$, $T_G' \in f^{-1}(T_H')$, every edge of $H$ on which $T_H, T_H'$ disagree (core edge or thread) forces at least one edge of disagreement between $T_G, T_G'$ in $G$, so $d(T_G, T_G') \geq \frac{1}{2}|T_H \triangle T_H'| \geq \rho$. As $d(T_G,T_G') \le \diam_2(\cT(G))$, we get $\rho \leq \diam_2(\cT(G))$.

    \emph{Step 2: counting vertices of $H$.} We say a vertex $v \in V(H)$ \emph{bad} if $v$ carries a loop of $H$ or $v \in V(R)$, and \emph{good} otherwise. We start with noting that every good vertex has degree at least $3$ in $T_H$: Every vertex in $H$ has degree at least $3$. If a good vertex $v$ has degree $\leq 2$ in $T_H$, this would imply that either an edge in $R$ is incident to $v$ or $v$ has a loop, which cannot happen since $v$ is a good vertex.  

    Now, consider any vertex $u$ in $H$ with degree at most $2$ in $T_H$. Since every vertex in $H$ has degree at least $3$, this implies that $u$ is adjacent to some edge in $R = E \setminus T_H$. Thus, $u $ belongs to $V(R)$. Since $|V(R)| \leq 2 \rho$, the number of vertices in $H$ with degree at most $2$ in $T_H$ is $\leq 2 \rho$. In particular, the number of leaves in $T_H$ is at most $2 \rho \leq 2 \diam_2(\cT(G))$. 

    The handshake lemma for trees implies that the number of vertices of degree $\ge 3$ in $T_H$, and hence the number of good vertices, is at most the number of leaves in $T_H$, which is in turn at most $2 \diam_2(\cT(G))$. Now, the number of bad vertices is at most $\ell + |V(R)| \leq \rho + 2\rho \leq 3 \diam_2(\cT(G))$.

    Combining the bounds on good and bad vertices, $|V(H)| \leq 5 \diam_2(\cT(G))$.

    \paragraph{Threads.} Consider any spanning tree $T_H \in \cT(H)$. Note $|E(T_H)| = |V(H)|-1$, so the number of thread edges selected in $T_H$ is at most $|V(H)|-1$. Let $t$ denote the total number of thread edges in $H$, and let $P_1, \dots, P_s$ denote the paths in $G$ that they represent; each path has length at least $2$. Consider any thread edge not present in $T_H$, corresponding to a path $P_i$. For any $T_G \in f^{-1}(T_H)$, $T_G$ must include all the edges in $P_i$ but one (as in the loop argument), and this excluded edge may be chosen independently for each such $i$. Hence we can define $T_G^1, T_G^2 \in f^{-1}(T_H)$ such that $|T_G^1 \triangle T_G^2| \geq 2\big(t -(|V(H)|-1)\big)$. Thus $t-|V(H)|+1 \leq \diam_2(\cT(G))$, which implies $t \leq \diam_2(\cT(G))+ |V(H)|-1 \leq 6 \diam_2(\cT(G))$.
\end{proof}

\begin{lemma} \label{lem:arbocity-bound}
Let $G=(V,E)$ be a simple connected graph on $n$ vertices with arboricity $\Arb(G)$. Let $H$ be the associated core multigraph. Then $ |E(H)| =\mathcal{O} \left( \Arb(G)\cdot \diam_2(\cT(G)) \right) $.
\end{lemma}

\begin{proof}
    Let $V_c$ be the set of core vertices of $G$. By \Cref{lem: structure-H} the number of vertices, loops and threads of $H$ is $\cO(\diam_2(\cT(G)))$; it remains to bound the number of core edges of $H$. Since core edges of $H$ are precisely the edges of $G$ whose two endpoints are both core vertices, if $G$ has at most one core vertex, the claim is trivial.

    Now assume $G$ has at least two core vertices. By the definition of arboricity, for every $U\subseteq V(G)$ with $|U|\geq 2$, $ |E_G(U)|\leq  \Arb(G)(|U|-1) $. Applying \Cref{def:arbocity} with $ U = V_c $ gives
    \[
        |E_{\mathrm{core}}(H)| = |E_G(V_c)| \leq \Arb(G) \cdot (|V_c|-1) = \Arb(G) \cdot(|V(H)|-1) = \mathcal{O}(\Arb(G) \cdot \diam_2(\cT(G)))  \; ,
    \]
    where we used \Cref{lem: structure-H} to bound $|V(H)|$. This completes the proof.
\end{proof}

\paragraph{Putting it all together.} We combine these ingredients into a $1/4$-approximation for DST. Fix a sufficiently large constant $c$.

\emph{Large-diameter regime.} For each $\eps>0$, there exists a constant $c$ such that if $\diam_2(\cT(G)) > c\log k$, then the large-diameter PTAS for Farthest Matroid Basis (\Cref{lem:fmb-ptas-large-diameter}) runs in $\poly(n,k)$ time and returns a $(1-\varepsilon)$-approximation; composing it with farthest insertion (\Cref{lem:gonzalez}) yields a $\tfrac{1-\varepsilon}{2}$-approximation for DST in $\poly(n,k,\log(1/\delta))$ time. Taking $\varepsilon=\tfrac12$ gives a $1/4$-approximation. 

\emph{Small-diameter regime.} Otherwise $\diam_2(\cT(G)) = \cO(\log k)$. By \Cref{lem: structure-H} and \Cref{lem:arbocity-bound}, and using $\Arb(G)=\cO(1)$, we have $|V(H)| = \cO(\log k)$ and $|E(H)| = \cO(\log k)$. Hence,
\[
    |\cT(H)| \; \le \; 2^{|E(H)|} \;=\; 2^{\cO(\log k)} \;=\; \poly(k) \; ,
\]
so $\cT(H)$ can be enumerated in $\poly(k)$ time. The remainder of the subsection handles this regime.

\paragraph{Structure of $f^{-1}(T_H)$.} Fix $T_H \in \cT(H)$ and let $G'$ be the subgraph of $G$ consisting of edges with at least one endpoint outside $V_c$. This is a edge-disjoint union of the thread paths and loop cycles. Any $T_G \in f^{-1}(T_H)$ is determined as follows:
\begin{enumerate}
    \item every core edge $e$ lies in $T_G$ if and only if $e \in T_H$; thus $T_G \setminus E(G')$ is fixed by $T_H$.
    \item for a path with corresponding thread edge in $T_H$, all its edges lie in $T_G$.
    \item for each cycle, $T_G$ contains all but exactly one of its edges.
    \item for a path with corresponding thread edge not in $T_H$, $T_G$ contains all but exactly one of its edges.
\end{enumerate}
The free choices are exactly one omitted edge per cycle in $G'$ (loop in $H$) and per unselected path in $G$ (thread in $H$), and any such choice yields a spanning tree. Hence $f^{-1}(T_H)$ is the set of bases of a partition matroid $M_{T_H}$ whose categories are these cycles and unselected paths, each with capacity $1$. We call $M_{T_H}$ the \emph{fiber} of $T_H$. Moreover, for $B,B' \in f^{-1}(T_H)$ the Hamming distance $d(B,B')$ in $G$ equals the base-exchange distance in $M_{T_H}$, since $B,B'$ agree outside $G'$. Particularly, we have as a consequence, the following claim. 

\begin{claim}\label{claim:lift}
For all $T,T'\in\cT(G)$,
\[
    d\big(f(T),f(T')\big)\ \le\ d(T,T').
\]
In particular, for every $T\in\cT(G)$ and every $S'\in\cT(H)$ there exists $\widehat T\in f^{-1}(S')$ with $d(T,\widehat T)=d\big(f(T),S'\big)$.
\end{claim}
\begin{proof}
The inequality $d\big(f(T),f(T')\big)\ \le\ d(T,T')$ simply follows from properties (1)–(4). Write $S=f(T)$. We can choose $\widehat T\in f^{-1}(S')$ to agree with $T$ on every core edge outside $S\triangle S'$ and, on each loop cycle and each thread not flipped between $S$ and $S'$, to omit the same edge as $T$. The only disagreements are then a single edge in $G$ for each core edge or thread in $S\triangle S'$, giving $d(T,\widehat T)=\tfrac12|S\triangle S'|=d(S,S')$. 

\end{proof}

\medskip
\paragraph{Partition-matroid sampler.} We use \Cref{cor:partition-diversity}: given a partition matroid $M$, a threshold $\lambda$, and a target $t$, if $\diam_t(M)\ge 2\lambda$ then in $\poly(n,t,\log(1/\eta))$ time one can, with probability $\ge 1-\eta$, output $t$ bases of $M$ with pairwise distance $\ge\lambda$.

\medskip
The algorithm iterates over all $\lambda \in \{n, n-1, \dots, 0\}$. For each $\lambda$ it (i) builds, by farthest point insertion over $\cT(H)$, a maximal set of \emph{centers} $\mathcal{C}\subseteq\cT(H)$ that have pairwise distance $\ge\lambda$, and (ii) from each center's fiber spanning trees with pairwise distance at least $\lambda$ using the sampler above. The union has pairwise distance at least $\lambda$ in $G$: within a fiber by construction, and across fibers because \Cref{claim:lift} gives $d_G(B,B')\ge d_H(f(B),f(B'))\ge\lambda$ for centers $\ge\lambda$ apart.
\medskip
\begin{mdframed}[frametitle={Algorithm: Diverse Spanning Trees in bounded-arboricity graphs (small-diameter regime)},nobreak=true]
\textbf{Input:} Connected graph $G$ with $\Arb(G)=\cO(1)$; integer $k\ge2$; failure probability $\delta$. \quad (\emph{Assumes} $\diam_2(\cT(G))=\cO(\log k)$.)\\
\textbf{Output:} $\cS\subseteq\cT(G)$ with $|\cS|=k$ and $\min_{T\ne T'\in\cS} d(T,T')\ge \tfrac14\,\diam_k(\cT(G))$.
\begin{enumerate}
    \item Construct the core multigraph $H$ and enumerate $\cT(H)$ \hfill ($|\cT(H)|=\poly(k)$).
    \item For $\lambda = n, n-1, \dots, 0$:
    \begin{enumerate}
        \item \emph{Centers.} Set $\mathcal{C}\leftarrow\{S_0\}$ for an arbitrary $S_0\in\cT(H)$. While some $S\in\cT(H)$ has $d(S,\mathcal{C}):=\min_{S'\in\mathcal{C}}d(S,S')\ge\lambda$, add such an $S$ to $\mathcal{C}$.
        Continue until no further center exists or if we have found $k$ centers.
        \item \emph{Lifting.} Set $\cS_\lambda\leftarrow\emptyset$. For each $S\in\mathcal{C}$: run the algorithm from \Cref{thm:special-vertex-tran} with threshold distance $\lambda$, failure parameter $\delta/(nk\,|\cT(H)|)$. If it outputs $k_S$ bases that are pairwise $\ge\lambda$-separated, add them to $\cS_\lambda$. Note that we will terminate this step as well as soon as we find $k$ or more bases. 
        \item If $|\cS_\lambda|\ge k$, return any $k$ elements of $\cS_\lambda$.
    \end{enumerate}
\end{enumerate}
\end{mdframed}

\begin{lemma}[Correctness]\label{lem:dst-correctness}
Suppose $\Arb(G)=\cO(1)$ and $\diam_2(\cT(G))=\cO(\log k)$. With probability at least $1-\delta$, the algorithm above outputs $k$ spanning trees of $G$ with pairwise distance at least $\tfrac14\,\diam_k(\cT(G))$, in $\poly(n,k,\log(1/\delta))$ time.
\end{lemma}
\begin{proof}
Let $D_k=\diam_k(\cT(G))$ and let $T_1^\star,\dots,T_k^\star\in\cT(G)$ be optimal, i.e.\ $d(T_i^\star,T_j^\star)\ge D_k$ for all $i\ne j$.
If $ D_k \leq 3 $, any $k$ distinct spanning trees have minimum distance at least 1, which is at least $D_k/4$. Now, we show the iteration with $\lambda=\lfloor D_k/4\rfloor$ returns a valid set. Consider step $2. (a)$. If this step outputs $k$ spanning trees, then, \Cref{claim:lift} implies that any lifts of these bases are also pairwise $\lambda$-separated, and hence, any two spanning trees in the set $S_\lambda$ are at least $\lambda$-apart. 

Now, we consider the case that step $2. (a)$ outputs less than $k$ spanning trees. Then the set $\mathcal{C}$ is maximal. Let $S_i^\star=f(T_i^\star)$. By maximality of $\mathcal{C}$, each $S_i^\star$ is within distance $<\lambda$ of some center $c(i)\in\mathcal{C}$. Using \Cref{claim:lift}, let $\widehat T_i\in f^{-1}(c(i))$ such that,
\[
    d(T_i^\star,\widehat T_i)=d(S_i^\star,c(i))<\lambda .
\]
For $i\ne j$, the triangle inequality gives
\[
    d(\widehat T_i,\widehat T_j)\ \ge\ d(T_i^\star,T_j^\star)-d(T_i^\star,\widehat T_i)-d(T_j^\star,\widehat T_j)\ >\ D_k-2\lambda\ \ge\ 2\lambda .
\]
Group the indices by center: for $S\in\mathcal{C}$ let $I_S=\{i:c(i)=S\}$ and $k_S=|I_S|$, so $\sum_S k_S=k$. The trees $\{\widehat T_i:i\in I_S\}$ lie in $f^{-1}(S)$ and have pairwise distance $\ge 2\lambda$, so $\diam_{k_S}(M_S)\ge 2\lambda$. By the partition matroid sampler, the lift at center $S$ returns at least $k_S$ bases pairwise $\ge\lambda$-separated (with probability at least $1-\delta$ after applying the union bound). Hence $|\cS_\lambda|\ge\sum_S k_S=k$.

Finally $\cS_\lambda$ has pairwise distance at least $\lambda$ in $G$: two trees drawn from the same fiber $f^{-1}(S)$ are $\ge\lambda$ apart by construction, and two trees from distinct centers $S\ne S'$ satisfy $d_G\ge d_H(f(\cdot),f(\cdot))=d(S,S')\ge\lambda$ by \Cref{claim:lift}, since centers are have pairwise distance $\ge\lambda$. Thus the returned set has $k$ trees with pairwise distance $\ge\lambda=\lfloor D_k/4\rfloor$.

\paragraph{Running time and success probability.} Enumerating $\cT(H)$, building $\mathcal{C}$, and the $\cO(n)$ outer iterations are $\poly(n,k)$. Each partition matroid sampler runs in $\poly(n,k,\log(1/\delta))$ time, and there are at most $|\cT(H)|=\poly(k)$ of them per iteration; a union bound over all $\poly(n,k)$ partition matroid sampler calls gives overall success probability $\ge1-\delta$.
\end{proof}

%% file: sections/hardness.tex
\section{Hardness Results}\label{sec:hardness}

In this section, we will prove hardness results for our problems.
We begin by restating the theorem statement.
\FstWHard*
For clarity of presentation, let us define the decision versions of our problem of interest:

\begin{center}
\fbox{
\begin{minipage}{0.95\textwidth}
\textbf{Problem:} \dfstprob \\
\textbf{Input:} An undirected unweighted graph $G$, $k$ spanning trees $T_1, \dots, T_k$, and target distance $d \in \N$ \\
\textbf{Question:} Is there a spanning tree $T$ of $G$ such that $\min_{i \in [k]} d(T, T_i) \geq d$?
\end{minipage}
}
\end{center}

Our main result in this section is the W[1]-hardness\footnote{We refer interested readers to~\cite{DowneyF13,CyganFKLMPPS15} and~\cite{Marx08,FeldmannSLM20} for detailed introduction to parameterized complexity and parameterized approximation algorithms (and hardness), respectively.} of \dfstprob under the standard parameterization (i.e. the distance $d$), as stated below. Note that our reduction also implies NP-hardness for the problem; however, as discussed in \Cref{subsec:w1-hardness} below, such a hardness result is known.

As shown by Marx~\cite{Marx08}, W[1]-hardness result of a problem under standard parameterization rules out possibility of any \emph{efficient polynomial-time approximation scheme (EPTAS)} for the optimization variant of the problem. Specifically, from \Cref{thm:fst-w1}, we obtain the following result.

\begin{corollary}
Unless W[1] = FPT, there is no approximation scheme with running time $f(d, 1/\eps) \cdot n^{\cO(1)}$ for FST (and FMB) for any function $f$.
\end{corollary}

In contrast, a PTAS with running time $k^{\cO(1/\eps)} \cdot n^{\cO(1)}$ or $n^{\cO(1/\eps)}$ is not ruled out by our result. It remains an interesting question whether such a PTAS exists (in comparison to our $(3 + \eps)$-approximation algorithm and QPTAS).

\subsection{W[1]-Hardness of Farthest String}
\label{subsec:w1-hardness}

As an intermediate result, we prove a W[1]-hardness result for the \emph{farthest string} problem (on non-binary alphabet) as formalized below. Here $d_H$ denote the Hamming distance.

\begin{center}
\fbox{
\begin{minipage}{0.95\textwidth}
\textbf{Problem:} \dfsprob \\
\textbf{Input:} An alphabet set $\Sigma$, length $\ell \in \N$, $k$ strings $s_1, \dots, s_k \in \Sigma^\ell$ and target distance $d \in \N$ \\
\textbf{Question:} Is there a string $s \in \Sigma^\ell$ such that $\min_{i \in [k]} d_H(s, s_i) \geq d$?
\end{minipage}
}
\end{center}

While NP-hardness of both FST and FMB follows directly from that of \dfsprob with binary alphabet (i.e. $\Sigma = \{0, 1\}$), the latter is FPT in the optimum $d$. 
In this subsection, we show that, unlike the  binary alphabet case, \dfsprob with unbounded alphabet is W[1]-hard when parameterized by the optimum $d$. 

\begin{lemma} \label{lem:fs-w1-nonbinary}
\dfsprob is W[1]-hard parameterized by $d$ and $\ell$.
\end{lemma}

We note that, while such a W[1]-hardness result is known for the \emph{closest} string variant \cite{FellowsGN02}, it does not immediately imply the W[1]-hardness for the \emph{farthest} string in the non-binary setting since non-binary alphabet does not allow for ``taking complement''.

Our reduction is from $k$-Clique, a well known W[1]-hard problem We recall this problem below.

\begin{center}
\fbox{
\begin{minipage}{0.95\textwidth}
\textbf{Problem:} \kclique \\
\textbf{Input:} An undirected unweighted graph $G$, and positive integer $k \in \N$ \\
\textbf{Question:} Is there a set of $k$ vertices of $G$ that form a clique?
\end{minipage}
}
\end{center}

\begin{proof}[Proof of \Cref{lem:fs-w1-nonbinary}]
Assume w.l.o.g. that $k \geq 4$.
Let $(G = (V, E), k)$ be the input of $k$-Clique. We construct an instance for \dfsprob as follows. First, let the alphabet $\Sigma$ be $V \cup \{\sigma_{\dummy}\}$ and let $\ell = k, d = \ell -1$. We create $\binom{k}{2} \cdot (n^2 - 2|E|)$ strings, as follow. For every $\{i, j\} \subseteq [k]$ and every pair of (not necessarily distinct) vertices $u, v \in V$ that do not form an edge, create $s^{i, j, u, v}$ where
\begin{align*}
(s^{i, j, u, v})_t =
\begin{cases}
\sigma_{\dummy} &\text{ if } t \notin \{i, j\} \\
u &\text{ if } t = i \\
v &\text{ if } t = j \\
\end{cases}
\end{align*}

\paragraph{Completeness.} When there are $k$ vertices $v_1, \dots, v_k \in V$ that forms a $k$-clique in $G$, let $s = (v_1, \dots, v_k)$. It is simple to see that $s$ is of distance at least $k - 1 = d$ from all strings constructed above.

\paragraph{Soundness.} Suppose contrapositively that there is a string $s$ that is of distance at least $k - 1 = d$ from all strings constructed above. Consider the following three cases based on how many coordinates of $s$ is equal to the dummy character $\sigma^{\dummy}$:
\begin{itemize}
\item \textbf{Case I:} All coordinates are equal to $\sigma^{\dummy}$. In this case, the distance to any of the constructed string is $2 < d$, a contradiction.
\item \textbf{Case II:} At least one but not all coordinates are equal to $\sigma^{\dummy}$. Let $i^{\dummy}$ be an arbitrary coordinate that is set to $\sigma^{\dummy}$. Since not all coordinates are set to $\sigma^{\dummy}$, there exist another coordinate $i$ whose value is some $v \in V$. Let $j$ be any coordinate apart from these two. Consider the string $s^{i, j, v, v}$. We have $d_H(s, s^{i, j, v, v}) \leq \ell - 2 < \ell - 1$,  a contradiction.
\item \textbf{Case III:} None of the coordinates is equal to $\sigma^{\dummy}$. In this case, $s = (v_1, \dots, v_k)$ for some vertices $v_1, \dots, v_k \in V$. We claim that this must be a $k$-clique. If not, then there exists $v_i, v_j$ for some $i \ne j$ that do not form an edge. However, this implies that $d_H(s, s^{i, j, v_i, v_j}) = \ell - 2$, a contradiction. 
\qedhere
\end{itemize}
\end{proof}

\subsection{W[1]-Hardness of Farthest Spanning Tree}

We are now ready to prove our main theorem of this section (\Cref{thm:fst-w1}).

\begin{proof}[Proof of \Cref{thm:fst-w1}]
We reduce from \dfsprob. Let $(\Sigma, \ell, d, s_1, \dots, s_k)$ be an instance of \dfsprob. We construct the graph $G = (V, E)$ as follows:
\begin{itemize}
\item Let the vertex $V$ consists of $2\ell + 1$ vertices: a hub $u$, and $\ell$ vertices $v_i$ for $i \in [\ell]$.
\item For each $i \in [\ell]$, let there be $|\Sigma|$ parallel edges between $u$ and $v_i$.
\item Assume w.l.o.g. that $\Sigma = \{1, \dots, |\Sigma|\}$. There is a bijection $\psi$ between $\Sigma^\ell$ an spanning trees of $G$: for any $s \in \Sigma^\ell$, $\psi(s)$ corresponds to the spanning tree where the $s_i$-th edge between $u$ and $v_i$ is included.
\item Let $T_1 = \psi(s_1), \dots, T_k = \psi(s_k)$.
\item Finally, let $d$ remains the same as before.
\end{itemize}
The completeness and soundness of the reduction is obvious since $\psi$ preserves\footnote{Namely, for every $s_1, s_2 \in \Sigma^\ell$, we have $d_H(s_1, s_2) = d(\psi(s_1), \psi(s_2))$.} the distance.
\end{proof}

%% file: sections/appendix.tex
\section{Missing Proofs}\label{app:proofs}

\lambdaAsymptotic*
\begin{proof}
First, note that $F$ is continuous, and
\[
    F'(\lambda) = \log\prn{\frac{\frac{1}{4}-\lambda}{1-\lambda}} - \frac{3}{4\lambda} < 0,
\]
since $\lambda \in (0, 1/4)$. Also $F(\lambda) \to +\infty$ as $\lambda \to 0^+$, while $F(1/8) > 0$ and $F(1/4) < 0$. Hence $F$ has a unique root in $(0,1/4)$ which is approximately $\lambda_F \approx 0.13525 > 2/15$.

It remains to show the claimed inequality for all $\ell \geq 4$ and $\lambda < 2/15$. Since the left-hand side is non-increasing in $\lambda$ and the right-hand side is increasing in $\lambda$, it suffices to prove the
inequality at $\lambda^* = 2/15$. Thus, we show that
\[
    \prod_{k = 0}^{j-1} \frac{\floor{(13/15)\ell-k}}{\ceil{(9/7)\ell-k}} \geq \frac{(2\ell/15)^j}{j!}.
\]
Let $N = \floor{13\ell/15}$ and $D = \ceil{9\ell/7}$. Since $k$ is an integer, we have that $\floor{(13/15)\ell-k} = N-k$ and $\ceil{(9/7)\ell-k} = D-k$, and the left-hand side is the falling-factorial ratio
\[
    \prod_{k = 0}^{j-1}\frac{N - k}{D - k}.
\]
The inequality can be computationally verified for $4 \leq \ell \leq 141$, with the minimum ratio of the left-hand side to the right-hand side in this range occurring for $\ell = 8$ and $j = 6$, where it equals approximately $1.058$.

It remains to show the inequality for $\ell\geq 142$, which implies $j\geq 107$. Since $j = \ceil{3\ell/4}$, we have
\[
    \frac{4(j-1)}{3} < \ell \leq \frac{4j}{3}.
\]
The left-hand side is at least
\[
    \prod_{k = 0}^{j-1}\frac{N - k}{D - k} \geq \prod_{k = 0}^{j-1}\frac{\frac{13\ell}{15} - 1 - k}{\frac{9\ell}{7} + 1 - k} \geq \prod_{k = 0}^{j-1}\frac{\frac{52(j-1)}{45} - 1 - k}{\frac{36 j}{21} + 1 - k} \geq \prod_{k = 0}^{j-1}\frac{\frac{52j - 97}{45} - k}{\frac{12 j + 7}{7} - k} = \prod_{k = 0}^{j-1}\frac{A - k}{C - k},
\]
where $A = (52j - 97)/45$ and $C = (12j + 7)/7$. Also, the right-hand side is at most
\[
    \frac{(2\ell/15)^j}{j!} \leq \frac{(8j/45)^j}{j!}.
\]
Thus, it suffices to show that
\begin{equation}\label{eq:appendix-derived-inequality}
     \prod_{k = 0}^{j-1}\frac{A - k}{C - k} \geq \frac{(8j/45)^j}{j!} \iff \log \prod_{k = 0}^{j-1}\frac{A - k}{C - k} \geq \log \frac{(8j/45)^j}{j!}.
\end{equation}
Note that all factors of the product are positive, since $A-j = (7j - 97)/45 > 0$ and $C - j = (5j + 7)/7 > 0$, and that each factor is decreasing in $[0, j]$. Thus, for all $x \in [k, k+1]$ and $k \in [j-1]$,
\[
    \log\frac{A-x}{C-x} \leq \log\frac{A-k}{C-k}.
\]
Integrating both sides from $k$ to $k+1$ and then summing over all $k \in [j-1]$ yields
\[
    F(j) \coloneqq \log\prod_{k = 0}^{j-1} \frac{A(j) - k}{C(j) - k} \geq \int_0^j \log\frac{A-x}{C-x} \dif x.
\]
The right-hand side integral can be evaluated as
\[
    I(j) \coloneqq \int_0^j \log\frac{A-x}{C-x} \dif x = A\log A-(A-j)\log(A-j)-C\log C+(C-j)\log(C-j).
\]
By Stirling's lower bound for the factorial function, we have that $j! \geq \sqrt{2\pi j} (j/e)^j$, or equivalently $\log(j!) \geq (1/2) \log(2 \pi j) + j (\log j - 1)$. Thus, the right-hand side of \eqref{eq:appendix-derived-inequality} is at most
\[
    \log \frac{(8j/45)^j}{j!} \leq j\prn{1+\log\frac{8}{45}} - \frac{1}{2} \log(2\pi j).
\]
It remains to show that, for all $j \geq 107$,
\[
    H(j) \coloneqq I(j) - j \prn{1 + \log\frac{8}{45}} + \frac{1}{2} \log(2\pi j) > 0.
\]
The second derivative of $H$ is
\[
    H''(j) = - \frac{21840j^4+793344j^3+2701535j^2+839244j+461041}{2j^2(5j+7)(7j-97)(12j+7)(52j-97)},
\]
which is negative for $j \geq 107$. Also,
\[
    \lim_{j \to \infty} H'(j) = \log\prn{{13^{52/45}2^{278/315}3^{2/7}5^{5/7}7^{38/45}}{288}} - 1 \approx 0.0194 > 0.
\]
Notice that, since $H''(j) < 0$ and the limiting value of $H'$ is positive, $H'(j) > 0$ for all $j \geq 107$. Finally, we have that $H(107) > 0.010$, which implies that $H(j) > 0$ for every $j \geq 107$, and the claim follows.
\end{proof}

\section{Efficient Sampling}
\label{app:efficient-sampling}

The objective of this Appendix is to provide a proof of  Lemma~\ref{lem:efficient-base-sampling}.

A \emph{coupling} of two probability distributions $\mu$ and $\nu$ is a pair of random variables $(X,Y)$ defined on a single probability space such that the marginal distribution of $X$ is $\mu$ and the marginal distribution of $Y$ is $\nu$.
The following identity relates couplings of $\mu$ and $\nu$ to their total variation distance $d_{TV}(\mu,\nu)$.

\begin{lemma}[Proposition~4.7+Remark~4.8 in~\cite{levin-peres-wilmer}]
\label{lem:coupling-tv}
Let $\mu$ and $\nu$ be probability distributions on a finite set $\Omega$.
Then
\[
d_{\mathrm{TV}}(\mu,\nu)
=
\inf \left\{\PP[X\neq Y] : (X,Y) \text{ is a coupling of } \mu \text{ and } \nu \right\},
\]
and there is a coupling for which equality holds.
\end{lemma}

The previous lemma gives the following estimate for products of distributions, which we will use to analyze the total variation distance of multiple samples.

\begin{lemma}
\label{lem:product-coupling-tv}
Let $\mu_i$ and $\nu_i$ be probability distributions on finite sets $\Omega_i$ for $i\in[L]$.
Then
\[
d_{\mathrm{TV}}\left(\bigotimes_{i=1}^L \mu_i,\bigotimes_{i=1}^L \nu_i\right)
\leq
\sum_{i=1}^L d_{\mathrm{TV}}(\mu_i,\nu_i).
\]
\end{lemma}
\begin{proof}
The proof is a standard consequence of Lemma~\ref{lem:coupling-tv}, we include it for completeness.

By Lemma~\ref{lem:coupling-tv}, for each $i\in[L]$ there is a coupling $(Y_i,Z_i)$ of $\mu_i$ and $\nu_i$ such that $\PP[Y_i\neq Z_i]= d_{\mathrm{TV}}(\mu_i,\nu_i)$.
Take these couplings independently over all $i\in[L]$.
Then $(Y_1,\ldots,Y_L)$ has distribution $\bigotimes_{i=1}^L\mu_i$, while $(Z_1,\ldots,Z_L)$ has distribution $\bigotimes_{i=1}^L\nu_i$.
Applying Lemma~\ref{lem:coupling-tv} once more and then a union bound gives
\[
d_{\mathrm{TV}}\!\left(\bigotimes_{i=1}^L \mu_i,\bigotimes_{i=1}^L \nu_i\right)
\leq
\PP\left[(Y_1,\ldots,Y_L)\neq (Z_1,\ldots,Z_L)\right]
\leq
\sum_{i=1}^L \PP[Y_i\neq Z_i]
\leq
\sum_{i=1}^L d_{\mathrm{TV}}(\mu_i,\nu_i).\qedhere
\]
\end{proof}

We have the following consequence for sampling multiple bases with and without replacement.

\efficientsampling*
\begin{proof}
Let $\mathcal B=\mathcal B(M)$ and let $\nu$ denote the uniform distribution on $\mathcal B$.

We first prove the first assertion.
For each $i\in[L]$, invoke the sampler of Lemma~\ref{lem:single-base-sampling} with accuracy $\delta/L$, independently of all other invocations.
Let $\nu_i$ be the distribution of the $i$th output.
Then $d_{\mathrm{TV}}(\nu_i,\nu)\le \delta/L$ for every $i$.
By Lemma~\ref{lem:product-coupling-tv},
\[
d_{\mathrm{TV}}\!\left(\bigotimes_{i=1}^L \nu_i,\nu^{\otimes L}\right)
\le \sum_{i=1}^L d_{\mathrm{TV}}(\nu_i,\nu)
\le \delta .
\]
Thus the ordered $L$-tuple produced by these independent calls is within total variation distance $\delta$ of $L$ independent uniformly random bases.
The time needed is $L\cdot \poly(n,\log(L/\delta))$.

We now prove the second assertion.
We will use the first assertion as a black box with a larger number of samples.

First we note that, since $M$ has at least $L$ bases, we can deterministically find $L$ distinct bases using $\poly(n,L)$ independence-oracle calls.
Indeed, find one base greedily.
Then perform a breadth-first search in the basis-exchange graph: from a discovered base $B$, test all sets of the form $B-e+f$, where $e\in B$ and $f\in E\setminus B$, and add those that are bases.
Thus, we can find $L$ distinct bases in $\poly(n,L).$

Set $T=\lceil 2L\log(4L/\delta)\rceil$.
Consider first the ideal experiment in which $X_1,\ldots,X_T$ are sampled independently from $\nu$.
If the sequence contains at least $L$ distinct bases, output the first $L$ distinct bases in their order of first appearance.
If not, output the fixed ordered $L$-tuple of distinct bases found by the deterministic subroutine above.

Let $\mathcal E$ be the event that $X_1,\ldots,X_T$ contain at least $L$ distinct bases.
Conditioned on $\mathcal E$, the ordered tuple of first appearances is uniformly distributed over all ordered $L$-tuples of distinct bases.
This follows by symmetry: the distribution of the sequence $(X_1,\ldots,X_T)$ and the event $\mathcal E$ are invariant under every permutation of the set $\mathcal B$.

It remains to bound $\PP[\mathcal E^c]$.
Write $N=|\mathcal B|$, so $N\ge L$.
After $s<L$ distinct bases have appeared, the probability that the next sample is new is
\[
1-\frac{s}{N}\ge 1-\frac{s}{L}.
\]
Therefore,
\[
\PP[\mathcal E^c]
\le
L\left(1-\frac1L\right)^T
\le
L e^{-T/L}
\le
\frac{\delta}{2}.
\]
Let $\mu_{\neq}$ denote the uniform distribution over ordered $L$-tuples of distinct bases.
The ideal experiment therefore outputs a distribution within total variation distance at most $\delta/2$ of $\mu_{\neq}$.

We now implement the ideal experiment approximately.
By the first assertion, applied with $T$ in place of $L$ and accuracy $\delta/2$, we can generate an ordered $T$-tuple of bases whose joint distribution is within total variation distance $\delta/2$ of $\nu^{\otimes T}$ using $T\cdot\poly\bigl(n,\log(T/\delta)\bigr)\leq\poly\bigl(n,L,\log(1/\delta)\bigr)$
time.
Apply to this $T$-tuple the same deterministic post-processing rule: output the first $L$ distinct bases if they exist, and otherwise output the fixed fallback tuple.
Note that this changes the output distribution by at most another $\delta/2$.
Thus, the implemented distribution is within total variation distance at most $\delta$ of $\mu_{\neq}$.

The fallback to $L$ fixed bases uses only $\poly(n,L)$ additional time, so the total time is $\poly(n,L,\log(1/\delta))$.
\end{proof}

%% file: sections/acknowledgments.tex
\section*{Acknowledgments}
This work was supported by the following grants:
\begin{itemize}
    \item Mayank Goswami and GiBeom Park: Supported by NSF grant CCF-2503086
    \item Vasilis Livanos: Supported in part by the Air Force Office of Scientific Research under award number FA9550-24-1-0261.
    \item Arturo Merino: Supported by ANID FONDECYT Iniciación No. 11251528.
    \item Adarsh Srinivasan: Supported by NSF CCF award 2313372 and NSF CAREER award 2443697.
\end{itemize}